\documentclass[11pt]{article}

\usepackage{amsmath,amssymb}
\usepackage{calc}
\usepackage{jheppub} 

\usepackage{bm,amsmath,amssymb,slashed,graphicx,%
            enumerate,alltt,xspace,multirow,xcolor,mathrsfs}
\usepackage{fancyvrb}
\usepackage{booktabs}
\usepackage{graphicx}
\usepackage{subcaption}
\usepackage{xspace} 
\usepackage[utf8]{inputenc}
\usepackage{xfrac}
\usepackage{float}
\usepackage{placeins}
\usepackage{url}
\usepackage{hyperref}

\usepackage{color}
\definecolor{darkgreen}{rgb}{0,0.5,0}
\definecolor{darkblue}{rgb}{0,0,0.7}
\definecolor{darkred}{rgb}{0.5,0,0.0}
\definecolor{darkorange}{rgb}{0.8,0.4,0.0}

\newcommand{\Ym}{Y$_{\text{m}}$-Splitter\ }
\newcommand{\Cf}{\text{C}_{\text{F}}}
\newcommand{\sd}{\text{d}}

\newcommand{\zetacut}{\zeta_{\text{cut}}}
\newcommand{\rhomin}{\rho_{\text{min}}}

\makeatletter
\g@addto@macro\bfseries{\boldmath}
\makeatother
\title{\textbf{Investigating top tagging with \Ym and N-subjettiness.}}
\author{Mrinal Dasgupta and Jack Helliwell}
\affiliation{Lancaster-Manchester-Sheffield Consortium for Fundamental Physics, School of Physics
  \& Astronomy, University of Manchester, Manchester M13 9PL, United Kingdom}
\emailAdd{mrinal.dasgupta@manchester.ac.uk, jack.helliwell@manchester.ac.uk}


\abstract{We study top-tagging from an analytical QCD perspective focussing on the role of two key steps therein : a step to find three-pronged substructure and a step that places constraints on radiation. For the former we use a recently introduced modification of Y-Splitter, known as \Ym, and for the latter we use the well-known N-subjettiness variable. We derive resummed results for this combination of variables for both signal jets and background jets, also including pre-grooming of the jet. Our results give new insight into the performance of top tagging tools in particular with regard to the role of the distinct steps involved.}

\begin{document}
\maketitle
\section{Introduction}\label{sec:intro}
The past decade has seen the emergence of jet substructure as one of the key areas of LHC phenomenology {\cite{Seymour:1993mx, Butterworth:2002tt, Butterworth:2008iy,
  Ellis:2009me, Ellis:2009su, Krohn:2009th, Abdesselam:2010pt, Altheimer:2012mn, Altheimer:2013yza,Adams:2015hiv,Larkoski:2017jix}}. The main impetus driving this emergence has come from studies involving boosted heavy particles, after pioneering work in the context of Higgs searches revealed the clear potential of substructure based analyses \cite{Butterworth:2008iy}. Following early studies, a rapid proliferation of tools and methods followed, mainly aiming at enhancing the discriminating power of substructure methods in various contexts including methods for tagging of two-pronged (e.g. W/Z/H decays) and three-pronged (i.e. top decay) jet substructure \cite{Butterworth:2002tt, Butterworth:2008iy, Ellis:2009me, Ellis:2009su, Krohn:2009th, Kaplan:2008ie,
  Plehn:2009rk,CMS:2009lxa,CMS:2014fya,CMS-PAS-JME-13-007,CMS-PAS-JME-09-001,Brooijmans:2008zza,Thaler:2008ju} as well as for quark-gluon discrimination \cite{Gras:2017jty,Frye:2017yrw,Larkoski:2019nwj}. For in-depth reviews of these topics and further references we refer the reader to the review articles Refs.~\cite{Marzani:2019hun,Larkoski:2017jix,Kogler:2018hem}.

While the rapid development of substructure methods often resulted in novel powerful techniques, many of which are currently still in use, some key questions also emerged about the robustness of the methods being employed. Such questions were concerned, for instance, with the accuracy to which Monte Carlo event generators provide a reliable description of substructure observables and about the dependence of tagger performance on poorly understood physics aspects like non-perturbative effects in QCD. This led to a parallel effort to better understand jet substructure as relevant to tagging and grooming of jets originating from boosted particles, from the first principles of QCD theory \cite{Dasgupta:2013ihk, Dasgupta:2013via,Larkoski:2013eya,Larkoski:2014wba,Larkoski:2015kga,Dasgupta:2015lxh,Dasgupta:2015yua,Dasgupta:2016ktv,Salam:2016yht,Larkoski:2017iuy,Larkoski:2017cqq,Dasgupta:2018emf,Napoletano:2018ohv}.  As a consequence it was possible to identify flaws in existing tools \cite{Dasgupta:2013ihk, Dasgupta:2013via}, design superior tools which remove some of the main flaws thus identified \cite{Dasgupta:2013ihk,Larkoski:2014wba} , and shed light on the factors that influence performance including the role of non-perturbative effects \cite{Dasgupta:2013ihk, Dasgupta:2015yua,Dasgupta:2016ktv,Salam:2016yht,Dreyer:2018nbf,Dreyer:2020brq}. \footnote{In this context a notable feature that has often emerged in a variety of contexts is the presence of a trade-off between performance and resilience to non-perturbative effects \cite{Dasgupta:2016ktv,Dreyer:2018nbf,Dreyer:2020brq}}

The more recent advent of machine-learning (ML) tools to study jets has also yielded impressive performance gains with ML based taggers shown to often significantly outperform standard (``QCD-based'') tagging algorithms \cite{deOliveira:2015xxd, Baldi:2016fql, Barnard:2016qma,Komiske:2017aww,Kasieczka:2017nvn, Butter:2017cot, Macaluso:2018tck, Moreno:2019bmu, Qu:2019gqs, Dreyer:2020brq,
Lim:2020igi}. Nevertheless the questions raised for earlier tagging methods in terms of exclusive reliance on parton showers to study performance and the issue of performance gains originating in non-perturbative effects remain in the ML case, and are indeed potentially re-enforced. Here one can mention studies that have investigated the resilience of Lund-plane based ML \cite{Dreyer:2018nbf,Dreyer:2020brq} against non-perturbative effects, finding that eliminating the non-perturbative region results in a marked decrease in performance. Furthermore new research on parton showers has revealed flaws in the \emph{perturbative structure} of dipole showers including a failure to reproduce the QCD double emission matrix-element for soft emissions strongly ordered in angle \cite{Dasgupta:2018nvj,Dasgupta:2020fwr}, in principle a crucial regime for meaningfully describing jet substructure.

Given all of the above, it therefore remains of importance to continue to develop the program of understanding jet substructure taggers via perturbative QCD. While much success has been obtained in analytic understanding of the impact of taggers and groomers on signal and background for two-pronged decays, there is a more limited understanding of top tagging which is a somewhat more complicated problem owing in part to the coloured parton initiating the signal jet. In terms of tools, various methods for finding three-pronged jet substructure have been introduced in the literature including the early ATLAS top tagger \cite{Brooijmans:2008zza} based on Y-splitter, as well as the CMS top tagger \cite{CMS:2009lxa,CMS:2014fya,CMS-PAS-JME-09-001} conceptually related to the mMDT/Soft Drop procedure\cite{Dasgupta:2013ihk,Larkoski:2014wba}.  Other widely used methods for top tagging include the Johns Hopkins top tagger \cite{Kaplan:2008ie} and the HEP top tagger \cite{Plehn:2009rk}, shower deconstruction \cite{Soper:2012pb} and template tagging  \cite{Almeida:2010pa}. 

Amongst methods aiming at constraining radiation around three hard prongs, the N-subjettiness ratio $\tau_{32}$ \cite{Thaler:2010tr} and Energy Correlation Function ratios \cite{Larkoski:2014zma} have been actively studied. Combinations of these tools with grooming have also been studied and exploited in experimental analyses. For example Refs.~\cite{Aad:2014xra,ATLAS:2015nkq} makes use of trimmed jets with a top tagging procedure involving a combination of Y-splitter and the N-subjettiness ratios $\tau_{32}$ and $\tau_{21}$ while Ref.~\cite{CMS:2016tvk} reports, amongst other studies, combinations of the CMS top tagger with a $\tau_{32}$ cut. Such combinations are similar in essence to the combinations we shall study in the present particle, though various details differ.
 
A first analytical study of the impact of prong-finding methods for top tagging supplemented with grooming, in the high $p_T$ limit i.e. with $p_T$ in the TeV range, was carried out in Ref.~\cite{Dasgupta:2018emf}.  This work included the study of IRC safe extensions of the IRC unsafe CMS top tagger as well as studying an adaptation of the Y-splitter method, \Ym. 

In this article we extend the work of Ref.~\cite{Dasgupta:2018emf} by combining prong-finding with \Ym, with an additional radiation constraint coming from a $\tau_{32}$ cut. Further, we account for the impact of pre-grooming with Soft Drop (SD) and mMDT. We begin with a set of Monte Carlo studies that motivate the use of this particular combination of methods as well as indicate optimal values for the $\tau_{32}$ cut,  $\tau \sim 0.2$. 
Next we obtain resummed analytic results for QCD background jets for \Ym with the $\tau_{32}$ cut, in the small $\tau$ limit. These results are obtained in a modified leading-logarithmic approximation where other than capturing all leading-logarithmic (LL) terms, one also retains important classes of next-to-leading-logarithmic (NLL) terms such as those from hard-collinear emission. Following the treatment of Ref.~\cite{Napoletano:2018ohv} we then extend our results to include finite $\tau$ effects which are in general non-negligible even for our typical value of $\tau \sim 0.2$. We study both the un-groomed case as well as consider pre-grooming with Soft Drop ($\beta=2$) and the mMDT. Despite this rather complex combination of methods leading to a highly non-trivial observable, we find that our results are in broad agreement with those from parton shower studies, with remaining moderate differences consistent with the expected size of omitted (beyond LL) terms.

Next we study signal jets on a similar footing. We start with a simple situation with only a mass window cut and compare the resulting Sudakov form factor to results from Pythia, finding excellent agreement. This step is useful in order to test the validity of our simplifying assumptions about radiation in a top initiated jet. We then extend our studies to include \Ym in addition to the mass window cut, also accounting for pre-grooming using both mMDT and SD ($\beta=2$). Our results here improve upon previous work by accounting for a previously neglected situation where one of the prongs found by \Ym can be a soft gluon rather than one of the top decay products. The inclusion of this correction term brings our results for the signal into substantially better agreement with Pythia simulations, than was seen in previous studies \cite{Dasgupta:2018emf}. We then account for the  impact of $\tau$ cut in the signal case. Although our treatment of finite $\tau$ corrections for the signal is not as accurate as the corresponding treatment for the QCD background, we obtain a good description of the $\tau$ dependence of the result especially for the un-groomed case and for pre-grooming with SD($\beta=2$). 

The layout of this paper is as follows: We start in section \ref{sec:deffs} by recalling the definitions of the \Ym tagger, N-subjettiness including our choice of axes, and Soft Drop grooming. In section \ref{sec:pheno} we  report results from initial Monte Carlo studies which help lay the ground for our subsequent analytical investigation. In section \ref{sec:background} we carry out our calculations for \Ym with a $\tau_{32}$ cut for QCD background jets. Here we discuss in detail the small $\tau$ limit as well as accounting for finite $\tau$ effects, studying both the differential distribution and the cumulant. We close this section by including grooming with both mMDT and SD ($\beta=2$) and comparing our results to those from Herwig and Pythia showers.  Section \ref{sec:signal} is devoted to signal jets where we first study the effect of a mass window cut alone followed by studies of \Ym including grooming and finally the inclusion of a $\tau_{32}$ cut. Section \ref{sec:mass} discusses our analytical results in terms of the understanding gained for the performance of the \Ym, $\tau_{32}$ and grooming combinations in terms of the interplay between the $\tau$ and mass window cuts,  and reports further comparisons to parton showers. Our conclusions are summarised in section \ref{sec:conclusions}.

\section{Tagger definitions}\label{sec:deffs}
The primary step involved in top-tagging is the identification of three-pronged jet substructure that characterises top-decay. There are various methods that have been suggested in the literature for the identification of three-pronged substructure within a fat jet, some of which have also been used for phenomenology. Examples of prong finding methods include the early CMS and ATLAS top taggers \cite{CMS-PAS-JME-09-001,ATL-PHYS-PUB-2009-081,Brooijmans:1077731} and \Ym, an adaptation of Y-splitter introduced in Ref.~\cite{Dasgupta:2018emf} which we shall use for our analytical studies here. Additionally jet shape variables such as N-subjettiness \cite{Thaler:2010tr}, which we also use here, are known to be powerful methods that quantify the N-pronged nature of a jet through placing constraints on radiation from N identified prongs within a fat  jet. Techniques combining prong-finding methods with jet shape variables are also known to give rise to important performance gains, have been used in experimental studies \cite{Aad:2014xra,ATLAS:2015nkq, CMS:2016tvk, CMS-PAS-JME-15-002} and motivate our desire to better understand such combinations. We define in more detail below  all the specific methods that we use in this article.
\begin{enumerate}
\item \Ym

The \Ym method for top tagging \cite{Dasgupta:2018emf} takes a jet clustered with the gen-$k_{t}(p=\frac{1}{2})$ algorithm (referred to as gen-$k_t$ from here on) and performs the following steps:
\begin{enumerate}
\item Undo the last clustering, to give two sub-jets, both of which are examined for the condition $p_{t,i}>\zeta p_{t,\text{jet}}$. If either sub-jet fails the $\zeta$ condition, the jet is rejected.
\item Check which sub-jet produces the larger gen-$k_{t}$ distance when de-clustered, and undo the last clustering of this sub-jet. Check whether the resulting sub-jets from this de-clustering pass the $\zeta$ condition. If either the de-clustering or the $\zeta$ condition fail, the jet is rejected.
\item Find the pairwise masses of the three final sub-jets, and require that $\min(m_{12},m_{13},m_{23})>m_{\text{min}}$. If this condition is not met, the jet is rejected.
\end{enumerate}

\item N-subjettiness

We will use the N-subjettiness ratio variable $\tau_{32}=\frac{\tau_{3}}{\tau_{2}}$, where
\begin{equation}
 \tau_{N}^{(\beta)}=\frac{1}{p_{t,\text{jet}}R^{\beta}}\sum_{i \in \text{jet}}p_{t,i}\min\left((\Delta R_{1,i})^{\beta},(\Delta R_{2,i})^{\beta},...,(\Delta R_{N,i})^{\beta}\right),
\end{equation}
with the sum over $i$ running over the jet constituents, $\Delta R_{ij}=\sqrt{(\Delta y_{ij})^{2}+(\Delta \phi_{ij})^{2}}$, and the $N$ partition axes are labelled $1  \cdots N$. There are various options for defining the partition axes, for instance finding the axes which minimise $\tau_{N}$ (optimal axes). Throughout this work we use $\beta=2$ as this facilitates our analytical studies and we make use of the gen-$k_t$ axes with $p=1/2$. These axes are obtained by clustering the jet with the gen-$k_t(p=\frac{1}{2})$ algorithm and identifying the axes with the $N$ exclusive sub-jets resulting from $N-1$ de-clusterings. For $\tau_{2}$ with $\beta=2$, these have been shown to be very close to the optimal axes \cite{Napoletano:2018ohv}. For $\tau_{3}$ these axes are exactly the three prongs returned by \Ym which is helpful in facilitating the resummation of the tagged fraction of events.

\item Soft Drop

Soft Drop takes a jet, re-clusters it with the Cambridge/Aachen (C/A) algorithm \cite{Wobisch:1998wt,Dokshitzer:1997in} and performs the following steps:
\begin{enumerate}
\item Undo the last clustering, to give two sub-jets.
\item Examine the lower $p_T$ sub-jet for the condition $p_{t,i}>z_{\text{cut}}(\frac{\Delta R}{R})^{\beta} (p_{t,i}+p_{t,j})$.
\item If this condition is not met this sub-jet is removed from the jet and the groomer goes back to step (a). If it is met, the groomer stops and this is the final jet.
\end{enumerate}
Throughout this work we set $z_{\text{cut}}=\zeta$, the \Ym parameter.
\end{enumerate}

\section{Monte-Carlo study}\label{sec:pheno}
In this section we investigate the performance of various tagging procedures based around the N-subjettiness variable $\tau_{32}$ as well as how they are impacted by hadronisation, ISR, and MPI. The tagging procedures considered all have the restriction that the jet mass is between $160$\,GeV and $225$\,GeV, corresponding to a window around the top mass, and are studied as a function of the cut on $\tau_{32}$. After examining N-subjettiness cuts with and without pre-grooming we combine these cuts with the \Ym  method which we again investigate with and without pre-grooming. Two pre-grooming options are considered, SD $(\beta=2)$ and mMDT (equivalent to SD with $\beta=0$).

\begin{figure}[h]
\centering
\begin{subfigure}[t]{0.45\textwidth}
\centering
\includegraphics[width=1.1\textwidth]{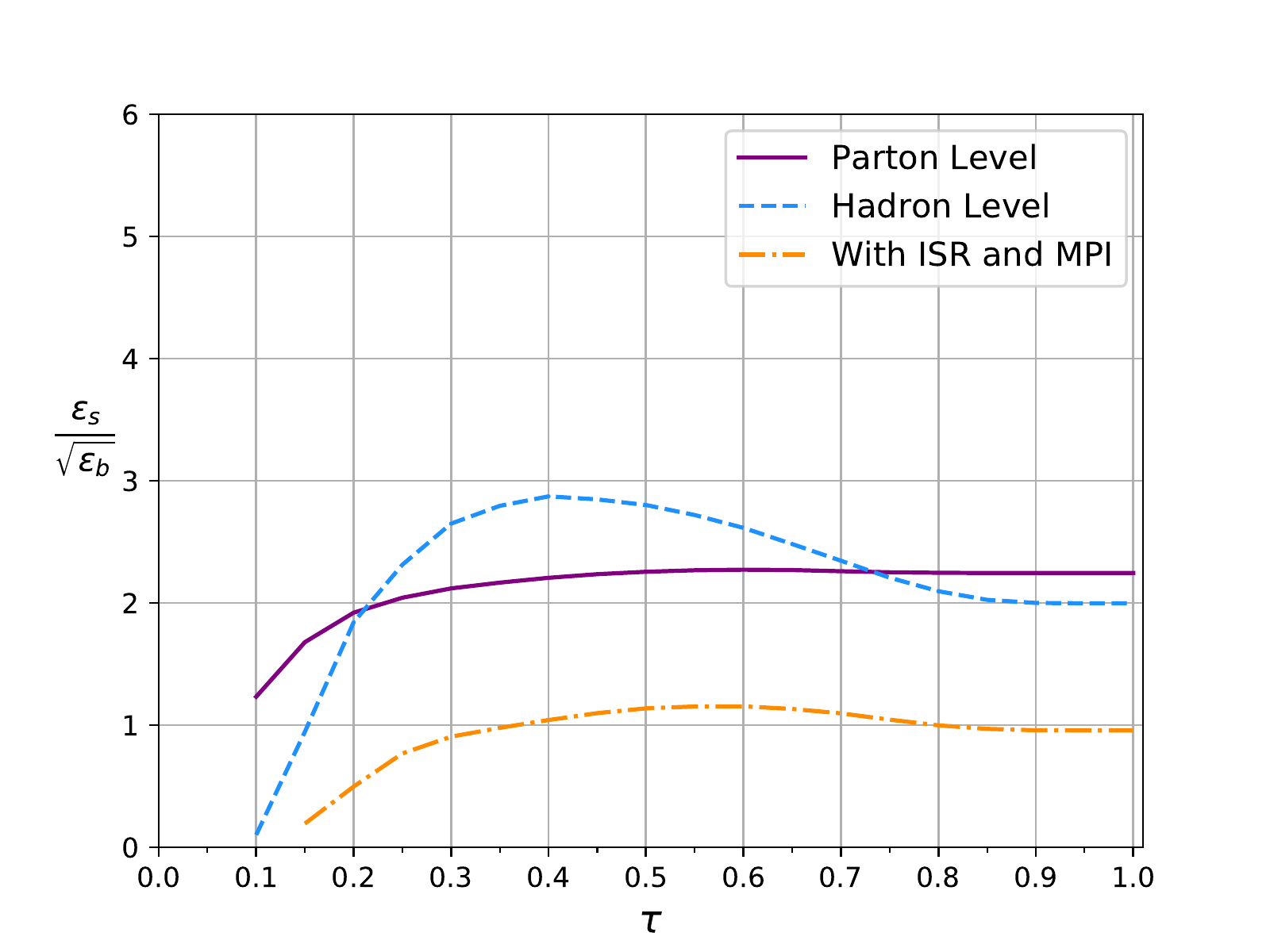}
\caption{Cut on $\tau_{32}$}
 \label{fig:tISR}
\end{subfigure}%
~
\begin{subfigure}[t]{0.45\textwidth}
\centering
\includegraphics[width=1.1\textwidth,]{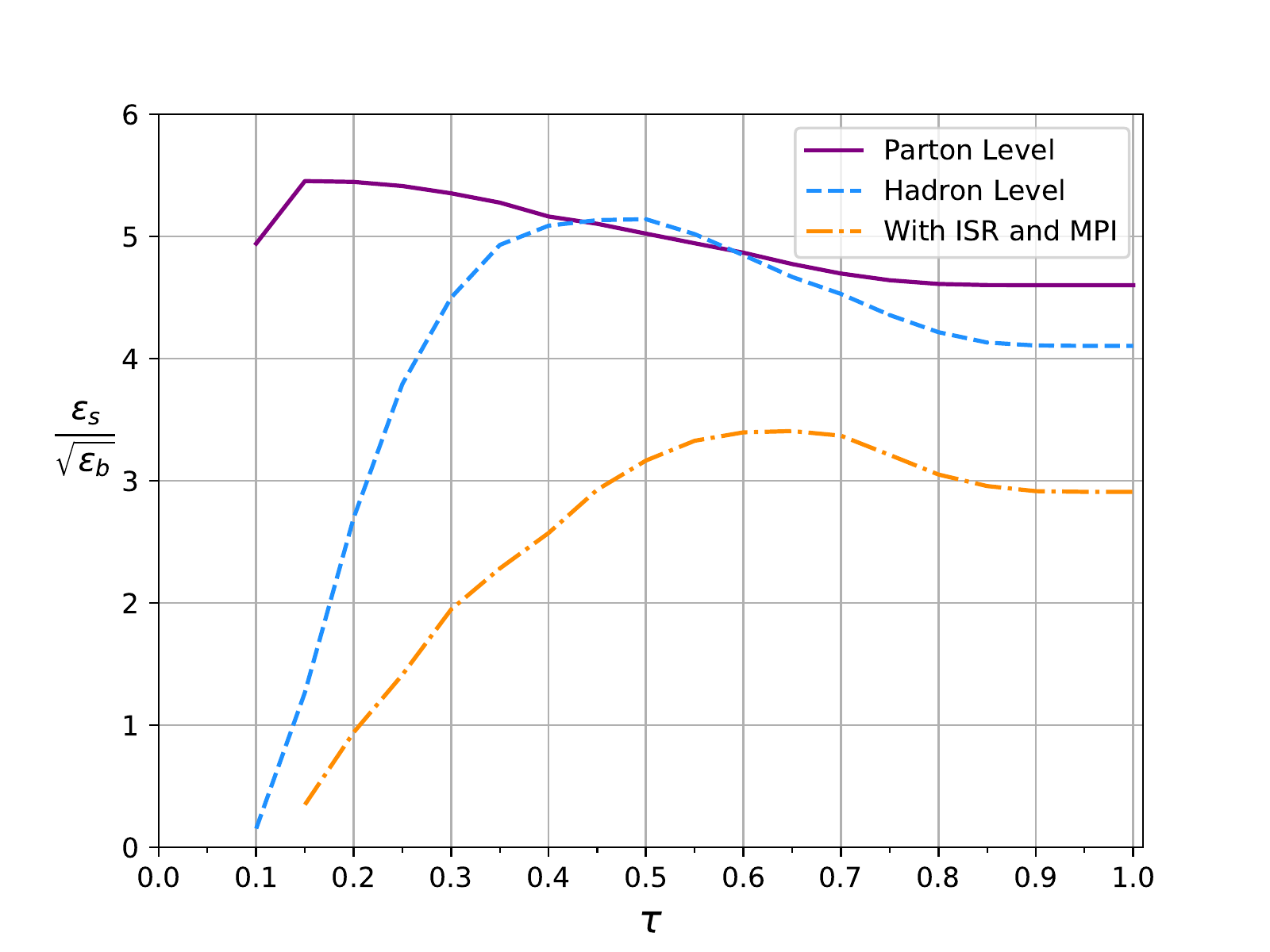}
\caption{Cut on $\tau_{32}$ after application of \Ym.}
 \label{fig:tMmdtISR}
\end{subfigure}
\begin{subfigure}[t]{0.45\textwidth}
\centering
\includegraphics[width=1.1\textwidth,]{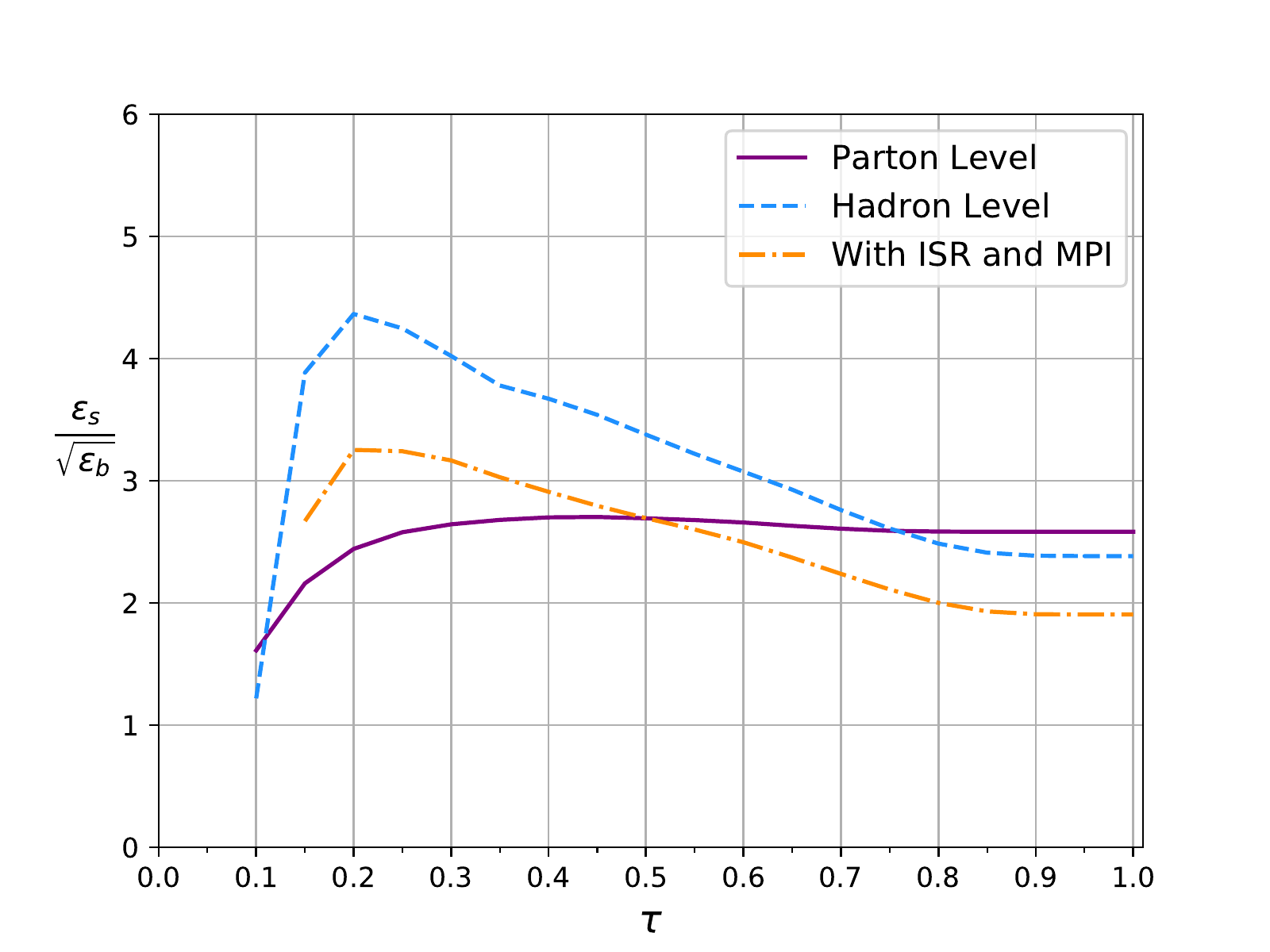}
\caption{Cut on $\tau_{32}$ after application of Soft Drop ($\beta=2$).}
 \label{fig:SDnoym}
\end{subfigure}%
~
\begin{subfigure}[t]{0.45\textwidth}
\centering
\includegraphics[width=1.1\textwidth]{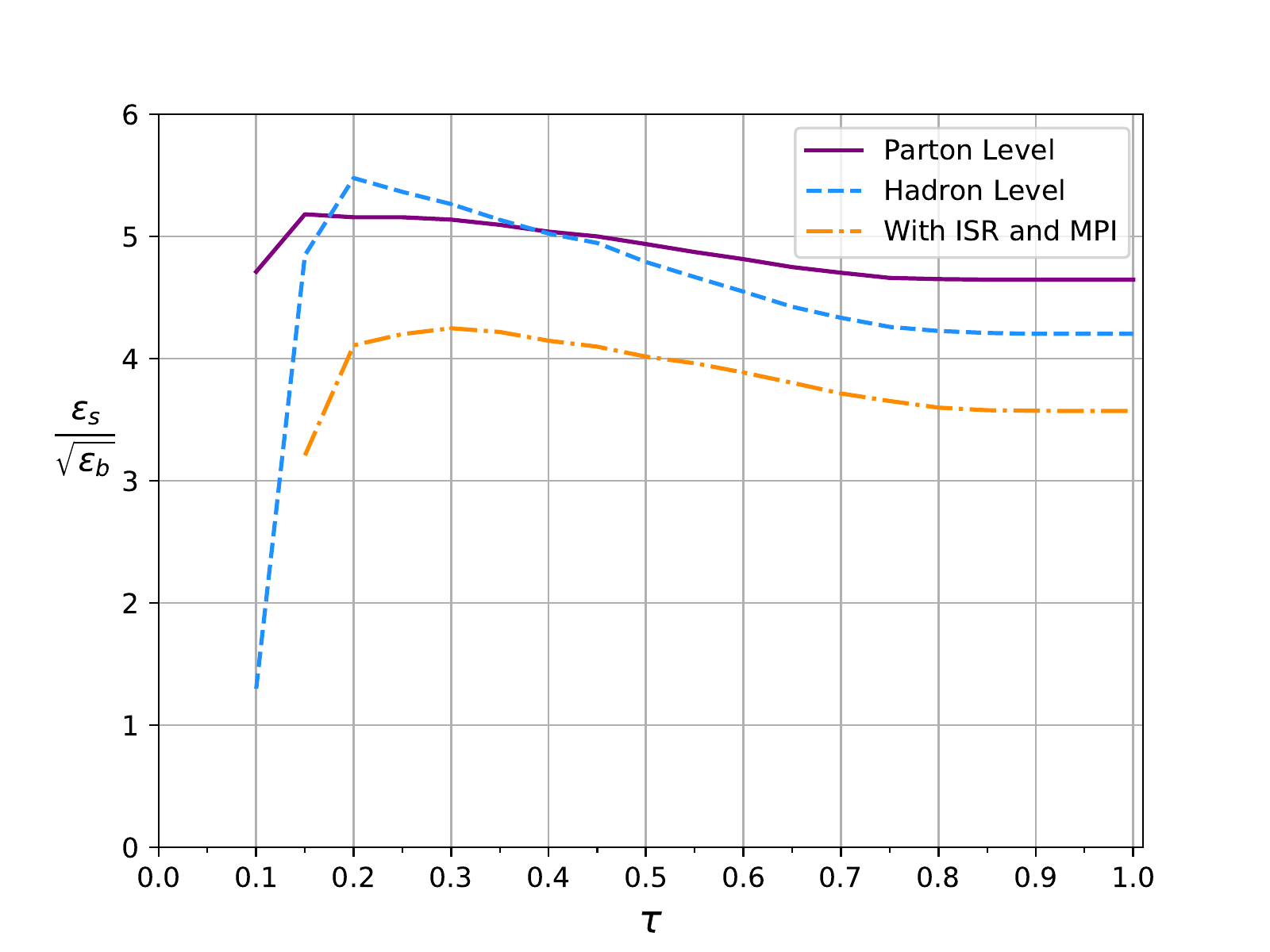}
\caption{Cut on $\tau_{32}$ after application of \Ym and Soft Drop ($\beta=2$).}
 \label{fig:SDym}
\end{subfigure}
\begin{subfigure}[t]{0.45\textwidth}
\centering
\includegraphics[width=1.1\textwidth]{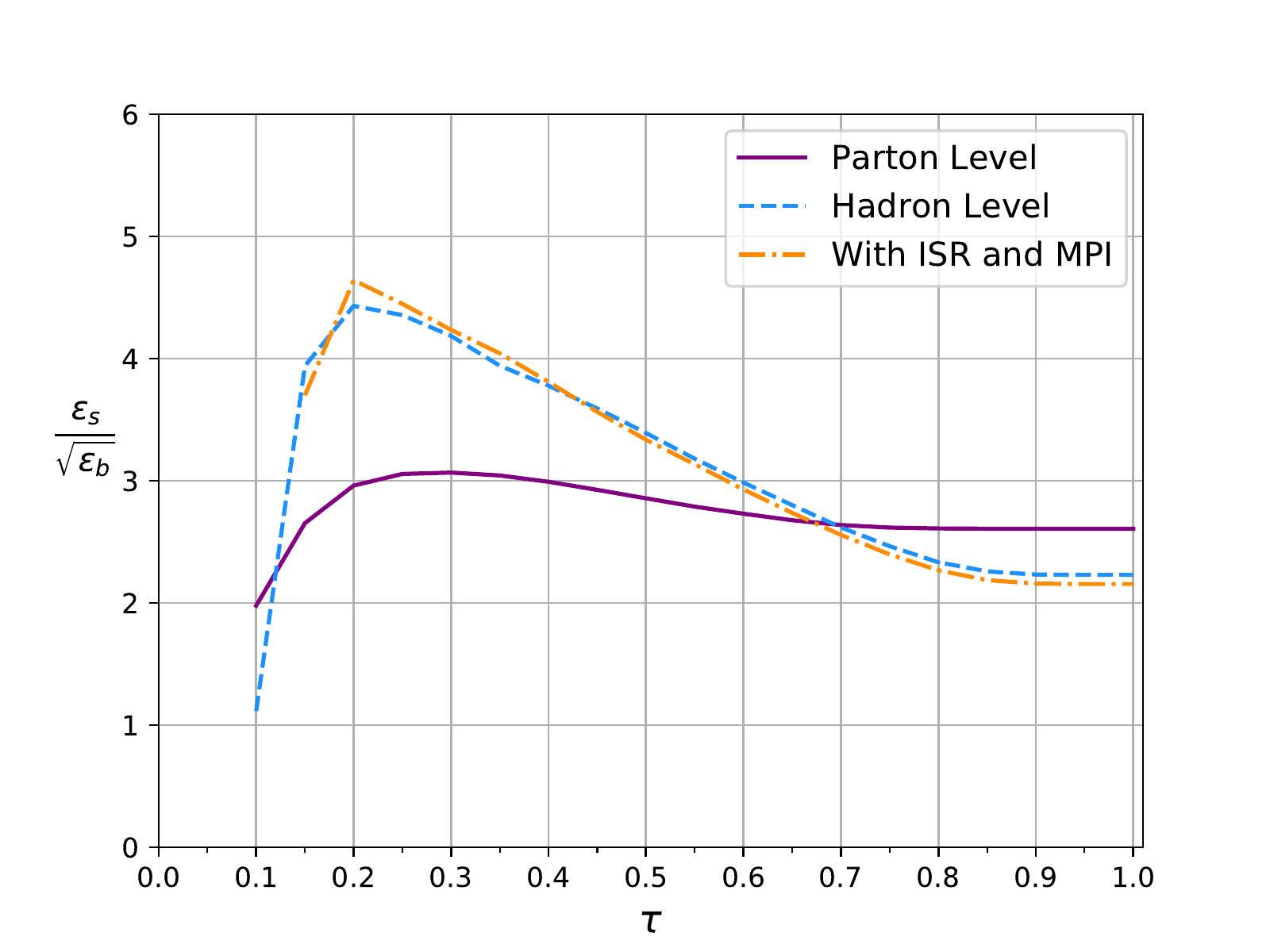}
\caption{Cut on $\tau_{32}$ after application of mMDT.}
\label{fig:SBtaummdt}
\end{subfigure}%
~
\begin{subfigure}[t]{0.45\textwidth}
\centering
\includegraphics[width=1.1\textwidth]{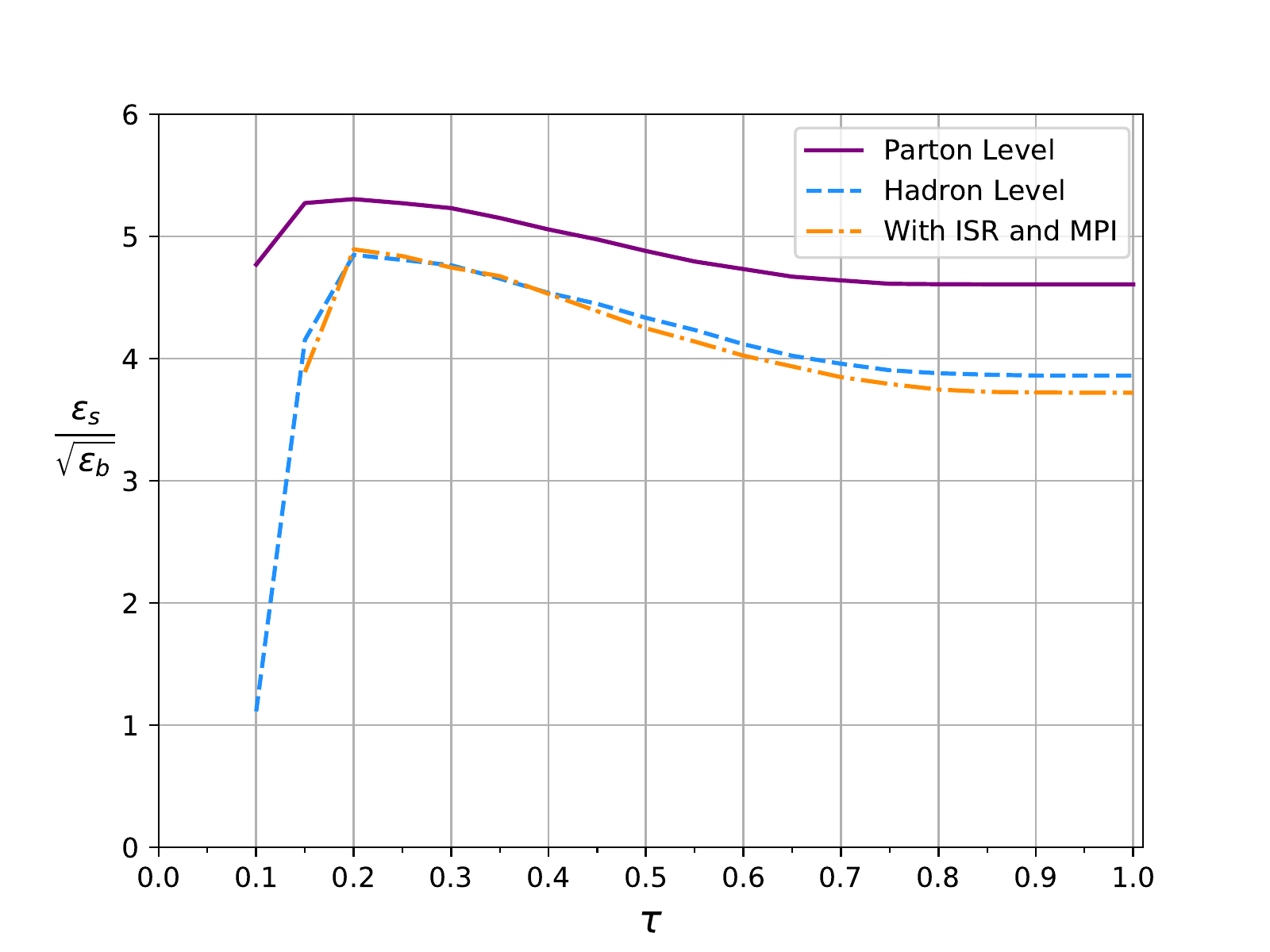}
\caption{Cut on $\tau_{32}$ after application of \Ym and mMDT.}
\label{fig:qMmdtISR}
\end{subfigure}
\caption{Plots showing the signal to square-root background of the four variants of the tagging procedure at parton level, hadron level, and with ISR and MPI activated at hadron level.}
\label{fig:StoB}
\end{figure}

We start by generating $1$ million $t\overline{t}$ and $q\overline{q}$ events with Pythia \footnote{We have studied the impact of including gluon jets and found that in the $p_t$ range under consideration they do not significantly modify the distributions presented.}. ISR, MPI and hadronisation were initially deactivated, and a generation cut of $p_{t}>1600$ GeV was applied. Jets were clustered with the Cambridge/Aachen algorithm with $R=1$ and $p_{t ,\text{min}}=2$\,TeV using Fastjet 3 \cite{Cacciari:2011ma}, as was the case for the studies in Ref.~\cite{Dasgupta:2018emf}.
\FloatBarrier
Where jets are groomed we use $z_{\text{cut}}=0.05$. $\tau_{32}$ is calculated using the N-subjettiness fastjet contrib \cite{Thaler:2011gf}, and where \Ym is used we choose $m_{\text{min}}=50$ \, GeV and $\zeta=0.05$. This information is then used to construct the tagged fraction of events and the signal to square-root--background as a function of a cut on $\tau_{32}$. The same procedure is used both with only hadronisation, and then hadronisation, ISR and MPI activated to assess their impact.\\

To discuss the features that emerge from our Monte Carlo studies let us first examine the top row of Fig.~\ref{fig:StoB}, i.e. Figs.~ \ref{fig:tISR} and \ref{fig:tMmdtISR}, which show the signal significance as a function of the $\tau_{32}$ cut, $\tau$, without any grooming and without \Ym on the left and with \Ym on the right. It is clear that in the absence of a grooming step ISR and MPI significantly damage performance in each case, although the inclusion of \Ym results in a higher signal significance after all effects are considered.

Next we come to the plots involving the application of grooming i.e. Figs. \ref{fig:SDnoym} and \ref{fig:SDym} for SD ($\beta=2$) pre-grooming and  Figs. \ref{fig:SBtaummdt} and \ref{fig:qMmdtISR} in the  bottom row for the mMDT. From these one notes that grooming, especially with mMDT, is an effective method to significantly mitigate ISR and MPI. When combining grooming with \Ym we observe that both hadronisation and ISR+MPI are significantly reduced, resulting in high performance with an optimal value of $\tau \sim 0.2$ emerging for mMDT pre-grooming and $\tau \sim 0.3$ for SD ($\beta=2$). The best performance, i.e. highest signal significance, comes with mMDT pre-grooming and \Ym applied in addition to the $\tau$ cut, as shown in Fig.~\ref{fig:qMmdtISR}. This combination is also more resilient to ISR and all non-perturbative effects at the same time. In contrast although pre-grooming jets and cutting on $\tau_{32}$ without \Ym (see figure \ref{fig:SBtaummdt}) gives good performance at hadron level, the discrepancy with the parton level result indicates that the performance of this procedure cannot necessarily be understood from perturbative QCD arguments alone and may be more susceptible to mis-modelling of non-perturbative effects in parton showers \footnote{A possible reason for this might be that a pure $\tau_{32}$ cut is not IRC safe and is instead only Sudakov safe \cite{Larkoski:2013paa,Larkoski:2015lea} while the application of \Ym prior to the subjettiness cut prevents $\tau_2$ from vanishing, resulting in an IRC safe quantity.}. 

In summary, applying \Ym to pre-groomed jets with cuts on $\tau_{32}$ and the jet mass is a high performing method for tagging hadronically decaying high-$p_T
$ top quarks.\footnote{We find that, for comparable signal significance, these methods appear, in the high $p_T
$ region, to outperform the dense neural net and boosted decision tree used by ATLAS in \cite{Aaboud:2018psm}, although it should be noted that the two studies are perhaps not equivalent, as no attempt was made here to examine detector effects, which were included in the ATLAS study.}
The performance is also well described by parton level predictions and is therefore reasonably robust against effects which are less well theoretically understood in this context. These observations provide some of the main motivation for detailed theoretical studies using perturbative QCD, which will be the subject of the next two sections.

\section{\Ym splitter with a $\tau_{32}$ cut : QCD jets} \label{sec:background}
We start by examining the impact of a $\tau_{32}$  cut on QCD jets after applying \Ym. Analytical studies for  \Ym as applied to top-tagging, with and without pre-grooming, have already been carried out in Ref.~\cite{Dasgupta:2018emf}. These studies derived  results for the jet mass distribution and consequently the efficiency for QCD jets tagged with \Ym using the technique of QCD resummation. Resummation is required in order to address the multi-scale nature of the problem. Crucially the highly boosted limit implies that the invariant jet mass $m^2 \ll p_T^2$, with $m^2 \sim m_t^2$ and $p_T$  values in the TeV  range, which leads to large logarithms in $\rho = m^2/R^2p_T^2$. A good description of the jet-mass distribution then requires resummation of the logarithms in $\rho$. Additionally for \Ym we have $\rho_{\mathrm{min}} = m_\text{min}^2/p_T^2  R^2 \ll 1$ and a further small scale $\zeta p_T$, the minimum energy of an emission that passes the $\zeta$ condition, with $\zeta \ll 1$. Large logarithms are then expected and do arise in $\rho$, $\rho_{\mathrm{min}}$, $\zeta$ and in $\rho_{\text{min}}/\rho$. In Ref.~\cite{Dasgupta:2018emf} a modified leading logarithmic resummation was performed which included all double-logarithmic terms and a subset of single-logarithmic terms such as those arising from hard-collinear emissions. The logarithms that are most crucial to resum are those in the smallest parameters $\rho$ and $\rho_{\text{min}}$. Typical values of $\zeta \sim 0.05$ and $\rho_{\text{min}} /\rho \sim m_W^2/m^2_{\text{top}}$ are larger and hence we only aim to retain logarithms in these parameters at leading double-logarithmic accuracy.

Here, relative to previous work \cite{Dasgupta:2018emf} we shall additionally include the $\tau_{32}$ cut, considering the possibility that $\tau_{32}$ is not small. In doing so we shall follow closely the treatment of Ref.~\cite{Napoletano:2018ohv} for resummation of jet mass with a $\tau_{21}$ cut.

\subsection{Leading-order result}\label{sec:LO}

We start with the leading-order result, computed in the soft and collinear approximation which yields the leading logarithmic terms. For \Ym this starts at order $\alpha_s^2$ for QCD jets, since one requires at least two emissions within the jet (i.e. at least three partons) in order to be accepted by \Ym. Since for three partons $\tau_{3}$ vanishes, a cut requiring $\tau_{32} < \tau$ is trivially satisfied. Therefore the leading-order result is unchanged from the pure \Ym case of  Ref.~\cite{Dasgupta:2018emf}. For the case of a quark initiated jet and in the abelian $C_F^2$ channel it is given by \footnote{We shall define all our angles to correspond to the actual angles rescaled by the jet radius $R$.}

\begin{multline}\label{eq:LLbasic}
\frac{1}{\sigma}\left(\frac{d\sigma}{d\rho}\right)^{\mathrm{LO, soft-collinear}} =  \bar{\alpha}^2 \int \frac{dz_1}{z_1}
\frac{dz_2}{z_2} \frac{d\theta_1^2}{\theta_1^2}
\frac{d\theta_2^2}{\theta_2^2}   \times\Theta(\theta_2^2<\theta_1^2<1)\, \delta(\rho-\text{max}(z_1 \theta_1^2, z_2 \theta_2^2))  \\
\Theta(z_1>\zetacut)\,\Theta(z_2>\zetacut)\, \Theta( \text{min}(z_2 \theta_2^2,z_1z_2\theta_1^2) > \rho_{\mathrm{min}} ),
\end{multline}
where we defined $\bar{\alpha} = \frac{C_F \alpha_s}{\pi}$, taking for definiteness the case of a quark initiated jet. In deriving the above result we have taken a strongly-ordered in angle configuration with $\theta_2 \ll \theta_1$, made a leading logarithmic approximation that the jet mass is dominated by the emission that makes the larger contribution, and imposed the tagger conditions by requiring both emissions to pass the 
$\zetacut$ and implemented the  $\rho_{\mathrm{min}}$ condition in the strongly-ordered limit where $\theta_{12}  \sim \theta_1$. 

We then obtain:
\begin{align}\label{eq:cmsll}
  \frac{\rho}{\sigma}\left(\frac{d\sigma}{d\rho}\right)^{\mathrm{LO,soft-collinear}}
  & \overset{\frac{\rho_\text{min}}{\rho}<\zetacut}{=}
    \bar{\alpha}^2 \ln^2 \frac{1}{\zetacut} \ln \frac{\rho}{\rhomin},\\
  & \overset{\frac{\rho_\text{min}}{\rho}>\zetacut}{=}
    \bar{\alpha}^2 \ln^2\frac{\rho}{\rhomin}
    \left(\frac{3}{2}\ln \frac{1}{\zetacut} -\frac{1}{2} \ln\frac{\rho}{\rhomin}\right).\nonumber
\end{align}
A similar result is obtained for the $C_F C_A$ colour factor while in the $C_F  T_R n_f$ channel the result is one logarithm down due to the lack of a soft enhancement in the $p_{qg}$ splitting function. For future convenience we note that the leading-order result can also be expressed in terms of the highest-mass emission $\rho_a$ and the next-highest--mass emission $\rho_b$. Written in these terms we have 
\begin{multline}\label{eq:LLbasicAlt}
\frac{1}{\sigma}\left(\frac{d\sigma}{d\rho}\right)^{\mathrm{LO, soft-collinear}} =  \bar{\alpha}^2 \int \frac{dz_a}{z_a}
\frac{dz_b}{z_b} \frac{d\rho_a}{\rho_a}
\frac{d\rho_b}{\rho_b}  \,  \delta(\rho-\rho_a)  \Theta \left(\rho_a>\rho_b\right)\Theta(z_a>\zetacut)\,\Theta(z_b>\zetacut)\\
\times \Theta( \text{min}\left \{\rho_b, z_a z_b \max(\theta_a^2,\theta_b^2) \} > \rho_{\mathrm{min}} \right),
\end{multline}
where in the  $\rho_{\mathrm{min}}$ condition we used strong angular ordering to replace $\theta^2_{ab}$ by $\max(\theta_a^2,\theta_b^2)$. Finally, we note that beyond double logarithmic accuracy a more precise result at order $\alpha_s^2$ can be achieved by considering three collinear partons within a jet without imposing strong ordering between the partons. Such configurations are described by triple-collinear splitting functions and calculations implementing the triple-collinear result were included in the studies of \Ym carried out in Ref.~\cite{Dasgupta:2018emf}.

\subsection{Resummed results}
Now we turn to the resummed result. We first consider the case where $\tau_{32}<\tau \ll 1$. Then we shall lift the requirement that $\tau \ll 1$ i.e. we shall account for finite $\tau$ effects. 
\subsubsection{The small $\tau$ limit}\label{sec:smalltau}
For the case  of \Ym one  considers, as in Ref.~\cite{Dasgupta:2018emf}, two real emissions that pass the tagger cuts accompanied by an ensemble of soft and collinear emissions which are constrained to set a smaller gen-$k_t$ distance (i.e. mass) than either of the  two leading emissions. This constraint on real emissions produces a Sudakov form factor. In the current case the emissions are additionally constrained by the $\tau$ cut. Here we shall derive the Sudakov form factor at leading-logarithmic (LL) accuracy, capturing all double-logarithmic terms including those in $\tau$ and running coupling effects, and also include some important single logarithmic effects such as accounting for hard-collinear radiation.\\

For the two emissions accounted for at leading-order, Eq.~\eqref{eq:cmsll},  we shall again label $\rho_a$ as the emission that sets the larger mass and $\rho_b$ the smaller mass. Consider first all  subsequent {\emph{primary}} emissions,  i.e. emissions from the hard parton initiating the jet. These emissions must not give rise to larger mass (gen-$k_t$) values than the first two emissions de-clustered by \Ym  \emph{and} they must set  a value of $\tau_{32}<\tau$. Recall that the contribution of an emission $i$ to $\tau_N$ is given by  $z_i\min(\theta_{i1}^2,...,\theta_{iN}^2)$. As was the case for $\tau_2$ \cite{Dasgupta:2015lxh}, the limit of strong angular-ordering ensures that, for emissions coming from a leg lying along one of the N-subjettiness axes, the smallest of the  $\theta_{ia}$ angles is either the angle between the emission and its emitter, or can be approximated by this angle to LL accuracy. For a primary emission this implies that the contribution to $\tau_3$, $\tau_{3i}= z_i  \theta_i^2$ where $z_i$ is the energy fraction and $\theta_i$ is  the angle of the emission wrt the hard initial parton. The  value of $\tau_2$ on the other hand  is dominated, to LL accuracy, by the second highest mass emission $\rho_b$, due to the strong ordering in masses relevant at LL accuracy. \footnote{For going beyond the small $\tau$ limit and including finite $\tau$ effects we shall, in the next  subsection, lift this requirement of strong ordering in masses as was done in Ref.~\cite{Napoletano:2018ohv}.} The condition on primary emissions then reads:
\begin{equation}
\Theta \left(\frac{\sum_{i=3}^{\infty} z_i\theta_i^2}{\rho_b}<\tau\right) \prod_{i=3}^{\infty}\Theta\left(z_i\theta_i^2<\rho_b \right ).
\end{equation}
The first step function reflects the condition on $\tau_{32}$ while the second condition reflects the constraint on mass which gives the primary emission Sudakov form factor for \Ym in Ref.~\cite{Dasgupta:2018emf}, i.e. that none of the emissions $i$ have a gen-$k_t$ distance larger than $\rho_b$ by assumption. Since $\tau <1$ the second condition is automatically satisfied and the condition on primary emissions is just given by the stronger constraint $\Theta \left(\sum_{i=3}^{\infty} z_i\theta_i^2 < \rho_b \tau\right)$. The primary emission Sudakov factor then arises from a veto on any emissions violating this condition. More precisely it takes the form $S=e^{-R^{(\text{primary)}}}$
with the ``radiator" $R^{\left(\text{primary} \right)}$ given by
\begin{equation}
R^{\text{(primary)}} \left(\tau \rho_b  \right)= \frac{C_R}{2\pi} \int\alpha_s(z^2\theta^2 R^2p_T^2)p(z) \sd z\frac{\sd \theta^2}{\theta^2}\Theta(z\theta^2>\rho_b \tau),
\end{equation}
where $C_R$ is a colour factor that depends on the identity of the initiating jet i.e. $C_F$ for a quark and $C_A$ for a gluon jet, and $p(z)$ is the QCD splitting function describing collinear emission from a quark ($p(z) =p_{gq}(z)$) or gluon $(p(z)=p_{gg}(z))$. For the argument of the running coupling we have used the $k_t$ of the emission (in terms of $z$ and $\theta$) as required in the soft and collinear limit.

As well as vetoing primary emissions from the parton initiating the jet, the overall Sudakov factor must also account for a veto on secondary emissions which would set a value of $\tau_{32}$ larger than $\tau$ from either of the two emissions included in the leading order pre-factor. In the case of soft secondary emissions the angle of emission $\theta_i$ is limited by angular-ordering to be less than the angle of the parent $\theta_a$ or $\theta_b$. Apart from this constraint, for emissions off parton $a$ we  have the same constraint as for primary emissions and hence we obtain, for emissions off parton  a:
\begin{equation}
\label{eq:secondaryymtau}
R^{\text{(secondary,a)}}\left(\tau,\rho_a,\rho_b  \right) = \frac{C_A}{2\pi} \int\alpha_s(z^2 z_a^2 \theta^2 R^2p_T^2)p_{gg}(z)\sd z\frac{\sd \theta^2}{\theta^2}\Theta(z z_a \theta^2>\rho_b \tau)\Theta(\theta^2<\theta_a^2),
\end{equation}
where we note that $z$ represents the energy  fraction of parton $a$'s energy carried by the soft secondary emission. We also note that for secondary emissions, the gen-$k_t$ distance, entering the veto condition above, differs from the mass even in the soft limit, involving one less factor of $z_a$. A similar equation gives the result for emissions off parton $b$, with the obvious replacement  of $z_a$ and $\theta_a$ by $z_b$ and $\theta_b$.

The overall  result can be written as a Sudakov form factor weighting the leading-order, order $\alpha_s^2$ result which serves as a pre-factor.  For simplicity if one retains just the leading-logarithmic expression for the pre-factor reported in Eq.~\eqref{eq:LLbasicAlt}, we can obtain the resummed result by inserting the factor $S=e^{-R}$ in  the integrand  in Eq.~\eqref{eq:LLbasicAlt} where $R =  R^{\text{(primary)}}+R^{\text{(secondary,a)}}+R^{\text{(secondary,b)}}$.  While our results include the full running coupling and hard-collinear effects we report below a simplified result for the Sudakov factor $S$ in the limit of a fixed coupling and retaining only the soft-collinear behaviour i.e. replacing $p(z)$ and $p_{gg}(z)$ by the soft limit expression $2/z$:
\begin{equation}
\label{eq:LLfixed-coupling}
 S^{\text{(fixed-coupling, soft)}} = \exp \left[-\frac{C_R  \alpha_s}{2\pi} \ln^2\frac{1}{\tau \rho_b} -\frac{C_A \alpha_s}{2\pi} \ln^2\frac{\rho_a}{\tau \rho_b} -\frac{C_A \alpha_s}{2\pi} \ln^2 \frac{1}{\tau}\right],
\end{equation}
where  the term involving $\ln^2 1/\rho \tau_b$ comes from primary emissions, the term involving $ \ln^2\frac{\rho_a}{\tau \rho_b} $ comes from vetoing emissions from emission $a$ and finally the suppression involving just $\ln^2 1/\tau$ comes from vetoing emissions from emission $b$. The difference between primary and secondary emissions arises entirely from angular-ordering and the ensuing limitation on emission angle we mentioned previously. Although the logarithms present in $S$ are written above in terms of $\rho_a$ and $\rho_b$, these will eventually be related to logarithms of $\rho$, $\rho_{\text{min}}$ and ratios thereof once the integrals in the pre-factor a carried out.  In the limit $\tau \to 1$ of the above result we obtain the pure \Ym result of Ref.~\cite{Dasgupta:2018emf}.

A couple of further remarks are in order concerning the result in Eq.~\eqref{eq:LLfixed-coupling}. First of all the result captures leading double logarithms in $\tau$ in addition to the logarithms involved in the resummation of plain \Ym \cite{Dasgupta:2018emf}. Including hard-collinear emission via using the full splitting functions, rather than just their soft limit, and using the running coupling helps to improve the result beyond double-logarithmic accuracy. The result indicates that the effect of the  N-subjettiness cut is to produce an extra suppression relative to the case of \Ym \cite{Dasgupta:2018emf} just by changing the scale $\rho_b$ to the smaller scale $\tau \rho_b$ and the extra secondary suppression factor we get from emissions off parton $b$. This suppression of the background is of course desirable but choosing a small $\tau$ value potentially also suppresses the signal, which in this case is a coloured particle namely the top quark. Also as is well-known from several prior applications \cite{Thaler:2011gf,Thaler:2010tr} and additionally emerges in the Monte Carlo studies we reported in section \ref{sec:pheno}, optimal values of $\tau$ do not necessarily satisfy $\tau \ll1$, so that finite $\tau$ effects generally need to be considered \cite{Napoletano:2018ohv} in addition to the resummation of logarithms of $\tau$. The inclusion of finite $\tau$ corrections is thus the topic of the next subsection.

\subsection{Finite $\tau$ corrections}\label{sec:finite_tau}

To obtain an insight into the role of the $\tau_{32}$ cut in a phenomenological context, one has to address values of $\tau \sim 1$. From the viewpoint of resummation this has implications identical to those first pointed out in the $\tau_{21}$ case \cite{Napoletano:2018ohv}. The small $\tau$ limit resummation of the previous subsection is designed to fully capture double logarithmic terms of the form $\alpha_s^n L^{2n}$ where, for power counting purposes, we use the symbol $L^{2n}$ to denote double logarithms in any of $\ln \rho$, $\ln \rho_{\mathrm{min}}$, $\ln \frac{\rho}{\rho_{\text{min}}}$, $\ln \tau$ or any combination of them. From the fixed-coupling Sudakov form factor, Eq.~\eqref{eq:LLfixed-coupling}, written in terms of $\rho_a, \rho_b$ and $\tau$ we note that we obtain terms that are single logarithmic in jet masses (and jet mass ratios) but double logarithmic overall due to the role of $\ln \tau$ i.e. terms of the form $\alpha_s \ln \rho_b \ln \tau$ and $\alpha_s \ln \frac{\rho_a}{\rho_b} \ln \tau.$ Beyond the small $\tau$ limit we need to account for such terms beyond just their $\ln \tau$ dependence i.e. obtain the full function $f_{\tau}$ that multiplies single logarithms in jet mass. However, given that single logarithms in jet mass ratios i.e. $\alpha_s \ln \rho_b/\rho_a$ are smaller, we do not attempt to obtain the finite $\tau$ corrections for such terms which is substantially more involved and accordingly retain their small $\tau$ form only. In the Sudakov form factor with inclusion of running coupling, we therefore wish to control terms of the form $\alpha_s^n L_\rho^n f_n(\tau)$ (where $L_\rho$ generically  denotes logarithms in jet masses but not ratios), while the small $\tau$ resummation accounts only for terms that approximate $f_n(\tau)$ by its leading small $\tau$ behaviour  $\sim \ln^{n} \tau$.

While in the small $\tau$ limit resummation of the previous subsection we assumed that emissions were strongly ordered in terms of their contribution to the jet mass (in addition to strong ordering in angle), in order to achieve resummation of terms $\alpha_s^n L_\rho^n$ with their accompanying $\tau$ dependence we no longer assume strong ordering in jet masses. In particular we assumed that $\tau_2$ was set by a single emission $b$, and hence its value was taken to be $\rho_b$.  Beyond the small $\tau$ limit, we must account for the fact that $\tau_2$ receives a contribution from all emissions in the jet except emission $a$. Since we still desire terms that  are at least single logarithmic in jet masses, we continue to assume that emissions are strongly ordered in angle. This approximation of emissions ordered in angle but not in mass is the same as was made in the case of $\tau_{21}$ studies for two-pronged decays \cite{Napoletano:2018ohv}, to obtain the finite $\tau$ correction to the small $\tau$ results \cite{Dasgupta:2015lxh}.

We can then write
\begin{equation}
\tau_{32} = \frac{\rho-\rho_a-\rho_b}{\rho-\rho_a},
\end{equation}
where the numerator comes from $\tau_3$ being given by the sum of jet masses contributed by all emissions except $a$ and $b$, while the denominator is $\tau_2$ which is given by the sum of jet masses contributed by all emissions except emission $a$. The sum of all emissions' contributions to jet mass, including those of $a$ and $b$, just gives the total jet mass $\rho$.

\subsubsection{Differential distribution in $\tau$}\label{sec:tau_differential}
We begin by presenting a result for the joint distribution in jet mass $\rho$ and $\tau$ i.e. the quantity $\frac{\rho \tau}{\sigma}  \frac{d^2  \sigma}{d\rho d\tau}$, where $\tau$ is a set value of $\tau_{32}$. To begin with we shall consider primary emissions only, since it is straightforward to account for secondary emissions in the final result.

We first write the result for the cross-section differential in both $\tau$ and $\rho$, which accounts for the two emissions $a$ and $b$ included in the leading-order formula but now accompanied by an infinite number of additional emissions which are strongly ordered in angle. The strong ordering in angle ensures that these emissions are emitted independently from the hard initial parton which leads to the standard factorised formula for any number of emissions:
\begin{multline}\label{eq:finiteTauStart}
\frac{\rho \tau}{\sigma} \frac{\sd^2\sigma}{\sd \rho\sd\tau}=\bar{\alpha}^2 \int_\zeta^1\frac{\sd z_a}{z_a}\int_0^1 \frac{\sd\rho_a}{\rho_a}\int_\zeta^1\frac{\sd z_b}{z_b}\int_0^{\rho_a} \frac{\sd\rho_b}{\rho_b}\Theta(\min(\rho_b,z_az_b\max(\frac{\rho_a}{z_a},\frac{\rho_b}{z_b}))>\rho_{\text{min}}) \\ \exp\left[-\int_0^1 R'(\rho')\frac{\sd \rho'}{\rho'}\right] \sum_{p=1}^{\infty}\frac{1}{p!}\prod_{i=1}^p\int_{0}^{\rho_b}R'(\rho_i)\frac{\sd \rho_i}{\rho_i} \rho\delta(\rho-\rho_a-\rho_b-\sum_{i\neq a,b}\rho_i)\, \tau \delta\left(\tau-1+\frac{\rho_b}{\rho-\rho_a}\right).
\end{multline}
The above result is written using a fixed-coupling approximation for the emission of partons $a$ and $b$ though we shall account for the running of the coupling, with the $k_t$ of those emissions, in the pre-factor for our final results. It involves considering $p$ factorised real emissions, with a sum over all $p$, alongside a sum over all virtual corrections included via the exponential form factor. The factor $R'$, appearing  in both real and virtual terms above, stems from the integral over the emission probability for a single emission in the soft and collinear limit, at a fixed mass $\rho$ :
\begin{equation}
\label{eq:rprime}
R'{(\rho)} = C_R\int_0^1 \frac{\sd \theta^2}{\theta^2} \sd z \ p(z) \frac{\alpha_s(z\theta p_T R)}{2\pi} \rho\delta(z\theta^2-\rho) \overset{\text{f.c.}}{=} \frac{C_R \alpha_s}{\pi} \left(\ln \frac{1}{\rho}+B_i \right),
\end{equation}
where the RHS of the above equation gives the fixed-coupling result and we have replaced the splitting functions $p(z)$ by their soft piece $\propto 1/z$ and incorporated the effect of hard-collinear emissions by introduction of the $B_i$ terms, corresponding to inclusion of the hard-collinear piece of the splitting function to our accuracy. For quark and gluon jets respectively $B_q=\sfrac{-3}{4}$ and $B_g=\sfrac{(-11C_A+4n_f T_R)}{(12C_A)}$.

Integrating over $\rho_b$ in Eq.~\eqref{eq:finiteTauStart} using the delta function constraint involving $\tau$ allows us to set $\rho_b = \left(1-\tau \right )(\rho-\rho_a)$ which then leads to the result, assuming $\tau <1/2$,

\begin{multline}\label{eq:finiteTauTwo}
\frac{\rho \tau}{\sigma} \frac{\sd^2\sigma}{\sd \rho\sd\tau}=\bar{\alpha}^2 \frac{\tau}{1-\tau} \int_\zeta^1\frac{\sd z_a}{z_a} \int_\zeta^1\frac{\sd z_b}{z_b}  \int_0^\rho \frac{\sd\rho_a}{\rho_a} \Theta \left(\rho_a > \frac{1-\tau}{2-\tau}  \rho \right) \Theta_{\rho_\text{min}} \left(\rho_a,\tau,  \rho_{\text{min}} ,z_a,z_b\right) \\ \exp\left[-\int_0^1 R'(\rho')\frac{\sd \rho'}{\rho'}\right] \sum_{p=1}^{\infty}\frac{1}{p!}\prod_{i=1}^p\int_{0}^{\rho_b}R'(\rho_i)\frac{\sd \rho_i}{\rho_i} \, \rho\delta\left((\rho-\rho_a ) \tau -\sum_{i\neq a,b}\rho_i  \right),
\end{multline}
where we have used the shorthand notation $\Theta_{\rho_\text{min}}$ to denote the $\rho_{\min}$ condition and the $\rho_b$ that occurs as an upper limit in the $\rho_i$ integral is understood to be the value set by the delta function i.e $(\rho-\rho_a)(1-\tau)$. The other step function constraint on $\rho_a$ derives from the condition that $\rho_a >  \rho_b$ and the value of $\rho_b$ set  by the delta function integral we have performed. Finally we observe that we are left to evaluate the multiple emission contribution where the sum over the $\rho_i$ are constrained to be equal to $(\rho-\rho_a) \tau$. Additionally each emission $i$ is constrained so that $\rho_i < \rho_b=(1-\tau)(\rho-\rho_a)$, however for $\tau <1/2$ this condition is automatically met if the stronger condition on the sum of $\rho_i$ is satisfied. In what follows we restrict our attention to $\tau <1/2$ as this region is sufficient given the optimal value of $\tau$ that emerged from the Monte Carlo studies in section \ref{sec:pheno}. Finally we note that one can evaluate the multiple emission contribution on the second line of Eq.~\ref{eq:finiteTauStart} simply by using known results for the standard jet mass \cite{CATANI19933}, since the constraint on multiple emissions is the same as for the plain jet mass $\rho$ but with $\rho$ replaced by $(\rho-\rho_a)\tau$. Hence without needing to perform any further explicit calculation one can write:
\begin{multline}
\label{eq:taudist}
\frac{\rho \tau}{\sigma} \frac{\sd^2\sigma}{\sd \rho\sd\tau}\overset{\tau<1/2}{=} \bar{\alpha}^2 \frac{1}{1-\tau} \int_\zeta^1\frac{\sd z_a}{z_a} \int_\zeta^1\frac{\sd z_b}{z_b}   \int_0^\rho \frac{\sd\rho_a}{\rho_a} \frac{\rho}{\rho-\rho_a} \Theta \left(\rho_a > \frac{1-\tau}{2-\tau}  \rho \right) \Theta_{\rho_\text{min}} \left(\rho_a,\tau,  \rho_{\text{min}} ,z_a,z_b\right) \\  R'((\rho-\rho_a)\tau)\frac{\exp[-R((\rho-\rho_a)\tau)-\gamma_E R'((\rho-\rho_a)\tau)]}{\Gamma[1+R'((\rho-\rho_a)\tau)]}.
\end{multline}
The above result accounts for configurations where the three prongs tagged by \Ym are the hard parton which initiates the jet along with two gluons emitted independently from it, however, our final result also contains configurations where a gluon emitted from the hard parton branches, with the resulting three particles corresponding to the three \Ym prongs. 

For the region $\tau>1/2$ one could in principle follow the same method as outlined for the $\tau_{21}$ calculation in Ref.~\cite{Napoletano:2018ohv}, though given our immediate motivation we do not consider this region further in the present study. We could also have initially integrated over $\rho_a$ instead of 
$\rho_b$ to obtain the result in the equivalent form
\begin{multline}
\label{eq:taudistRhob}
\frac{\rho \tau}{\sigma} \frac{\sd^2\sigma}{\sd \rho\sd\tau}\overset{\tau<1/2}{=} \bar{\alpha}^2 \frac{1}{1-\tau} \int_\zeta^1\frac{\sd z_a}{z_a} \int_\zeta^1\frac{\sd z_b}{z_b}   \int_0^{\frac{\rho}{2}} \frac{\sd\rho_b}{\rho_b}  \frac{\rho}{\rho-\frac{\rho_b}{1-\tau}}  \Theta_{\rho_\text{min}} \left(\rho_b,\tau,  \rho_{\text{min}} ,z_a,z_b\right) \\  R'(\rho_b\frac{\tau}{1-\tau})\frac{\exp[-R(\rho_b\frac{\tau}{1-\tau})-\gamma_E R'(\rho_b\frac{\tau}{1-\tau})]}{\Gamma[1+R'(\rho_b\frac{\tau}{1-\tau})]}.
\end{multline}
 A key feature of our results is the presence of an overall $1/(1-\tau)$ factor as was also the case in the $\tau_{21}$ result of Ref.~\cite{Napoletano:2018ohv}.  Taking the small $\tau$ limit of Eq.~\eqref{eq:taudist} we should return to the small $\tau$ result we derived in the previous subsection, which is indeed the case up to subleading terms in the order $\bar{\alpha}^2$ pre-factor. To be more precise, in the previous subsection we had evaluated the pre-factor taking $\rho_a$ to dominate the jet mass, by using the condition $\delta(\rho-\rho_a) \Theta(\rho_a >\rho_b)$, which correctly captures all double logarithmic terms in the pre-factor. If instead one uses the more accurate condition $\delta(\rho-\rho_a-\rho_b) \Theta(\rho_a >\rho_b)$, then after integration over $\rho_b$ we obtain the same result as the small $\tau$ limit of Eq.~\eqref{eq:taudist}. Relative to the  strong ordering of emissions $\rho_a$ and $\rho_b$, using the exact jet mass conditions affects only single logarithmic terms $\bar{\alpha}^2 L^2$ in the pre-factor where  $L$  denotes logarithms in $\rho/\rho_{\min}$ or $\zeta$. Such terms are two logarithms below the leading $\bar{\alpha}^2 L^4$ terms in the pre-factor and only involve more modest logarithms than those in the jet mass. We can therefore consider such terms as negligible and hence the  small $\tau$ limit of  Eq.~\eqref{eq:taudist} is equivalent to the result of the previous subsection. For an explicit calculation demonstrating the argument above, we refer the reader to Appendix A.

Beyond the small $\tau$ limit the most crucial feature of the result is the overall $1/(1-\tau)$  factor in Eq.~\eqref{eq:taudist}. While there is additionally a $\tau$  dependence in the step functions in Eq.~\eqref{eq:taudist}, that is again responsible for introducing $\tau$ dependent terms of order $\bar{\alpha}^2 L^2$ in the pre-factor, and hence can be neglected to our accuracy as illustrated in Appendix A. In what follows we shall use the freedom to set $\tau$ to zero in the step function conditions to obtain an analytic form for the cumulative distribution.

\subsubsection{Cumulative distribution}\label{sec:tau_Cumulative}
It is of direct interest to also obtain the result for the differential distribution in the jet mass with a cut on $\tau_{32}$, $\tau_{32}<\tau$ rather than fixing a value for $\tau_{32}$ as required for the double differential distribution above. In order to do this one can integrate the differential distribution, Eq.~\eqref{eq:taudist}, between $0$ and $\tau$. Setting $\tau$ to zero in the step functions of the pre-factor, which has only a sub-leading impact as discussed before, we can write 
\begin{multline}
\left.\rho \frac{d\sigma}{d \rho}\right |_{\tau_{32}<\tau} \overset{\tau<1/2}{=} \bar{\alpha}^2  \int_\zeta^1\frac{\sd z_a}{z_a} \int_\zeta^1\frac{\sd z_b}{z_b}   \int_0^\rho \frac{\sd\rho_a}{\rho_a} \frac{\rho}{\rho-\rho_a} \Theta \left(\rho_a > \frac {\rho}{2} \right) \Theta_{\rho_\text{min}} \left(\rho_a,\rho_{\text{min}} ,z_a,z_b\right) \\ \int_0^\tau \frac{d\tau'}{\tau^{'}(1-\tau')} R'((\rho-\rho_a)\tau')\frac{\exp[-R((\rho-\rho_a)\tau')-\gamma_E R'((\rho-\rho_a)\tau')]}{\Gamma[1+R'((\rho-\rho_a)\tau')]},
\end{multline}
which corresponds to integrating the distribution up to some maximum value $\tau$ for $\tau_{32}$. To single-logarithmic accuracy, we can expand the radiator about some point $\tau_0$ to write:
\begin{equation}
R((\rho-\rho_a)\tau')\simeq R((\rho-\rho_a)\tau_0)-R'((\rho-\rho_a)\tau_0)\ln\left(\frac{\tau'}{\tau_0} \right) +\mathcal{O} \left(R^{''} \right),
\end{equation}
where $R^{\prime}(x) = -\frac{\partial R}{\partial \ln x}$.
With $\tau_0$ chosen such that in the small $\tau$ limit $\tau_0$ is of order $\tau$, and  given that the integral is dominated by values $\tau^{\prime} \sim \tau$, terms of order $R^{''}$ and beyond can be neglected as they are beyond single-logarithmic accuracy and we may replace $\tau'$ by $\tau_0$ in the $R^{\prime}$ terms to obtain:

\begin{multline}
\label{eq:tauint}
\left.\rho \frac{d\sigma}{d \rho}\right |_{\tau_{32}<\tau} \overset{\tau<1/2}{=} \bar{\alpha}^2  \int_\zeta^1\frac{\sd z_a}{z_a} \int_\zeta^1\frac{\sd z_b}{z_b}   \int_0^\rho \frac{\sd\rho_a}{\rho_a} \frac{\rho}{\rho-\rho_a} \Theta \left(\rho_a > \frac {\rho}{2} \right) \Theta_{\rho_\text{min}} \left(\rho_a,\rho_{\text{min}} ,z_a,z_b\right) \\ R'((\rho-\rho_a)\tau_0)\frac{\exp[-R((\rho-\rho_a)\tau_0)-\gamma_E R'((\rho-\rho_a)\tau_0)]}{\Gamma[1+R'((\rho-\rho_a)\tau_0)]} \times  I(R^{'},\tau,\tau_0),
 \end{multline}
where
\begin{equation}
\label{eq:tauintegral}
I(R^{'},\tau,\tau_0) = \int_0^\tau \frac{d\tau'}{\tau'(1-\tau')}  \exp \left[R^{'} \left(\left (\rho-\rho_a \right) \tau_0\right) \ln \frac{\tau'}{\tau_0} \right].
\end{equation}
Upon evaluating the integral over $\tau$ we obtain 
\begin{equation}
\label{eq:hyp}
I(R^{'},\tau,\tau_0) = \left(\frac{\tau}{\tau_0} \right)^{R'} \frac{{}_2 F_1(1,R',1+R',\tau)}{R'}, \, R'  \equiv R'((\rho-\rho_a)\tau_0).
\end{equation}

We then have the result for the cumulative distribution given by Eq.~\eqref{eq:tauint} with finite $\tau$ effects encoded in  the Hypergeometric function of Eq.~\eqref{eq:hyp} precisely as for the $\tau_{21}$  case \cite{Napoletano:2018ohv}.  The origin of the Hypergeometric factor is simply the extra overall factor of $1/(1-\tau)$ in the finite $\tau$ differential distribution. Without this factor we simply obtain the usual result for the cumulative (integrated distribution) up to terms involving $R''$ beyond our accuracy, as long as $\tau_0 \sim \tau$. In what follows we shall simply choose $\tau_0=\tau$ while noting that varying this choice by an 
$\mathcal{O}(1)$  factor will correspond to an effective resummation scale uncertainty on our results.

To obtain an alternate form of Eq. \eqref{eq:tauint} we could have integrated Eq. \eqref{eq:taudistRhob} over $\tau$ instead. Again, as before, we can drop any $\tau$ dependence in the pre-factor other than the overall $\frac{1}{1-\tau}$, which leads to a factor $\rho/(\rho-\rho_b)$, rather than $\rho/\left(\rho-\frac{\rho_b}{1-\tau}\right)$. This again leads one to consider only the overall $1/(1-\tau)$ factor together with the $\tau$ dependence in the exponent. Then integrating over $\tau$ using the same steps that gave Eq.\eqref{eq:tauint} we obtain
\begin{multline}\label{eq:CumulantRhoB}
\left.\rho \frac{d\sigma}{d \rho}\right |_{\tau_{32}<\tau}\overset{\tau<1/2}{=} \bar{\alpha}^2 \int_\zeta^1\frac{\sd z_a}{z_a} \int_\zeta^1\frac{\sd z_b}{z_b}   \int_{\rho_{min}}^{\frac{\rho}{2}} \frac{\sd\rho_b}{\rho_b}\frac{\rho}{\rho-\rho_b}  \Theta_{\rho_\text{min}} \left(\rho_b,\rho_{\text{min}} ,z_a,z_b\right)\\   {}_2F_1(1,R'(\rho_b\frac{\tau}{1-\tau}),1+R'(\rho_b\frac{\tau}{1-\tau}),\tau) \frac{\ \exp[-R(\rho_b\frac{\tau}{1-\tau})-\gamma_E R'(\rho_b\frac{\tau}{1-\tau})]}{\Gamma[1+R'(\rho_b\frac{\tau}{1-\tau})]},
\end{multline}
where we have again used the freedom to neglect factors of $\tau$ in the pre-factor which only introduce terms of order $\bar{\alpha}_s^2L^2$ and set $\tau_0$ to $\tau$.

While so far we have worked with a fixed-coupling approximation in our pre-factor, we now introduce the running of the coupling for ``emissions" $a$ and $b$. In order to do so we replace the $\bar{\alpha}^2$ term with $\bar{\alpha}(z_a \rho_a p_T^2 R^2) \times \bar{\alpha}(z_b (\rho-\rho_a)p_T^2 R^2)$ inside the integral of Eq.~\eqref{eq:tauint}. This corresponds to using the $k_t$ of each emission in the argument of the corresponding coupling factor, with neglect of a factor $1-\tau$ in the coupling associated to emission $\rho_b$, i.e. using $\rho_b = (\rho-\rho_a)$ instead of $(\rho-\rho_a)(1-\tau)$. The $1-\tau$ factor only results in sub-leading terms involving logarithms of $1-\tau$ which we neglect, consistent with our general treatment of the pre-factor.

Finally to include secondary emissions we use the full radiator including the secondary emission terms i.e. replace
\begin{equation}
R((\rho-\rho_a)\tau)\to R^{\text{(primary)}}((\rho-\rho_a)\tau)+R^{\text{(secondary,a)}}((\rho-\rho_a)\tau,z_a,\theta_a^2)+R^{\text{(secondary,b)}}((\rho-\rho_a)\tau,z_b,\theta_b^2),
\end{equation}
where $\theta_a^2 = \frac{\rho_a}{z_a}$, $\theta_b^2 =\frac{\rho_b}{z_b}$ and $\rho_b =(\rho-\rho_a)$.

\subsubsection{Pre-grooming with Soft Drop}\label{sec:grooming}
It is known that the Y-Splitter and \Ym methods need to be supplemented by some form of grooming in order to yield good performance for the signal significance (signal to square-root of background ratio) \cite{Dasgupta:2015yua, Dasgupta:2016ktv,Dasgupta:2018emf}.  In Ref.~\cite{Dasgupta:2016ktv} it was found, in the context of W/Z/H tagging, that pre-grooming jets with Soft Drop was optimal in terms of increasing performance while minimising the sensitivity to non-perturbative effects. Furthermore, in the context of top-tagging there is another advantage to pre-grooming,  namely that the pre-grooming procedure leads to a Sudakov form factor inherited from the groomer \cite{Dasgupta:2016ktv}. In other words for mMDT pre-grooming we obtain the mMDT Sudakov structure while for Soft Drop with non-zero $\beta$ we obtain the Soft Drop Sudakov for both signal and background jets. Given that a modest rather than strong Sudakov suppression was found to be beneficial for signal significance in top-tagging \cite{Dasgupta:2018emf}, pre-grooming with mMDT which has only a single-logarithmic Sudakov form factor, followed by \Ym, emerged as the most performant method as well as being resilient to non-perturbative effects. 

Here we consider QCD jets pre-groomed with mMDT as well as Soft Drop for $\beta=2$ . In Ref.~\cite{Dasgupta:2018emf}  a result was obtained for the jet mass distribution with Soft Drop pre-grooming followed by the application of \Ym i.e. without the additional $\tau$ cut involved here. As described in detail in Ref.~\cite{Dasgupta:2018emf}, three situations can arise : a) the largest gen-$k_t$ emission, i.e. $a$ in the present paper, stops the groomer, b) the next largest gen-$k_t$ emission, i.e. $b$ stops  the groomer and c) another emission stops the groomer. For the first situation the result obtained for the primary emission radiator, with mMDT grooming, was shown to be of the form:
\begin{equation}
\label{eq:mmdtplus}
 R^{(1),(\text{primary)}}(\theta_a,\rho_b)
  = R_{\text{mMDT}}(\rho_b) +R_{\text{mMDT}}^{\text{angle}}(\theta_a,\rho_b).
\end{equation}
This corresponds to the usual mMDT Sudakov at the scale $\rho_b$ but modified by the addition of an extra piece, $R_{\text{mMDT}}^{\text{angle}}$ that arises because emissions with angle below $\theta_a$ are not examined by the groomer and hence need to be vetoed (if they have mass above $\rho_b$) even if they have  $z< \zeta$. This extra contribution, at fixed-coupling and leading logarithmic accuracy, is given by \cite{Dasgupta:2018emf}:
\begin{equation}
\label{eq:rangle}
R_{\text{mMDT}}^{\text{angle}}(\theta_a,\rho_b) = \frac{C_R  \alpha_s}{\pi} \int \frac{dz}{z} \frac{d\theta^2}{\theta^2} \Theta\left(z<\zeta \right) \left( z\theta^2>\rho_b \right) \Theta \left ( \theta_a^2 > \theta^2\right) \ .
\end{equation}
In case b), where emission $b$ stops the tagger, one obtained instead just the standard mMDT result  $R_{\text{mMDT}}(\rho_b)$, while for case c) where an emission other than $a$ or $b$ stops the tagger, there is a complete cancellation against virtual corrections and hence no contribution.

For our current work where we apply also a $\tau$ cut, situation a) yields the result reported in Eq.~\eqref{eq:mmdtplus} but now the mass scale $\rho_b$ is replaced by $\tau (\rho-\rho_a)$ in both terms of Eq.~\eqref{eq:mmdtplus}. In the case b) where emission $b$ stops the tagger we now have to also account for the fact that while emissions with $z< \zeta$ and $\theta < \theta_b$ can never set a mass, or equivalently gen-$k_t$ distance, above $\rho_b$, they can set a mass larger than $\tau(\rho-\rho_a)$. This is disallowed by the $\tau$ cut and hence such emissions have to be vetoed which leads to the appearance of a term $R_{\text{mMDT}}^{\text{angle}}(\theta_b, \tau(\rho-\rho_a))$ , in addition to $R_{\text{mMDT}}(\tau (\rho-\rho_a))$, also in case b).

\begin{figure}[H]
\centering
\begin{subfigure}[b]{0.5\textwidth}
\centering
\includegraphics[width=1\textwidth,trim=180 580 180 60, clip]{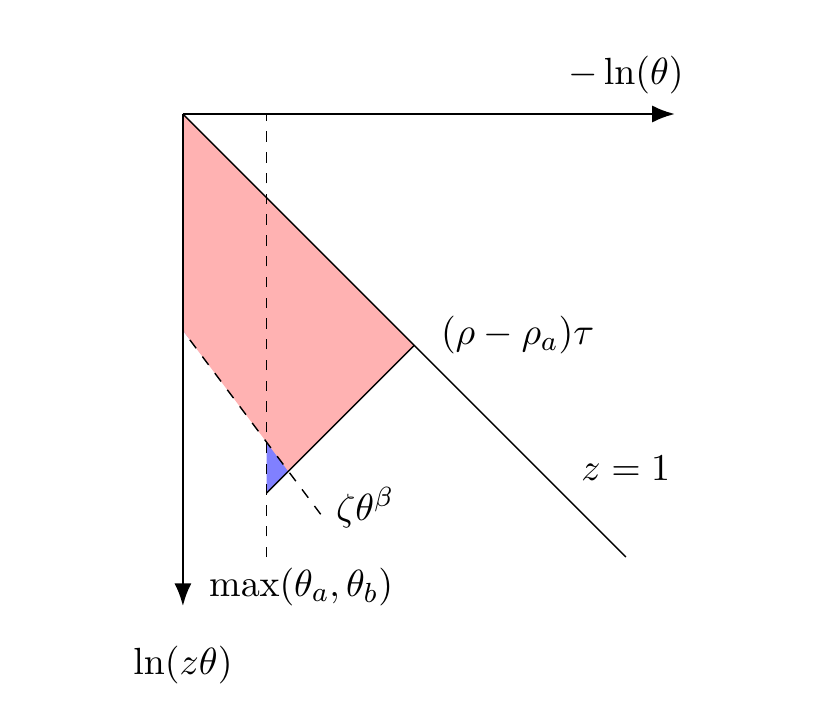}
\caption{$\max(\theta_a^{2+\beta},\theta_b^{2+\beta})\zeta>(\rho-\rho_a)\tau$}
\label{fig:RanglePlane}
\end{subfigure}%
~
\begin{subfigure}[b]{0.5\textwidth}
\centering
\includegraphics[width=1\textwidth,trim=180 360 180 260, clip]{figs/LundDiagrams}
\caption{$\max(\theta_a^{2+\beta},\theta_b^{2+\beta})\zeta<(\rho-\rho_a)\tau$}
\end{subfigure}
\caption{Lund diagrams showing the region of phase space vetoed for jets which are groomed with Soft Drop and tagged with \Ym and a cut on $\tau_{32}$.}
\label{fig:SDYmTauPlane}
\end{figure}

Taking into account hard-collinear emissions and the running of the coupling we can write our result in the form
\begin{multline}
\label{eq:mMDTradiator}
R^{\text{(primary)}}_{\text{groomed-mMDT}}((\rho-\rho_a)\tau,\theta_1)=\int \frac{C_R\alpha_s(z\theta p_T)}{\pi}\left(\frac{1}{z}+B_i\right)\sd z\frac{\sd\theta^{2}}{\theta^2}\Theta(z\theta^2>(\rho-\rho_a)\tau) \Theta(z>\zeta) +\\ +\int \frac{C_R\alpha_s(z\theta p_T)}{\pi}\left(\frac{1}{z}+B_i\right)\sd z\frac{\sd\theta^{2}}{\theta^2}\Theta(z\theta^2>(\rho-\rho_a)\tau) \Theta(z<\zeta) \Theta(\theta^2<\theta_1^2),
\end{multline}
where  the first line is just the standard mMDT result \cite{Dasgupta:2013ihk},  the second line is the extra $R^{\text{angle}}$ contribution and $\theta_1 = \text{max}(\theta_a,\theta_b)$ is the angle of the emission which stops the groomer. The basic form of the result is then that of the mMDT Sudakov evaluated at the scale $(\rho-\rho_a)\tau$, which corresponds to a single-logarithmic Sudakov suppression. In a fixed-coupling leading log approximation, the $R^{\text{angle}}$ term can be written as
\begin{equation}\label{eq:RangleFC}
R^{\text{angle}}(\theta_1,(\rho-\rho_a)\tau)=\frac{C_R  \alpha_s}{2\pi} \ln^2 \frac{\zeta \theta_1^2}{(\rho-\rho_a) \tau}\Theta(\theta_1^2\zeta>(\rho-\rho_a)\tau) \ ,
\end{equation}
where the logarithm involves a ratio of two small quantities similar to be behaviour obtained for secondary emission contributions. Overall therefore we retain the feature that pre-grooming with mMDT results in a reduced Sudakov suppression factor relative to the un-groomed case. The step function in eq. \eqref{eq:RangleFC} switches off the $R^{\text{angle}}$ contribution when $\theta_1^2<\frac{(\rho-\rho_a)\tau}{\zeta}$ leading to the two regimes shown on the Lund diagrams in figure \ref{fig:SDYmTauPlane}, where the $R^{\text{angle}}$ piece is active only in figure \ref{fig:RanglePlane} and is responsible for vetoing emissions in the region of phase space shown in blue. The standard soft drop Sudakov factor at the scale $(\rho-\rho_a)\tau$ is responsible for vetoing the region of phase space shown in red in figure \ref{fig:SDYmTauPlane}.

One can also consider pre-grooming with Soft Drop. Identical considerations to the mMDT case apply, with the only difference being in the grooming condition i.e. for an emission to pass the grooming one needs $z  > \zeta \theta^\beta$. We then obtain a result  along similar lines to that for the mMDT above, but with the Soft Drop Sudakov (i.e. radiator) replacing that for the mMDT and a corresponding $R^{\text{angle}}$ contribution whose fixed-coupling leading-log form is explicitly reported in Ref.~\cite{Dasgupta:2018emf}.

Secondary emissions are unaffected by grooming so the only change to the radiator, relative to the un-groomed case, arises from the primary emission term discussed above. The inclusion of finite $\tau$  effects is also unchanged relative to our previous discussions so that we still have the result Eq.~\eqref{eq:taudist} for the differential distribution and Eq.~\eqref{eq:tauint} for the cumulant but with the primary emission radiator replaced by that for the groomed case Eq.~\eqref{eq:mMDTradiator} for mMDT and its analogue for Soft Drop.

\subsubsection{Numerical implementation and parton shower studies}
For the rest of this section we focus on quark initiated jets, as in the jet $p_T$ range under consideration, these are the dominant background to top jets, though most of what follows could equally be applied to gluon initiated jets with minimal modifications.
The form of Eq. \eqref{eq:CumulantRhoB} is that of the leading $\mathcal{O}\left(\alpha_{s}^2 \right)$ result multiplied by a factor accounting for further emissions. We now perform a type of matching to improve the accuracy with which we calculate this leading order pre-factor. While we have mentioned in section \ref{sec:LO} that a more precise calculation of the leading order pre-factor based around the triple collinear splitting functions is possible, it was shown in \cite{Dasgupta:2018emf} that the numerical difference between such a calculation and one using a product of $1\to 2$ splitting functions, but the full phase-space, is slightly less than $10\%$ for a jet mass of $175 \, \text{GeV}$ and $m_{\text{min}}=50$ GeV. Further to this, when a pair of collinear emissions are strongly ordered in angle, as we have considered them to be throughout this work, the appropriate matrix element is a product of $1\to2$ splitting functions. We therefore choose to match our resummed calculation on to a pre-factor calculated by taking the matrix element to be a product of $1\to2$ splitting functions but still using the full three-particle phase-space in the collinear limit. This particular matching procedure also potentially serves to bring the effects included in our calculations more in line with what is captured by the parton showers which we will compare our calculations to, as while these may be expected to contain elements of the phase-space, they do not include the full triple collinear splitting functions. 

\begin{figure}[h]
\centering
\includegraphics[width=0.7\textwidth,trim=0 0 0 0, clip]{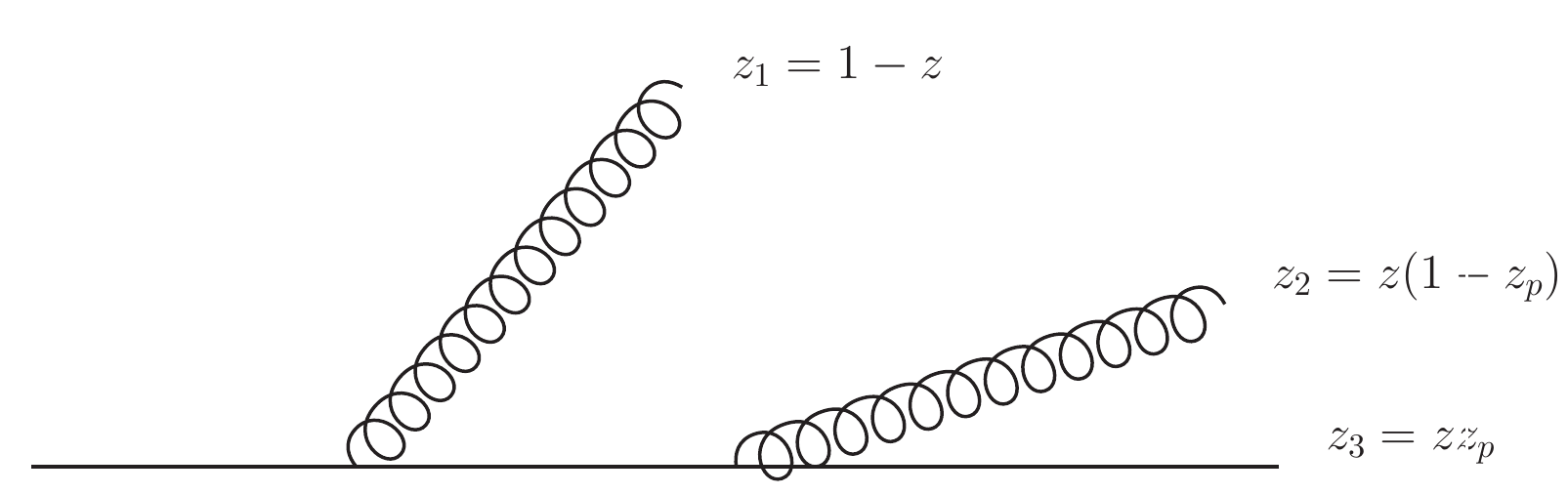}
\caption{Diagram showing the parametrisation of the energy fraction variables used along with our labelling of the partons in the $C_F^2$ channel.}
\label{fig:labeling}
\end{figure}

We now re-calculate the LO pre-factor at this higher level of accuracy, before showing how it is matched to the full resummation. As before, we use the $C_F^2$ channel for illustrative purposes, although our final results contain contributions from the $C_FC_A$ and $C_Fn_f$ colour channels where similar modifications can be made to those listed below. In what follows, the parton initiating the jet is labelled as parton $3$, with the emission at the widest angle to this parton labelled with $1$ and the smaller angle emission labelled $2$. So as to ensure that the variables appearing as the arguments of the factorised $1 \to 2$ splitting functions are defined appropriately, we work with the energy fraction variables $z$ and $z_p$ defined so that $z_1=1-z$ and $z_2=z(1-z_p)$ as illustrated in figure \ref{fig:labeling}.  At this level of accuracy, the leading-order calculation in section \ref{sec:LO} becomes
\begin{multline}\label{eq:HardCollinearPrefactor}
\frac{\sd \Sigma^{\text{LO}}(\tau,\rho)}{\sd \rho}= \\ \left(\frac{C_F\alpha_s}{2\pi}\right)^2 \int p_{gq}(z)p_{gq}(z_p)\sd z \sd z_p \Delta^{\sfrac{-1}{2}}\frac{\sd \theta_{12}^2}{\pi}\frac{\sd \theta_{13}^2}{\theta_{13}^2}\frac{\sd \theta_{23}^2}{\theta_{23}^2}\delta \left(\rho-\frac{s_{123}}{p_T^2R^2} \right)\Theta_\text{\Ym} \Theta_{\text{clust.}}\Theta(\theta_{13}>\theta_{23}) \
\end{multline}
where the Gram determinant is given by
\begin{equation}
\Delta=4\theta_{13}^{2}\theta_{23}^{2}-(\theta_{13}^{2}-\theta_{12}^{2}-\theta_{23}^{2})^{2},
\end{equation}
and
\begin{equation}
\Theta_{\text{\Ym}}= \Theta(\min(\rho_{12},\rho_{13},\rho_{23})>\rho_{\text{min}})\Theta(\min(z_1,z_2,z_3)>\zeta),
\end{equation}
where $\rho_{ij}=z_iz_j\theta_{ij}^2,$ encapsulating the conditions imposed by \Ym without approximating any particles as soft. Similarly, without approximating any particles as soft 
\begin{equation}
\label{eq:clustering}
\Theta_{\text{clust.}}=\sum_{i<j\neq k}\Theta(\theta_{ij}<\min(\theta_{ik},\theta_{jk}))\Theta(\theta_{ij}<R)\Theta(\theta_{ij,k}<R).
\end{equation}

Although the LO part of $\left.\frac{\rho}{\sigma} \frac{\sd\sigma}{\sd \rho}\right |_{\tau_{32}<\tau}$ is now calculated without approximating any of the three LO partons as soft, we cannot simply substitute it in place of the $\mathcal{O}(\alpha_s^2)$ part of Eq.\eqref{eq:CumulantRhoB}. We must first specify how the quantities appearing in the Sudakov factor of Eq.~\eqref{eq:CumulantRhoB}, which are defined in the soft and collinear limit, are related to the kinematic variables appearing in our improved LO pre-factor (Eq.~\ref{eq:HardCollinearPrefactor}).  For the $C_F^2$ channel we make the following prescription:
\begin{equation}\label{eq:defs}
\begin{split}
\rho_b=\min(\min(z z_p,(1-&z))\theta_{13}^2,z \min(z_p,1-z_p)\theta_{23}^2) \\
k_{t1}=\min(z,1-z)\theta_{13}Rp_T & \qquad k_{t2}=z\min(z_p,1-z_p)\theta_{23}Rp_T \\
\theta_{1}=\theta_{13} & \qquad \rho=\frac{s_{123}}{p_T^2R^2}+\sum_i\rho_i
\end{split}
\end{equation}
which we note that there is some freedom in choosing, the only constraint being that the correct result must be recovered in the soft and strongly-ordered limit.

Replacing the $\mathcal{O}(\alpha_s^2)$ part of equation \eqref{eq:CumulantRhoB} with Eq. \eqref{eq:HardCollinearPrefactor} and using the matching prescription given in Eq. \eqref{eq:defs} we can write:
\begin{multline}\label{eq:TC_result}
\rho\frac{\sd\Sigma^{\tau<\frac{1}{2}}(\tau)}{\sd \rho}= \left(\frac{C_F}{2\pi}\right)^2 \\ \int \alpha_s(k_{t1}) \alpha_s(k_{t2}) p_{gq}(z)p_{gq}(z_p) \Delta^{\sfrac{-1}{2}}\Theta(\theta_{13}>\theta_{23})\delta(\rho-\frac{s_{123}}{p_T^2R^2}) \Theta_\text{\Ym} \Theta(\theta_{13}>\theta_{23}) \\ \ _2F_1(1,R'(\rho_b\frac{\tau}{1-\tau}),1+R'(\rho_b\frac{\tau}{1-\tau}),\tau)\frac{e^{-R(\rho_b\frac{\tau}{1-\tau})-\gamma_E R'(\rho_b\frac{\tau}{1-\tau})}}{\Gamma[1+R'(\rho_b\frac{\tau}{1-\tau})]} \sd z \sd z_p  \frac{\sd \theta_{12}^2}{\pi}\frac{\sd \theta_{13}^2}{\theta_{13}^2}\frac{\sd \theta_{23}^2}{\theta_{23}^2} \ ,
\end{multline}
where the quantities $\rho_b$, $k_{t1}$, and $k_{t2}$ are as defined in Eq. \eqref{eq:defs} \footnote{In deriving equation \eqref{eq:TC_result} all emissions are considered to contribute to the jet mass as shown in eq. \eqref{eq:defs}. As well as allowing us to capture the function of $\tau$ multiplying single logarithms in mass scales, this also generates $\tau$ dependant terms which are beyond our accuracy. These terms are removed, as discussed in section \ref{sec:finite_tau}, by neglecting the $\tau$ dependence in the pre-factor beyond the leading $\sfrac{1}{1-\tau}$ term which leads to the hypergeometric function in eq. \eqref{eq:TC_result}. Specifically, we have set $\tau$ to zero inside the delta function, which would otherwise be written as $\delta \left(\rho-\frac{s_{123}}{p_T^2}+\mathcal{O}(\tau) \right)$.}. Eq. \eqref{eq:CumulantRhoB} can be recovered from Eq. \eqref{eq:TC_result} by replacing $s_{123}\to \rho_a+\rho_b$, neglecting the hard collinear part of the splitting functions, carrying out the $\theta_{12}$ integral (equivalent to an azimuthal integral) and changing phase-space variables back to $\rho_a, \rho_b, z_a\  \text{and} \ z_b$. For the sake of brevity, the above result is given only for the $C_F^2$ colour channel, however, our final results include the $C_FC_A$ and $C_Fn_f$ colour channels, where a single gluon is emitted and then decays as opposed to the two independent emissions shown above. We also include secondary Sudakov factors in our final result exactly as before. Our results for pre-groomed jets are obtained by replacing the primary radiator with the groomed variant as discussed in section \ref{sec:grooming}.
 
\begin{figure}[ht]
\centering
\begin{subfigure}[b]{0.32\textwidth}
\centering
\includegraphics[width=1\textwidth]{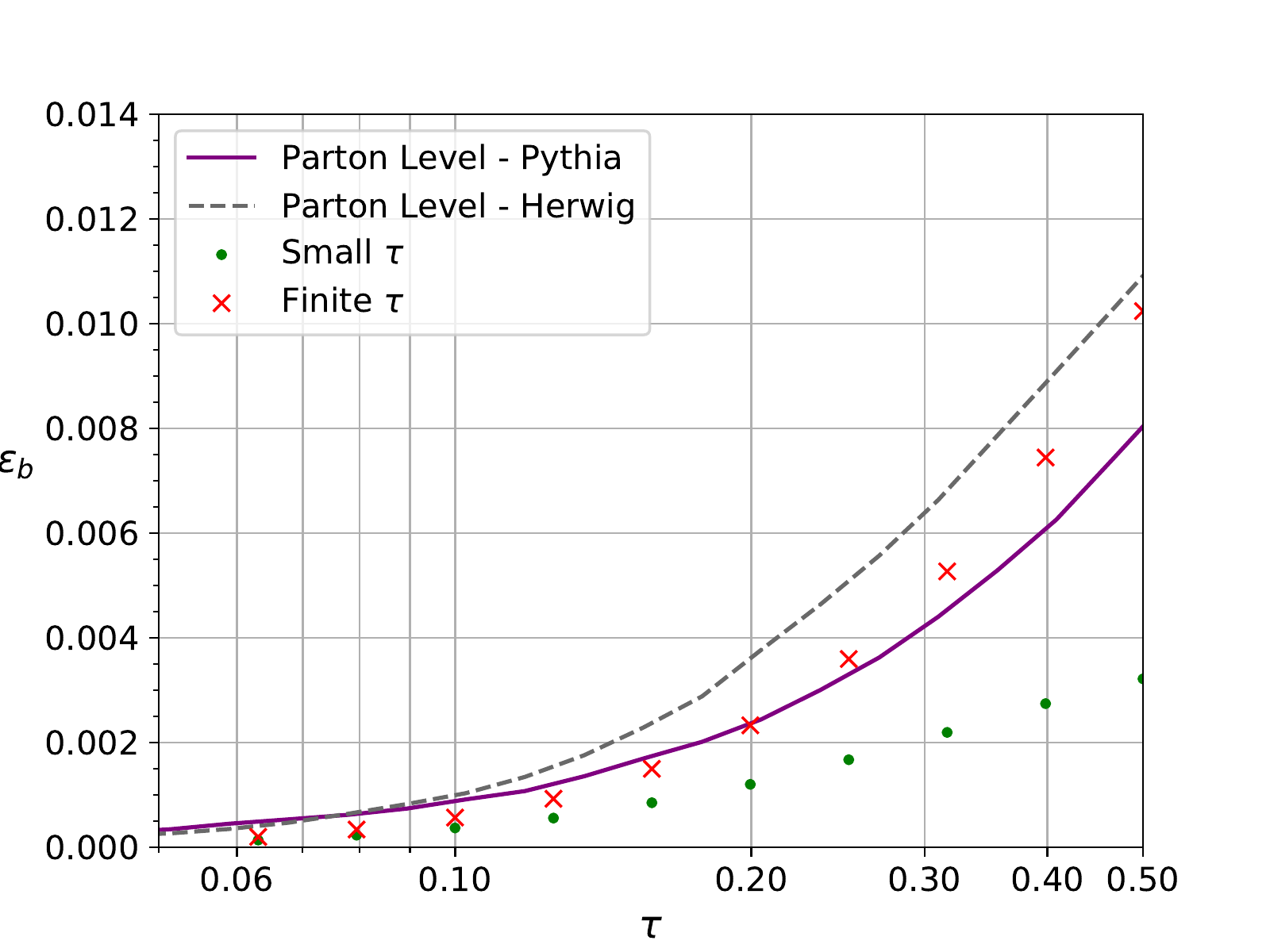}
\caption{No Grooming.}
\end{subfigure}%
~
\begin{subfigure}[b]{0.32\textwidth}
\centering
\includegraphics[width=1\textwidth]{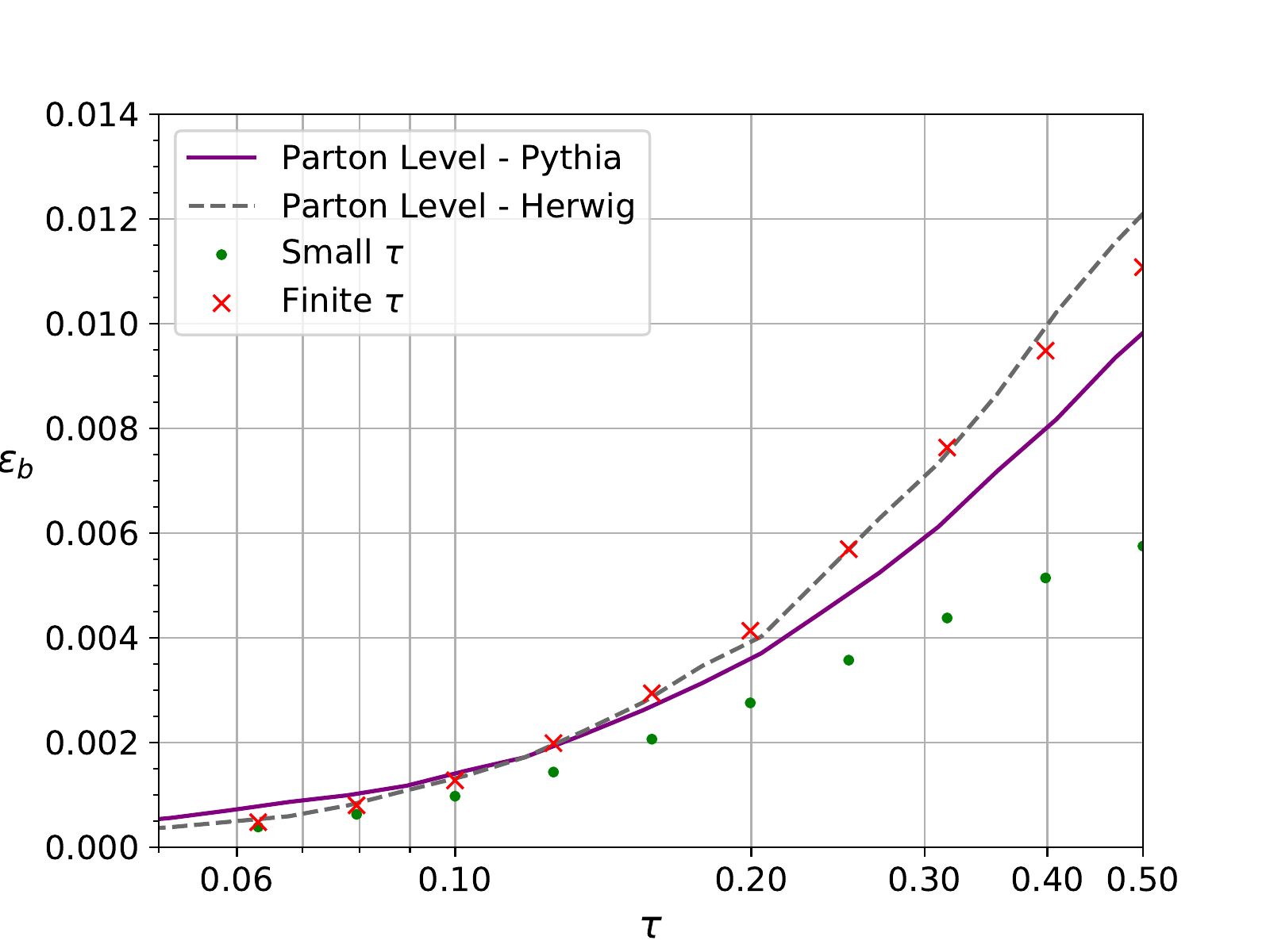}
\caption{With $\beta=2$ Soft Drop}
\end{subfigure}%
~
\begin{subfigure}[b]{0.32\textwidth}
\centering
\includegraphics[width=1\textwidth]{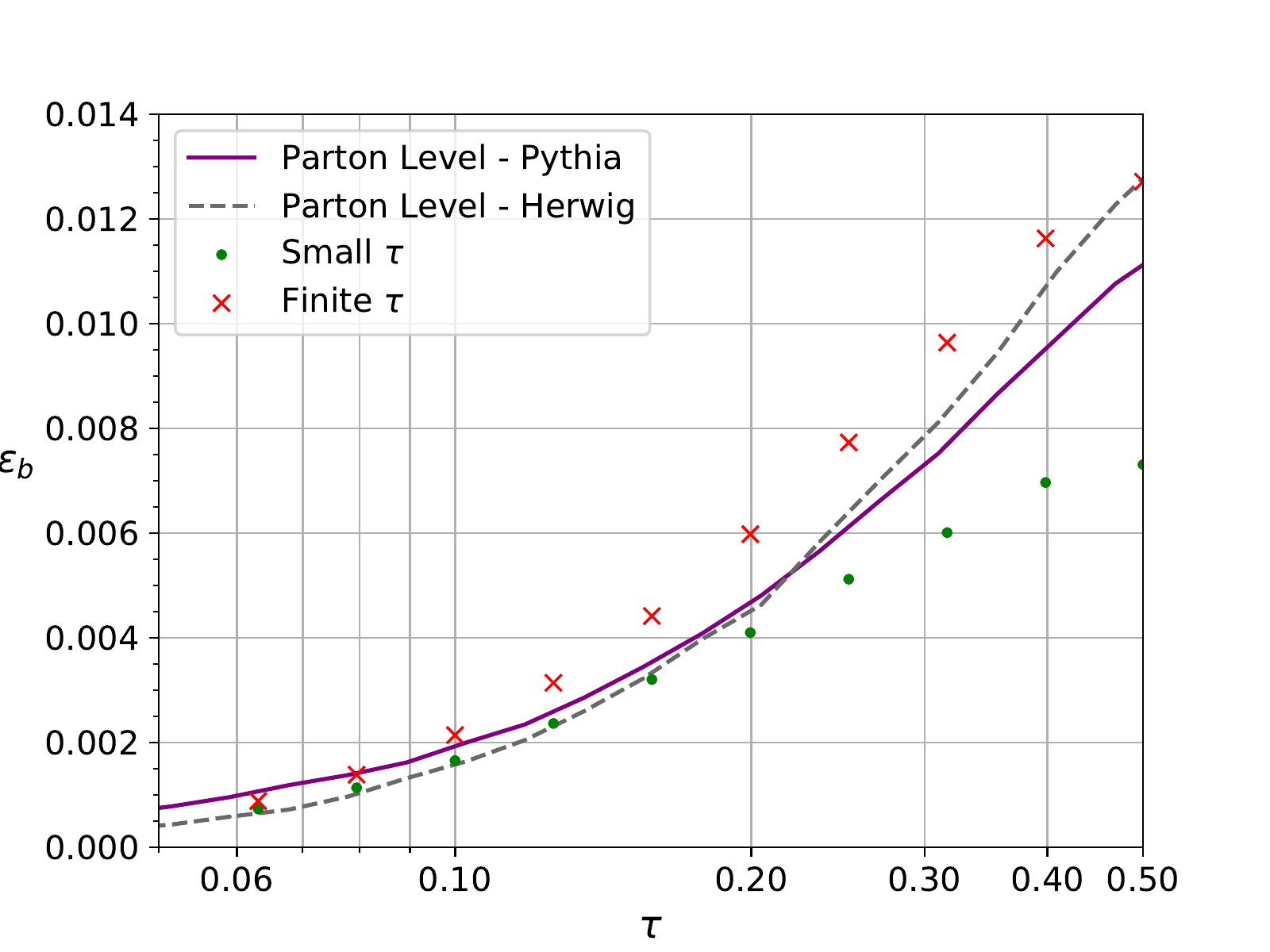}
\caption{With mMDT}
\end{subfigure}
\caption{A comparison of our calculation for background tagging rate against parton show Monte Carlo simulations for no grooming, grooming with Soft Drop $\beta=2$, and grooming with mMDT.}
\label{fig:BackgroundComparison}
\end{figure}

Eq. \eqref{eq:TC_result} is evaluated numerically using the Suave numerical integrator \cite{Hahn:2004fe} interfaced to Mathematica \cite{Mathematica}. As the cut on $\tau_{32}$ restricts emissions down to very low transverse momenta, we freeze the running coupling at $k_t=1.5$  \, GeV to prevent divergences due to the Landau pole. The tagged background fraction is constructed from Eq. \eqref{eq:TC_result} by integrating $\rho$ over the mass window. This is shown in figure \ref{fig:BackgroundComparison} along with the same quantity derived from parton shower simulations using both Pythia and Herwig \cite{Bahr:2008pv} for three variations on our calculation: no grooming, Soft Drop with $\beta=2$ pre grooming, and pre-grooming with mMDT.

In all cases one notes that our results are in reasonable agreement with parton shower predictions, given the uncertainties of the calculations and the shower predictions due to subleading terms present in each case, and also reflected in the difference between Herwig and Pythia showers.
In the case of no pre-grooming or grooming with $\beta=2$ Soft Drop, our finite $\tau$ calculation is clearly an improvement of the small $\tau$ calculation over a wide range of $\tau$ values. Where jets are pre-groomed with mMDT, the finite $\tau$ effects we include still have a sizeable impact and improve agreement with the parton showers as $\tau\to \frac{1}{2}$, however, at smaller values of $\tau$ it is not clear that agreement with the parton showers is improved by their inclusion. This is potentially due to the fact that the leading logs in this case are single logs and we do not include any sources of next-to--leading logarithms (or their interplay with the $\tau$ dependence), other than  the finite $\tau$ corrections we introduced here.

Figure \ref{fig:BackgroundComparison} also shows increasing differences between results from parton showers as the level of grooming decreases. Hence the mMDT result, involving more aggressive grooming, is in better agreement between the two shower descriptions over a wider range in $\tau$, while the un-groomed case shows the largest differences. This is likely due to the differences in the modelling of soft gluon effects between the two showers, which is ameliorated by grooming.

\section{Signal jets} \label{sec:signal}
Here we consider the action of the \Ym method with a $\tau_{32}$ cut on the top quark initiated signal jet. In Ref.~\cite{Dasgupta:2018emf} studies were carried out for top jets with a range of tagging methods including \Ym both with mMDT and Soft Drop pre-grooming.  Here one has to account, in principle, for gluon radiation in both the top production and top decay processes. In the highly boosted limit the top quark is similar to a light quark and the role of soft gluon radiation and its resummation therefore becomes as important as for the background QCD case.  In particular in the boosted limit one can ignore  the dead-cone effect \cite{Dokshitzer:1991fd}, which does not affect our logarithmic accuracy.  We shall also consider soft gluon energies well above the top width where we can neglect additional details of the soft gluon emission pattern studied for instance in \cite{Orr:1993kd}. In the region relevant to our studies we can therefore consider soft emissions as arising from a single fast moving colour charge aligned with the initial top quark direction. \footnote{To account for large logarithms with a $\tau_{32}$ cut we will also need to consider additional collinear radiation from the colour charges arising from $W$ decay within the top jet.}

In spite of these simplifying dynamical assumptions, for top jets, the resummation of large logarithms for the tagging and grooming combinations we consider is more complicated than for the case of background jets. In particular the three-pronged structure of the jet can arise in multiple ways including from the electroweak decay of the top system as well as from soft gluon emission effects. Therefore as in Ref.~\cite{Dasgupta:2018emf} our targeted accuracy will be lower for the signal case and shall omit double logarithms in $\zeta$, $\rho/\rho_{\text{min}}$ and other similar ratios. We shall mainly aim at capturing leading logarithms in $m^2/p_T^2$ where $m$ is a mass-scale which is at most of the order of the top mass.

\subsection{Jet mass distribution for top jets}

We start by computing the fraction of top jets tagged by simply requiring the invariant mass to be within some mass window. Radiation produced by the virtual top quark emerging from the hard process can be recombined with the final top decay products to form the final jet. Placing an upper limit on the jet mass therefore directly constrains this radiation and results in a Sudakov form factor precisely as for a light quark jet. We restrict ourselves to the case where the lower edge of the mass window is below the top mass, so that jets containing all of the top decay products will have mass larger than this. Of course, there will be some fraction of events where not all of the top decay products are reconstructed as a single jet, however such configurations as suppressed by a power of $\frac{m_t}{p_T}$ \cite{Dasgupta:2015yua} and hence can be neglected to our accuracy. We can then write the tagged fraction of events as
\begin{equation}
\label{eq:topMassDist}
\Sigma(\rho)=\frac{1}{\sigma_{0}}\int |M_{t\rightarrow bq\overline{q}}|^{2} \sd\Phi_{3} \delta\left(\frac{s_{123}}{R^2p_{t}^{2}}-\rho_{t}\right) \Theta_{\text{Clust}} \ S_{\text{QCD}},
\end{equation}
where $|M_{t\rightarrow bq\overline{q}}|^{2}$ is the squared matrix-element for the top decay, $\sd \Phi_{3}$ is the three-body phase-space in the collinear approximation and  $\Theta_{\text{Clust}}$ is the jet clustering condition as for the background case (see Eqs.~\eqref{eq:HardCollinearPrefactor} and \eqref{eq:clustering} ). The normalisation factor $\sigma_0$ is just  the result without considering QCD corrections i.e. 
the squared matrix-element for top decay integrated over the final state phase-space with the jet clustering requirement. The factor $S_{\text{QCD}}$ takes into account the constraint on QCD radiation through limiting the jet mass. Given that the jet mass can be expressed in terms of multiple soft gluon emissions such that $\rho = \rho_t+\sum_i \rho_i$, with $\rho_t = m_t^2/(R^2 p_T^2)$,  the constraint on $\rho_i$ just produces a Sudakov form factor which factorises from the integral over the top-decay phase-space to give:
\begin{equation}
\label{eq:signalsudakov}
\Sigma(\rho) = \ S_{\text{QCD}} =e^{-R\left(\rho-\rho_t \right)},
\end{equation}
where $R \left(\rho-\rho_t\right)$ is the standard jet mass Sudakov evaluated to NLL accuracy \cite{CATANI1991368} at the shifted scale $\rho-\rho_t$.\footnote{Although we use the full heavy jet mass radiator evaluated to NLL accuracy the result is only accurate to modified LL accuracy for our case. In particular we neglect non-global \cite{Dasgupta:2001sh} and clustering logarithms \cite{Banfi:2010pa} that are relevant here at NLL level.}

\begin{figure}[ht]
\centering
\includegraphics[width=0.9\textwidth]{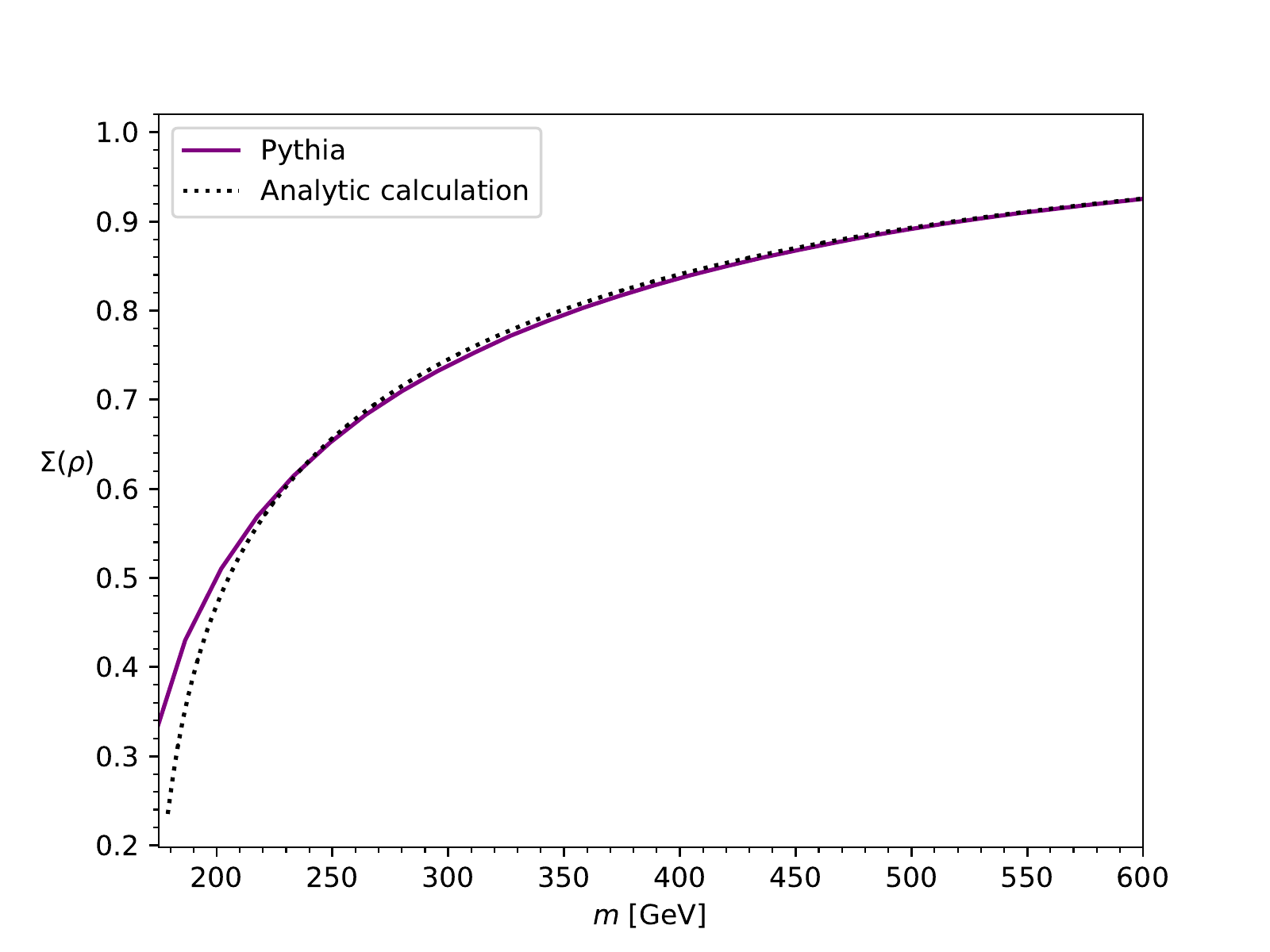}
\caption{A comparison between our analytical calculation of the cumulative jet mass distribution $\Sigma(\rho)$ for top quark initiated jets with $\rho=m^2/R^2 p_T^2$, with $p_T>2$ TeV and the same distribution derived from Pythia simulations.}
\label{fig:topMass}
\end{figure}

In order to test this result and the approximations inherent in deriving it, we compare our result to expectations from Pythia 8. For our Pythia 8 study we choose the lower edge of the mass window to be 10 GeV below the top mass which serves to further reject events where the top decay constituents are not recombined into the final jet. Effects contributing at the lower edge of the mass range should thus only differ from our result by numerically small effects. Pythia 8 was used to create a sample of 1 million $t\overline{t}$ events, with UE, MPI and hadronisation deactivated, and Fastjet \cite{Cacciari:2011ma} was then used to find CA jets with $R=1$. Figure \ref{fig:topMass} shows the integrated jet mass distribution as the upper limit on the mass range is varied. Our analytical estimate is in good agreement with the distribution obtained with Pythia and Fastjet. As $m$ approaches the top mass the agreement between our calculation and Pythia slightly worsens which is to be expected as effects which we neglect, including non-perturbative effects, become relevant for values of $m$ very close to the top mass.

\subsection{Top jets with \Ym}
Next we consider the application of \Ym to the tagging of top jets. This was already studied in Ref.~\cite{Dasgupta:2018emf} where it was noted that the signal case had a number of additional complications relative to the description of QCD background jets, which made the attainment of leading logarithmic accuracy in each of the parameters $\rho$, $\rho_{\text{min}}/\rho$ and $\zeta$ substantially harder. For this reason only basic leading logarithmic accuracy in $\rho$ (or equivalently in $\rho_{\text{min}}$) was targeted which allowed for a simplified treatment of the Sudakov form factor. Consistently with the accuracy goal of that article, complications including the possibility of soft gluon emissions giving one of the three prongs found by the tagger, and the interplay with the mass window constraint were neglected. The results, broadly speaking, gave a reasonable description of the main behaviour seen with parton showers, but the agreement was not as good as seen for QCD background jets.

Here, prior to discussing N-subjettiness, we shall attempt to at least partially address some of the complications that are mentioned above for pure \Ym.  In particular we now consider in the soft-collinear limit, the situation where a single soft emission can be de-clustered as one of the prongs found by the tagger in addition to the case where the de-clustered prongs arise from the electroweak top decay process.  Let us start by considering the result at leading-order i.e. neglecting all QCD radiative corrections.  Then we can write 
\begin{equation}\label{eq:LoTopYmDist}
\Sigma(\rho,\rho_{\text{min}},\zeta)=\frac{1}{\sigma_{0}}\int |M_{t\rightarrow bq\overline{q}}|^{2} \sd\Phi_{3} \delta\left(\frac{s_{123}}{R^2p_{t}^{2}}-\rho_{t}\right) \Theta_{\text{Clust}} \Theta_{\text{\Ym}},
\end{equation}
where
\begin{equation} 
\Theta_{\text{\Ym}}= \Theta(\text{Min}(\rho_{12},\rho_{13},\rho_{23})>\rho_{\text{min}})\Theta(\text{Min}(z_{1},z_{2},z_{3})>\zeta),
\end{equation}
and $1,2,3$ refer to the three prongs identified by \Ym. As there is no soft enhancement to the top decay matrix element, we use only a collinear approximation for the pairwise invariant masses, $\rho_{ij}=x_ix_j\theta_{ij}^2$ in our calculations of the leading-order top decay.\\ Next we consider QCD radiative corrections in the soft and collinear limit. We first take into account the situation that no soft gluon emissions are de-clustered as a prong by \Ym. This imposes a constraint on real emissions in addition to the constraint on jet mass, which comes from the requirement that the soft emission must set a smaller gen-$k_t$ distance than those set by the three-pronged top system. Labelling the soft emission by $i$ we then have that $\min(d_{i1},d_{i2},d_{i3})<\min(d_{12},d_{13},d_{23})$.  This complicated constraint simplifies in the soft and strongly-ordered limit responsible for the leading double logarithms we seek. To be more precise, the three-pronged top decay results in relatively energetic particles owing to the lack of soft enhancement in the electroweak decay. For a soft gluon emission to set a comparable gen-$k_t$ distance it must be emitted  at a relatively large angle compared to the opening angle between the top decay products, $1 \gg \theta_i^2 \gg \theta_{ij}^2$, where $\theta_i$ is the angle wrt the jet axis or equivalently the emitting top quark. In this region we can approximate the angle made by the soft emission with any given prong from the top decay simply by the angle wrt the jet axis which allows us to write the gen-$k_t$ distance for the gluon as $z_i \theta_i^2$.

In addition to the gen-$k_t$ distance, the soft emissions are also subject to the jet mass constraint as before. Therefore the argument of the Sudakov corresponds to whichever is the tighter constraint which gives 
\begin{equation}\label{eq:alpha0TopYm}
\Sigma^{(0)}(\rho,\rho_{\text{min}},\zeta)=\frac{1}{\sigma_{0}}\int |M_{t\rightarrow bq\overline{q}}|^{2} \sd\Phi_{3} \delta\left(\frac{s_{123}}{R^2p_{t}^{2}}-\rho_{t}\right) \Theta_{\text{Clust}}\Theta_{\text{\Ym}} \ e^{-R(\min(d_{12},d_{13},d_{23},\rho-\rho_t))},
\end{equation}
where by $\Sigma^{(0)}$ we mean the contribution where we enforce that no soft gluons can give one of the 3 prongs found by the tagger. 

Next we correct this picture by allowing a soft emission to form one of the prongs found by \Ym, a situation that can first arise at order $\alpha_s$. Consider a single gluon emerging from the de-clustering process before one of the top decay products, and thus being identified as a prong. This gluon is constrained so that it has energy fraction $z>\zeta$ and sets a minimum pairwise mass with the other prongs (labelled 1 and 2) of $m_{\text{min}}$, i.e. $\min(\rho_{1g}, \rho_{2g})>\rho_{\text{min}}$, where $g$ labels the gluon. The gluon must also not set a jet mass which pushes the jet outside of the mass window. The emission of a single soft gluon factorises from the top decay process and gives an order $\alpha_s$ contribution to the pre-factor. Subsequent gluon emissions are constrained by the requirement of not being de-clustered as a prong as well as being subject to the jet mass constraint and again give rise to a Sudakov suppression. Hence we obtain the result:
\begin{multline}\label{eq:alpha1TopYm}
\Sigma^{(1)}(\rho,\rho_{\text{min}},\zeta)=\frac{1}{\sigma_{0}}\int |M_{t\rightarrow bq\overline{q}}|^{2} \sd\Phi_{3} \delta\left(\frac{s_{123}}{R^2p_{t}^{2}}-\rho_{t}\right)\int \sd z \frac{d\theta^2}{\theta^2}  \frac{\alpha_{s}(z\theta p_{t}) \Cf}{\pi}p_{gq}(z)\Theta(z\theta^{2}<\rho-\rho_{\text{top}}) \\ \Theta_{\text{Clust}} \sum_{i<j\neq k}\bigg(\Theta(d_{ij}<\text{Min}(d_{ik},d_{kj}))\Theta(z\theta^2>d_{ij})\Theta(\text{Min}(\rho_{k(ij)}, \, z\theta^{2})>\rho_{\text{min}})\Theta(\text{Min}(z,z_{k},(z_{i}+z_{j})>\zeta) \\ e^{-R (\text{Min}(d_{k(ij)},z\theta^{2},\rho-\rho_t-z\theta^2))}\bigg).
\end{multline}
In the above result the first line gives the pre-factor which, aside from the usual squared matrix-element and phase-space integration for top decay, now also has the QCD pre-factor coming from real emission of the soft gluon. The three prongs are given by the soft gluon, a clustered pair of particles ($ij$) from the top decay and the remaining particle $k$ arising from the top decay. The condition $\Theta(z\theta^2>d_{ij})$ alongside the 
requirement that  $z>\zeta$ ensures that the soft gluon is de-clustered as a prong. \footnote{Note that  here we used the same leading-logarithmic simplification for the gen-$k_t$ distance for soft gluon emissions that led to the result in Eq.~\eqref{eq:LoTopYmDist}.} The condition $\Theta(\text{Min}(\rho_{k(ij)},z\theta^{2})>\rho_{\text{min}}$ is the $\rho_{\text{min}}$ condition where again we used the fact that at our accuracy we can replace the gluon angle wrt a given prong by that wrt the jet axis. Finally we discuss the Sudakov which has as argument $(\text{Min}(d_{k(ij)}, \, z\theta^{2}, \, \rho-\rho_t-z\theta^2)$, reflecting the competing constraints on subsequent soft emissions. Firstly we have that emissions must not set a gen-$k_t$ distance larger than the smallest gen-$k_t$ distance amongst the 3 prongs found by \Ym, given by $\text{Min}(d_{k(ij)}, \, z\theta^{2})$.
Secondly we have that  the soft emissions must not push the jet out of the mass window, i.e. the jet mass should be below $\rho$. Taking into account the additional soft emission we now have as a prong, this condition implies that for multiple subsequent emissions $i$ we must have $\sum_i \rho_i  < \rho-\rho_t-z \theta^2$. Taken together these conditions, on gen-$k_t$ and mass, produce the Sudakov in Eq.~\eqref{eq:alpha1TopYm}. 

It is additionally possible for two soft emissions to be resolved i.e. form two of the prongs found by \Ym. This occurs at order $\alpha_s^2$ with only modest logarithmic enhancements \footnote{We remind the reader that resolved emissions are constrained in several ways. They need to have energy larger than $\zeta$ as well as a mass large enough to satisfy the $\rho_{\text{min}}$ condition but not large enough to push the jet out of the mass window. These constraints lead to the appearance of only modest logarithmic contributions.} and hence such contributions are suppressed relative to the terms we include. We therefore omit them here. We also note that we have ignored soft emissions from the $q\bar{q}$ system produced by the splitting of the W boson. Soft emissions from this dipole are restricted in angle, by virtue of angular ordering, to have an angle less than that of the $q\bar{q}$ pair. Since they are part of the top system they also do not contribute to a shift in mass. Hence to our leading logarithmic accuracy they can also be ignored.

\begin{figure}[h]
\centering
\includegraphics[width=0.9\textwidth]{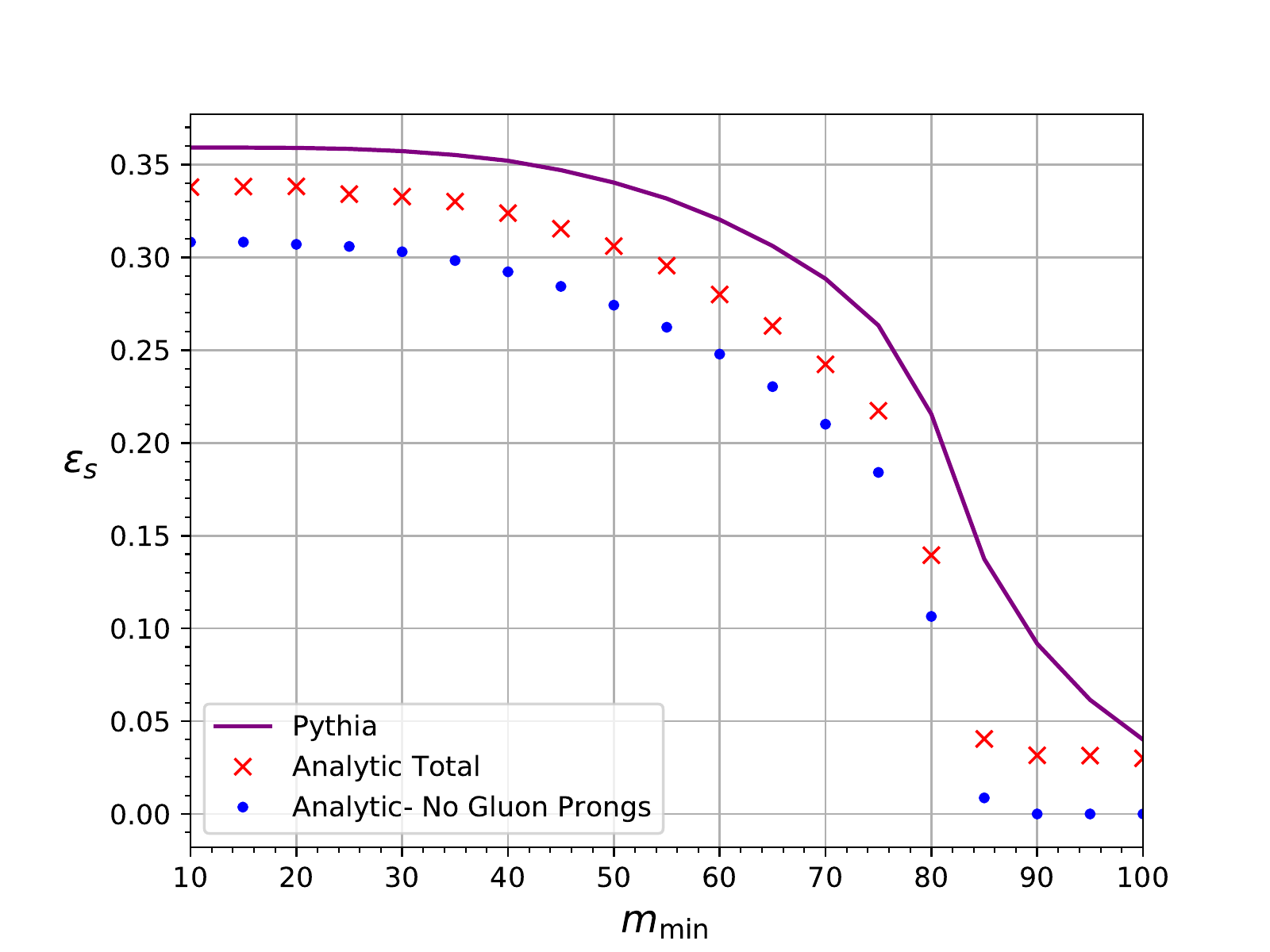}
\caption{A comparison between different levels of approximation in analytical calculations and a Pythia simulation for top jets tagged with \Ym in a mass range $163 \, \text{GeV}<m<225 \, \text{GeV}$ with $\zeta=0.05$ as a function of $m_{\text{min}}$.}
\label{fig:topYm}
\end{figure}

Our results are compared to Pythia 8 in Fig.~\ref{fig:topYm}, where we plot the signal efficiency as a function of $m_{\mathrm{min}}$ (c.f. similar plots in Ref.~\cite{Dasgupta:2018emf}). We show our results for both cases with (red crosses) and without (blue dots) a resolved gluon prong. Our analytics agrees in both cases with the general behaviour seen with Pythia and we note an improved agreement with Pythia when the $\Sigma^{(1)}$ contribution, amounting to an $\mathcal{O}(15 \%)$ correction, is included. As before we choose the lower limit of the mass window to be $10$ GeV below the top mass.

\subsection{\Ym with grooming for signal jets}
\label{sec:SignalYm}
Next we examine the impact of pre-grooming with Soft Drop on our results for \Ym applied to top jets. Relative to results from previous studies \cite{Dasgupta:2018emf} here we also account for the possibility of a resolved gluon prong as in the previous subsection. The result of pre-grooming with mMDT or Soft Drop is again to essentially replace the Sudakov for the un-groomed case by the Sudakov for the groomer i.e. we make the following replacements in the $\Sigma^{(0)}$ and $\Sigma^{(1)}$ terms of the un-groomed results (see Eqs.~\eqref{eq:alpha0TopYm} and \eqref{eq:alpha1TopYm}):
\begin{align}
e^{-R(\min(d_{12},d_{13},d_{23},\rho-\rho_t))} &\to e^{-R_{\text{mMDT/SD}}(\min(d_{12},d_{13},d_{23},\rho-\rho_t)}, \\ \nonumber
e^{-R(\text{Min}(d_{k(ij)},z\theta^{2},\rho-\rho_t-z\theta^2))} &\rightarrow e^{-R_{\text{mMDT/SD}}(\text{Min}(d_{k(ij)},z\theta^{2},\rho-\rho_t-z\theta^2)},
\end{align}
where the suffix mMDT or SD is used to indicate the grooming variant.  We note that unlike the case of the QCD background  jets, we have not included $R_{\text{angle}}$ terms in the signal case. Although such terms would in principle be present, the angular scales involved are of the order of the opening angles between top decay products. At such angular scales the radiation pattern becomes more complicated as one also needs to account for radiation from the $q\bar{q}$ dipole produced by the colour singlet W decay. Given that the terms produced are logarithms in the ratio of two small scales, i.e. of the same level of significance as $\ln \rho/\rho_{\text{min}}$ terms, they are beyond the accuracy we aim for in the case of signal jets.

\begin{figure}[ht]
\centering
\begin{subfigure}[b]{0.5\textwidth}
\centering
\includegraphics[width=1\textwidth]{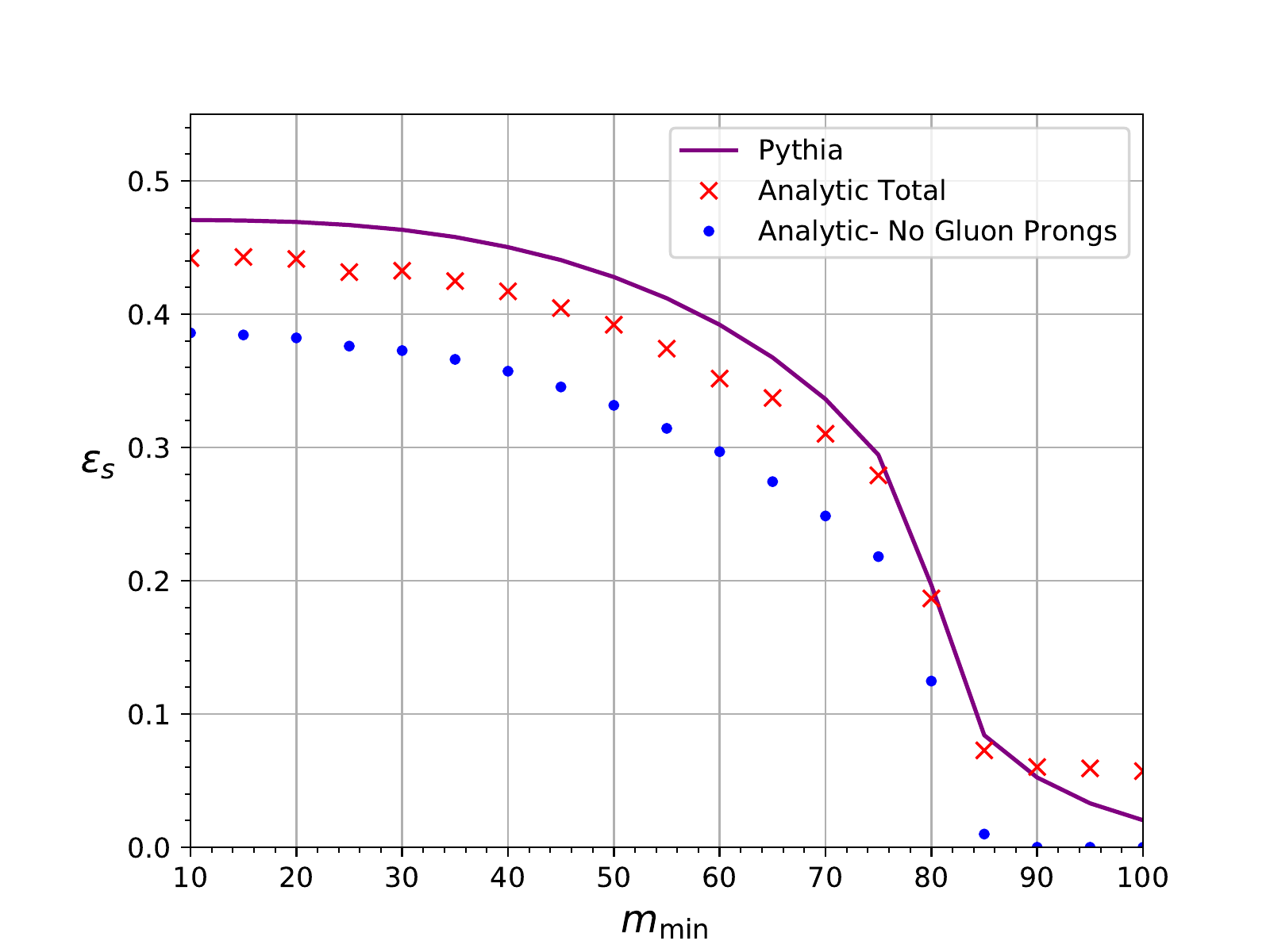}
\caption{Pre-grooming with Soft Drop ($\beta=2$)}
\end{subfigure}%
~
\begin{subfigure}[b]{0.5\textwidth}
\centering
\includegraphics[width=1\textwidth]{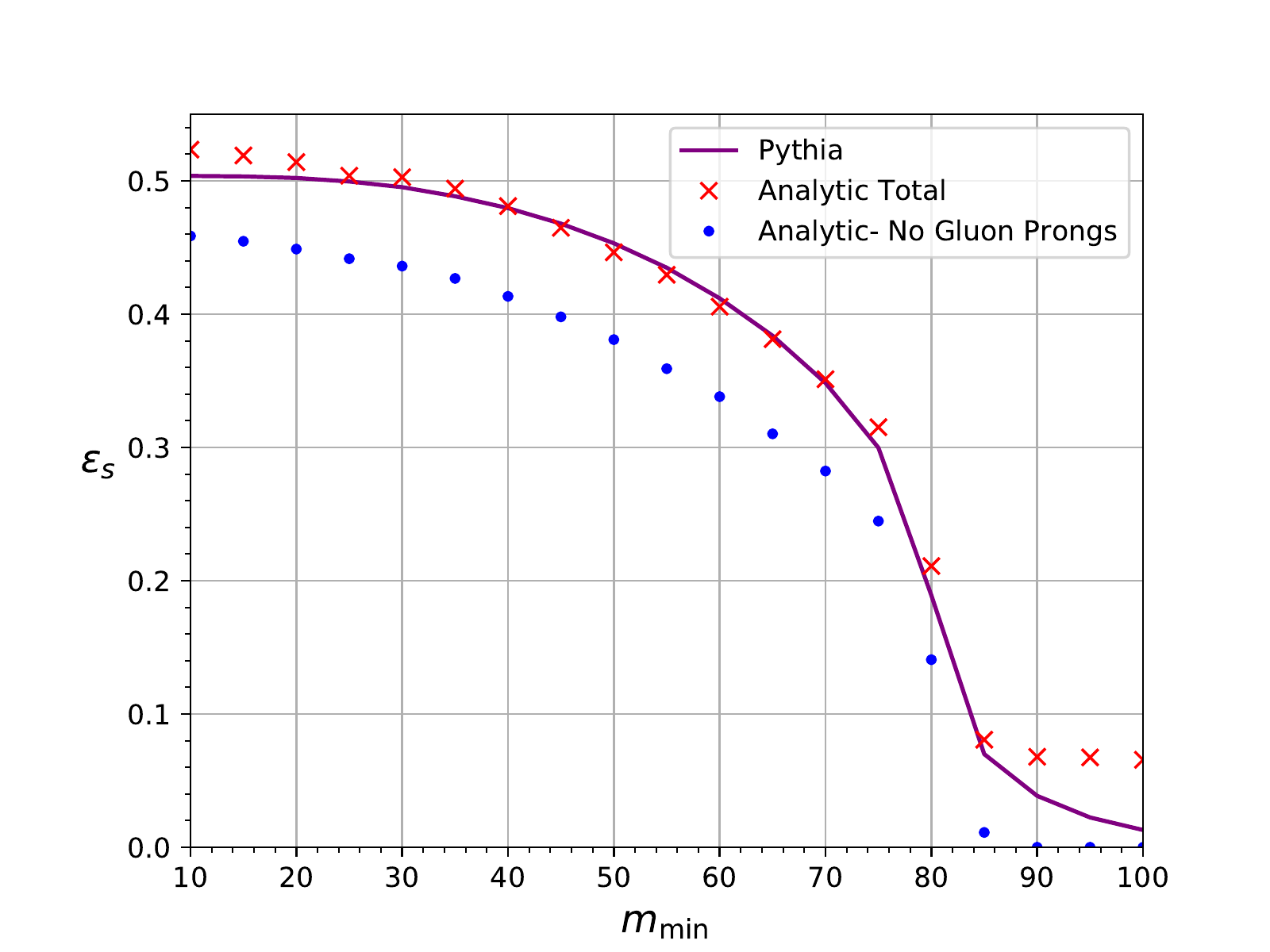}
\caption{Pre-grooming with mMDT}
\end{subfigure}
\caption{A comparison between our analytical calculations and a Pythia simulation for pre-groomed top jets tagged with \Ym in a mass range $163 \, \text{GeV}<m<225 \, \text{GeV}$ with $\zeta=0.05$ as a function of $m_{\text{min}}$.}
\label{fig:topMmdtYm}
\end{figure}

The tagged signal fraction, with our usual choice of parameter values, is compared to a Pythia simulation in Figure \ref{fig:topMmdtYm}, again showing the results with and without a resolved gluon prong and for grooming with SD (left) and mMDT(right). We see that except for the extreme region, where the tagged signal fraction is very small, our analytic results, especially after inclusion of the resolved gluon case, are in good overall agreement with the behaviour seen with Pythia.

\subsection{\Ym with $\tau_{32}$ and grooming for signal jets}
\label{sec:SignalYmtau}
We now wish to understand the effect of adding a cut on $\tau_{32}$ to the tagged signal distribution after application of \Ym. We shall first consider the un-groomed case and then include the effects  of grooming. We begin with the configuration where all three of the LO top decay products are identified as prongs by \Ym. With no additional emissions $\tau_3$ vanishes and hence a cut on $\tau_{32}$ has no impact. Adding a set of soft and collinear emissions, one has to consider how these emissions are constrained by the $\tau$ cut, the mass-window cut and the requirement that  they should not give a resolved prong on applying \Ym. 

We first introduce an approximation into our definition of $\tau_2$ which is valid to within the overall accuracy we can obtain with our current calculations for the signal case, i.e. LL accuracy in $\rho$ with neglect of logs in ratios of mass scales and $\zeta$. Consider the region of phase space where say $d_{12}<\min(d_{13},d_{23})$, so that the first de-clustering will lead to two gen-$k_t$ axes lying along $p_3$ and $p_1+p_2$. In this region of phase space, to leading order where there are no additional emissions, $\tau_2=z_1\theta_{1,12}^2+z_{2}\theta_{2,12}^2$. As the $p_1+p_2$ direction will be aligned more with the harder of partons $1$ and $2$ we make the approximation that the gen-$k_t$ axis is aligned with this parton, so that to LO we can approximate $\tau_2=\min(d_{12},d_{13},d_{23})$. As there is no logarithmic enhancement associated with the leading order decay of the top, this approximation will introduce an $\mathcal{O}(1)$ rescaling of the argument of the Sudakov factor, which is consistent with an NLL correction and hence beyond our LL accuracy.

When considering the role of additional soft emissions let us first consider, as in section \ref{sec:SignalYm} before, primary emissions at a large angle to the opening angles of the  top decay system. Regardless of which of the gen-$k_t$ axes these emissions are closer to, their contribution to $\tau_3$  and $\tau_2$ may always be approximated by $\sum_i \rho_i$, where $\rho_i = z_i \theta_i^2$ and $\theta_i$ is the emission angle wrt the emitting top quark direction. The constraint on emissions due to the $\tau_{32}$ cut is then $\tau_{32} \approx \frac{\sum_i \rho_i}{\min(d_{12},d_{13},d_{23})+\sum_i \rho_i} < \tau$ which gives the constraint $\sum_i \rho_i <\min(d_{12},d_{13},d_{23})\frac{\tau}{1-\tau}$. For $\tau <1/2$ this subjettiness constraint overcomes the constraint from \Ym, $\rho_i <  \min(d_{12},d_{13},d_{23})$ and  hence the argument of the primary emission Sudakov depends only on the competing subjettiness and jet mass constraints. 

Until now we have neglected the role of secondary radiation from the $q\bar{q}$ system (arising from W decay) since these emissions bring only enhancements in ratios of similar mass scales. If we wish to obtain a good description of the signal with a $\tau$ cut including also the region where $\tau \ll 1$, we need to consider all sources of double-logarithmic corrections in $\tau$. Secondary emissions are a source of such double-logarithmic terms  and hence we include them here. The secondary emission terms are given by taking into account soft and collinear emissions from the  $q$ and $\bar{q}$ with the constraint that the emission angle is smaller than $\theta_{q\bar{q}}$ the opening angle of the  $q\bar{q}$ dipole. This leads to results which have the same form as the corresponding results for the background case (see Eq.~\eqref{eq:secondaryymtau}) with $z_a$ replaced by $z_q$ and $\theta_a$ by $\theta_{q\bar q}$ for emission from $q$ and similarly for emission from the $\bar{q}$. We note that secondary emissions are part of the decaying top system and hence do not contribute to a shift in mass so that the jet mass constraint is irrelevant here. 

Thus we can write
\begin{multline}\label{eq:topTau}
\Sigma^{\tau<\frac{1}{2}}(\rho_{\text{min}},\zeta,\tau) = \frac{1}{\sigma_{0}}\int |M_{t\rightarrow bq\overline{q}}|^{2} \sd\Phi_{3} \delta\left(\frac{s_{123}}{R^2p_{t}^{2}}-\rho_{t}\right) \Theta_{\text{Clust}}\Theta_{\text{\Ym}} \frac{e^{-R -\gamma_E R'}}{\Gamma[1+R']},
\end{multline}
with 
\begin{multline}
R \equiv  R \left(\min(\frac{\tau}{1-\tau}\min(d_{12},d_{13},d_{23}),\rho_{\text{max}}-\rho_t)\right)+R^{\text{secondary}}\left(\frac{\tau}{1-\tau}\min(d_{12},d_{13},d_{23}),z_q,\theta_{q\overline{q}}^2\right)+ \\ +R^{\text{secondary}}\left(\frac{\tau}{1-\tau}\min(d_{12},d_{13},d_{23}),z_{\overline{q}},\theta_{q\overline{q}}^2 \right),
\end{multline}
where $\rho_{\text{max}}$ is the upper limit on the jet mass. Finally we account for the effect of grooming. To take this into account one makes the usual replacement of the primary emission radiator by its groomed counterpart. An additional subtlety that is present here is the existence of $R_\text{angle}$  terms (see Eq.~\eqref{eq:rangle}) which originate from emissions which are not visible to the groomer as they are shielded by larger angle emissions that stop the grooming. Such terms have been ignored for the signal since they are complicated to account for and produce only logarithms of mass ratios which we neglect. However in the presence of a $\tau$ cut such terms also induce double logarithms in $\tau$ as described by Eq.~\eqref{eq:mMDTradiator}. A consistent description of the double logs in $\tau$ should also include the double logarithm originating here while we can neglect all other details associated to this term.
Grooming is therefore included through the replacement of the radiator as
\begin{multline}
R(\min(\frac{\tau}{1-\tau}\min(d_{12},d_{13},d_{23}),\rho_{\text{max}}-\rho_t))\to \\ R_{\text{mMDT}}(\min(\frac{\tau}{1-\tau}\min(d_{12},d_{13},d_{23}),\rho_{\text{max}}-\rho_t))+R_{\text{angle}}(\tau),
\end{multline}
where, at  fixed coupling, $R_{\text{angle}}(\tau) = \frac{C_F \alpha_s}{2\pi} \ln^2 \tau$.

We have thus far not considered the case where a soft gluon is resolved as a \Ym prong, which we took into account in the previous subsections. For such a configuration, the effect  of the $\tau$ cut is actually to constrain the phase space of partons arising from the LO top decay. As the electroweak top decay is not logarithmically enhanced, the restriction from the $\tau$ cut leads to a suppression proportional to $\tau$. Given  that  the configuration with a resolved gluon prong is already suppressed by a power of $\alpha_s$ a further suppression with $\tau$ implies that we may ignore this term while still retaining a reasonable description of the overall behaviour.\footnote{We remind the reader that the value of $\tau$ that  gives the highest signal significance is $\tau \sim 0.2$. We have checked numerically, by studying specific configurations, that the power suppression with $\tau$ holds at leading order.}

\begin{figure}[ht]
\centering
\begin{subfigure}[b]{0.32\textwidth}
\centering
\includegraphics[width=1\textwidth]{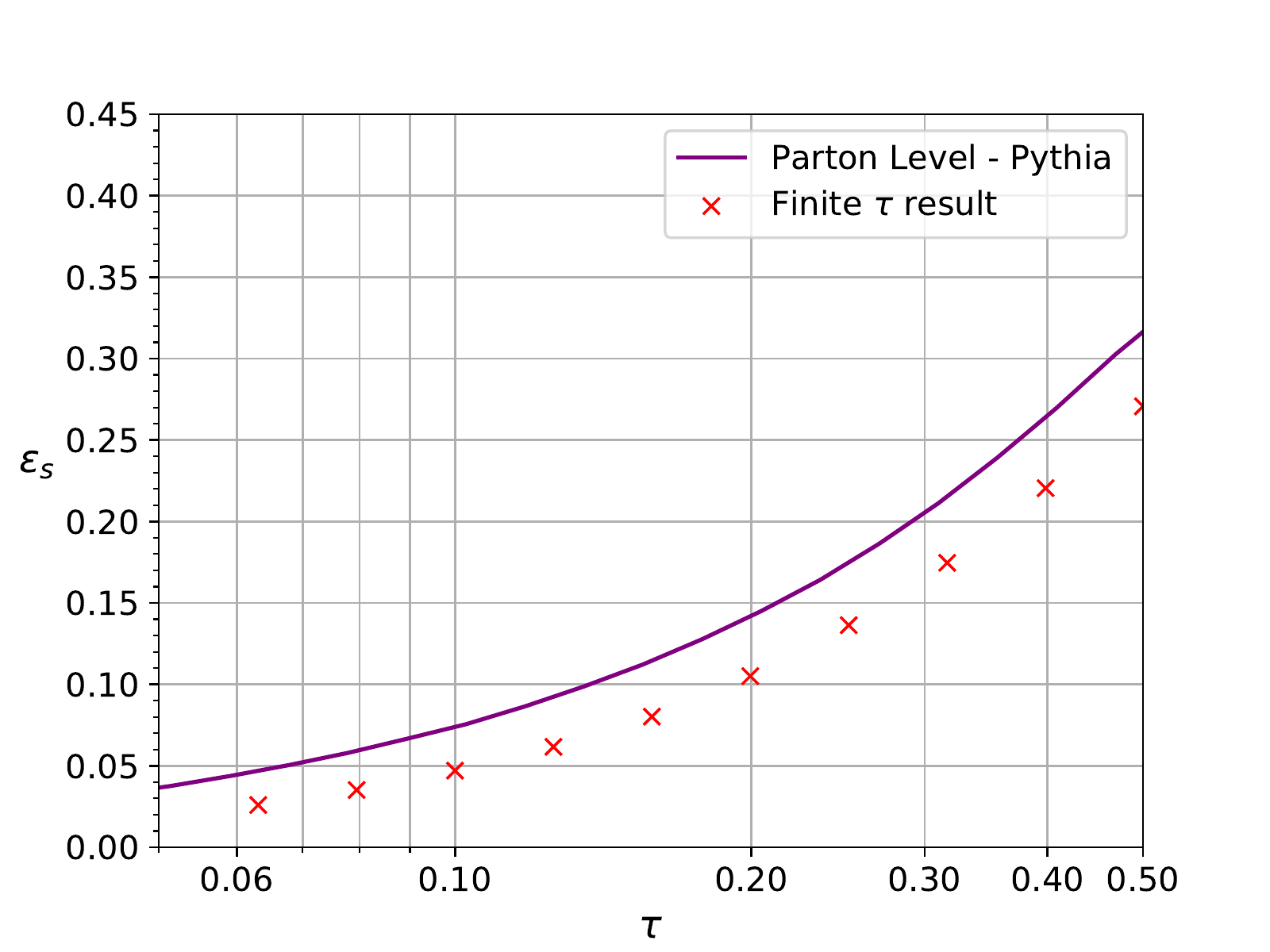}
\caption{No Grooming.}
\end{subfigure}%
~
\begin{subfigure}[b]{0.32\textwidth}
\centering
\includegraphics[width=1\textwidth]{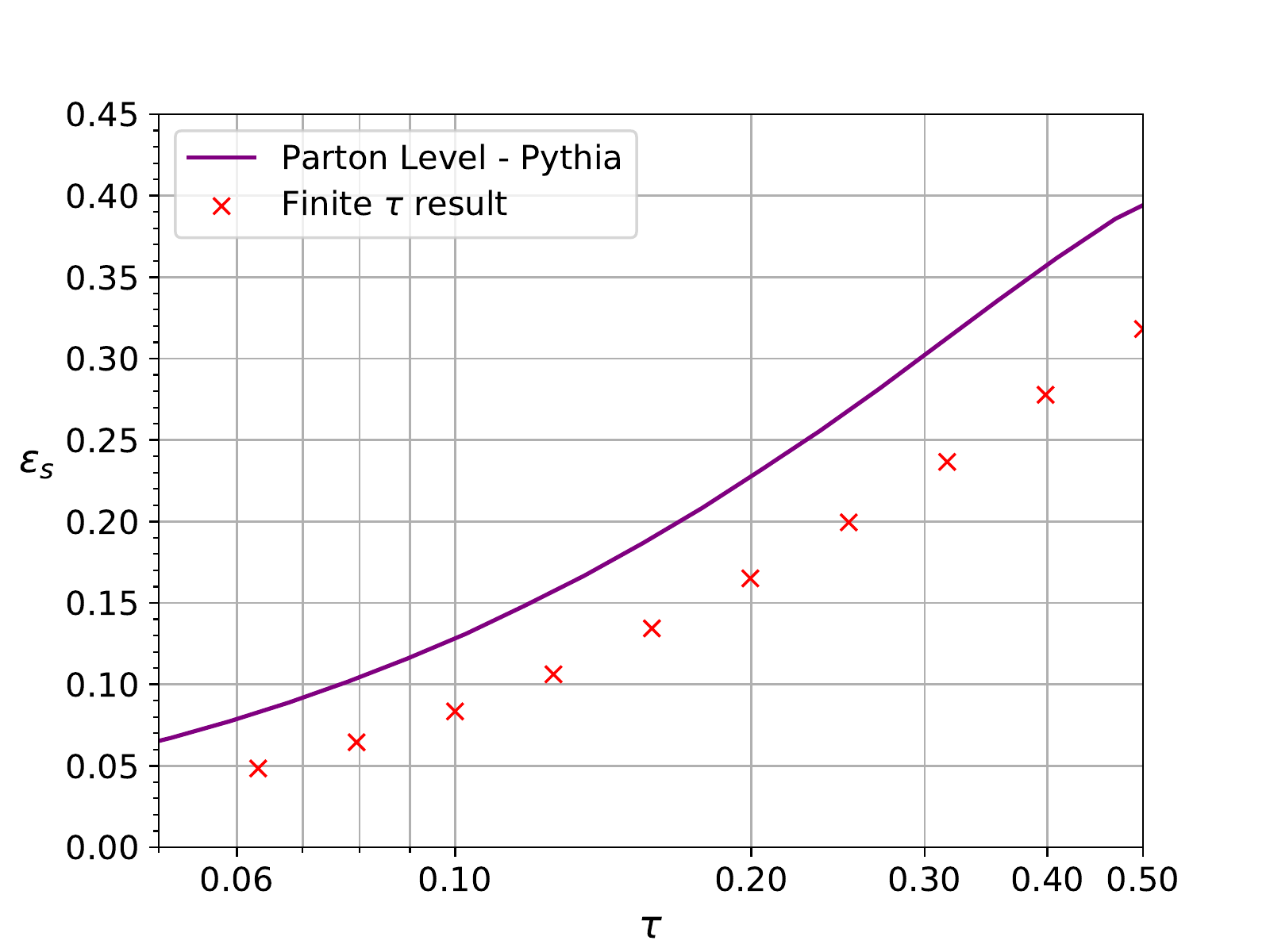}
\caption{With $\beta=2$ Soft Drop}
\end{subfigure}%
~
\begin{subfigure}[b]{0.32\textwidth}
\centering
\includegraphics[width=1\textwidth]{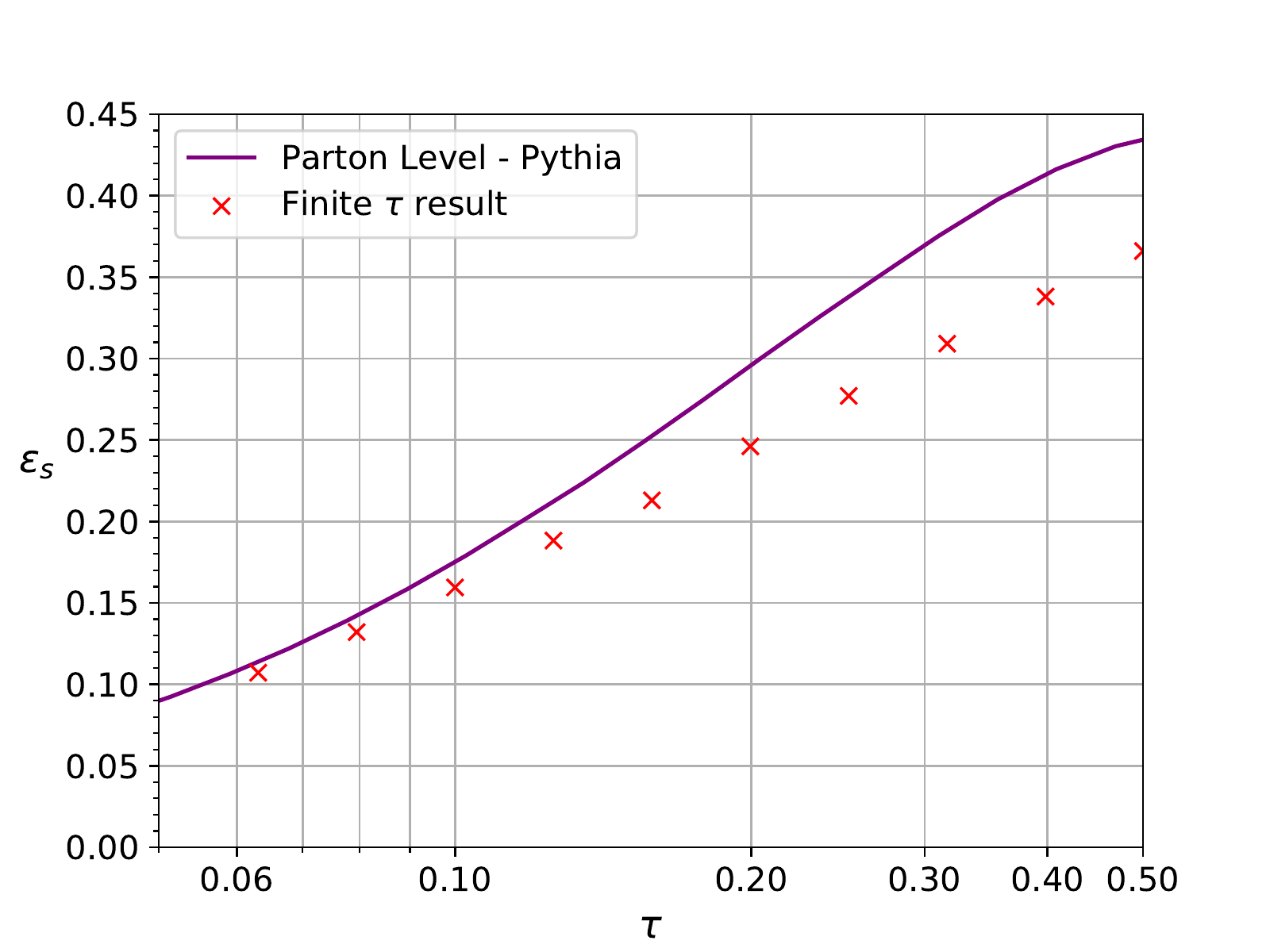}
\caption{With mMDT}
\end{subfigure}
\caption{Comparison between our analytic calculation (crosses) and Pythia for the tagged signal distribution as a function of the cut on $\tau_{32}$  for jets without pre-grooming (left) , with pre-grooming using Soft Drop (centre)  and with pre-grooming using the mMDT.}\label{fig:topTauAnalytic}
\end{figure}

 Equation \eqref{eq:topTau} is evaluated and compared to the same distribution derived from simulations using Pythia in figure \ref{fig:topTauAnalytic}. Although given the accuracy of the shower and the analytic calculations (each of which is leading-logarithmic albeit with inclusion of some key NLL effects), one would expect to see the moderate level of difference that can be observed in the figure, it is noticeable that the behaviour in $\tau$ is well captured by the analytics especially  for the un-groomed case and for pre-grooming with Soft Drop. For grooming with mMDT there is good agreement at smaller $\tau$ and a deviation at larger values of $\tau$. Here, given that the leading logarithms are single logarithms, the analytics and the shower would each only contain (at best) a correct leading-logarithmic description, but with potentially larger differences from spurious NLL effects in the shower and their interplay with $\tau$. Moreover our neglect of configurations where a gluon is one of the resolved prongs from \Ym would also lead to differences at larger values of $\tau$ where the power suppression with $\tau$, which was a factor in our neglecting this configuration, will be less pronounced. Neglect of such configurations may have more of an impact on the distributions where jets are pre-groomed, as they can allow the jet to be tagged even if one of the electroweak top decay products is groomed away.

We note that Eq.~\eqref{eq:topTau} for the signal case reflects a few features that are different to the corresponding results for the QCD background. In particular for signal jets there is a lack of soft and collinear enhancements in the pre-factor resulting in the absence of the Hypergeometric function. Also, to our accuracy, the jet mass constraint does not affect the distribution for small enough $\tau$ cuts, or large enough $\rho_{\text{max}}$, as a result of the fixed invariant mass of the leading-order system. This is clear from the argument of the Sudakov factor in equation \eqref{eq:topTau} which contains a competition between the $\tau$ cut and the mass-window. For a given $\tau_{32}$ cut we can estimate the threshold below which $m_{\text{max}}$ should be taken if varying it is to have an effect on the tagged signal fraction:
\begin{equation}\label{eq:TauToMass}
m_{\text{max}}^{2}<m_t^2+ p_T^2 \frac{\tau}{1-\tau}\min(d_{12},d_{13},d_{23}).
\end{equation}

For top jets, where $\min(d_{12},d_{13},d_{23})p_T^2$ may be roughly approximated by the W boson mass squared, we estimate that, for $\tau=0.3$ and $m_{t}=173$ GeV, the jet mass constraint will not significantly affect the signal efficiency unless $m_{\text{max}}\lesssim181$ GeV. In reality there will not be a hard threshold but some range of parameters over which the Sudakov suppression transitions from being due to the cut on $\tau_{32}$ to being due to the jet mass constraint. The application of this will be discussed further in the next section.\\

\section{Exploiting jet mass cuts}\label{sec:mass}

\begin{figure}[ht]
\centering
\begin{subfigure}[t]{0.48\textwidth}
\centering
\includegraphics[width=1\textwidth]{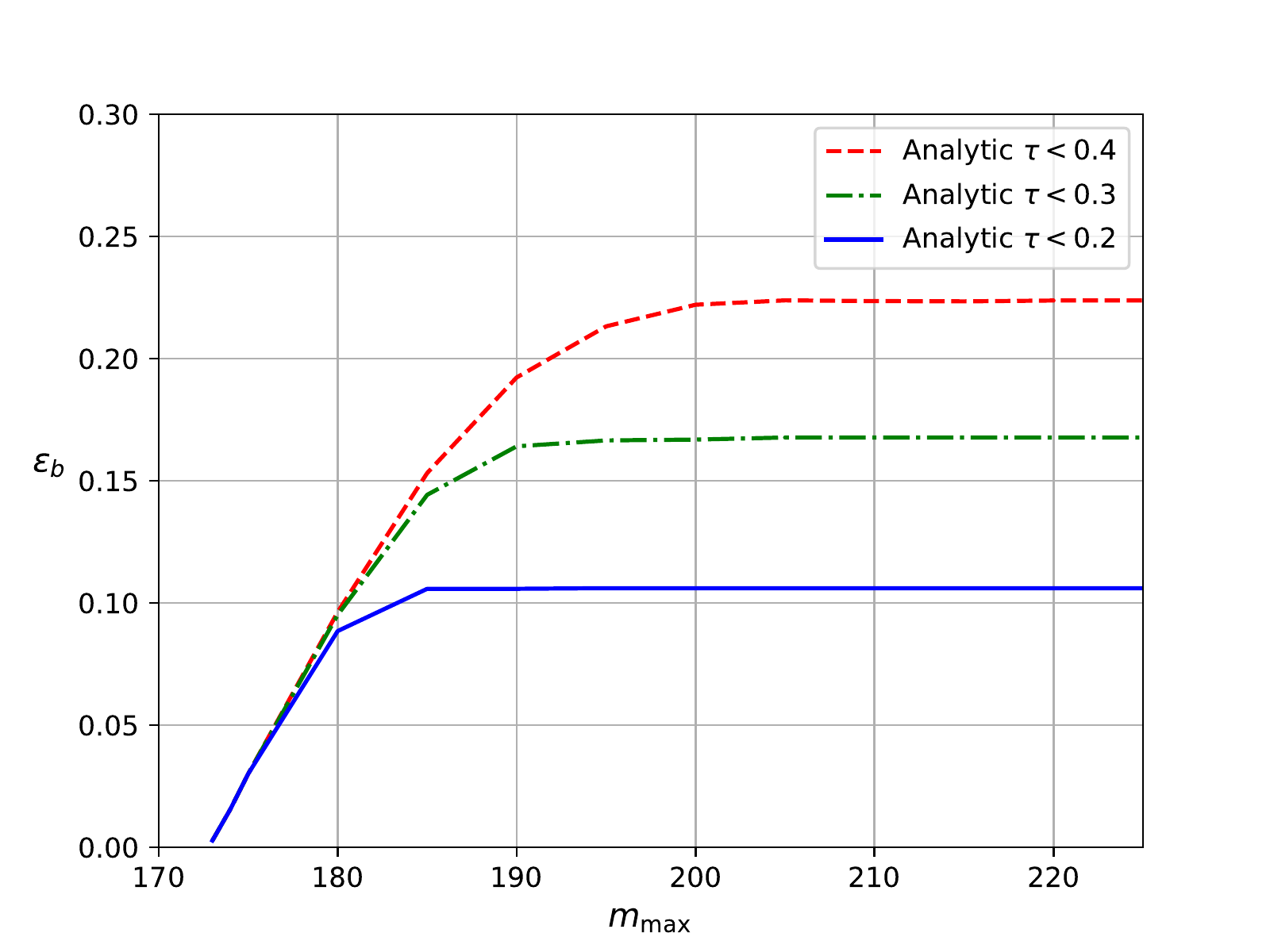}
\caption{Analytic, un-groomed.}
\end{subfigure}%
~
\begin{subfigure}[t]{0.48\textwidth}
\centering
\includegraphics[width=1\textwidth,]{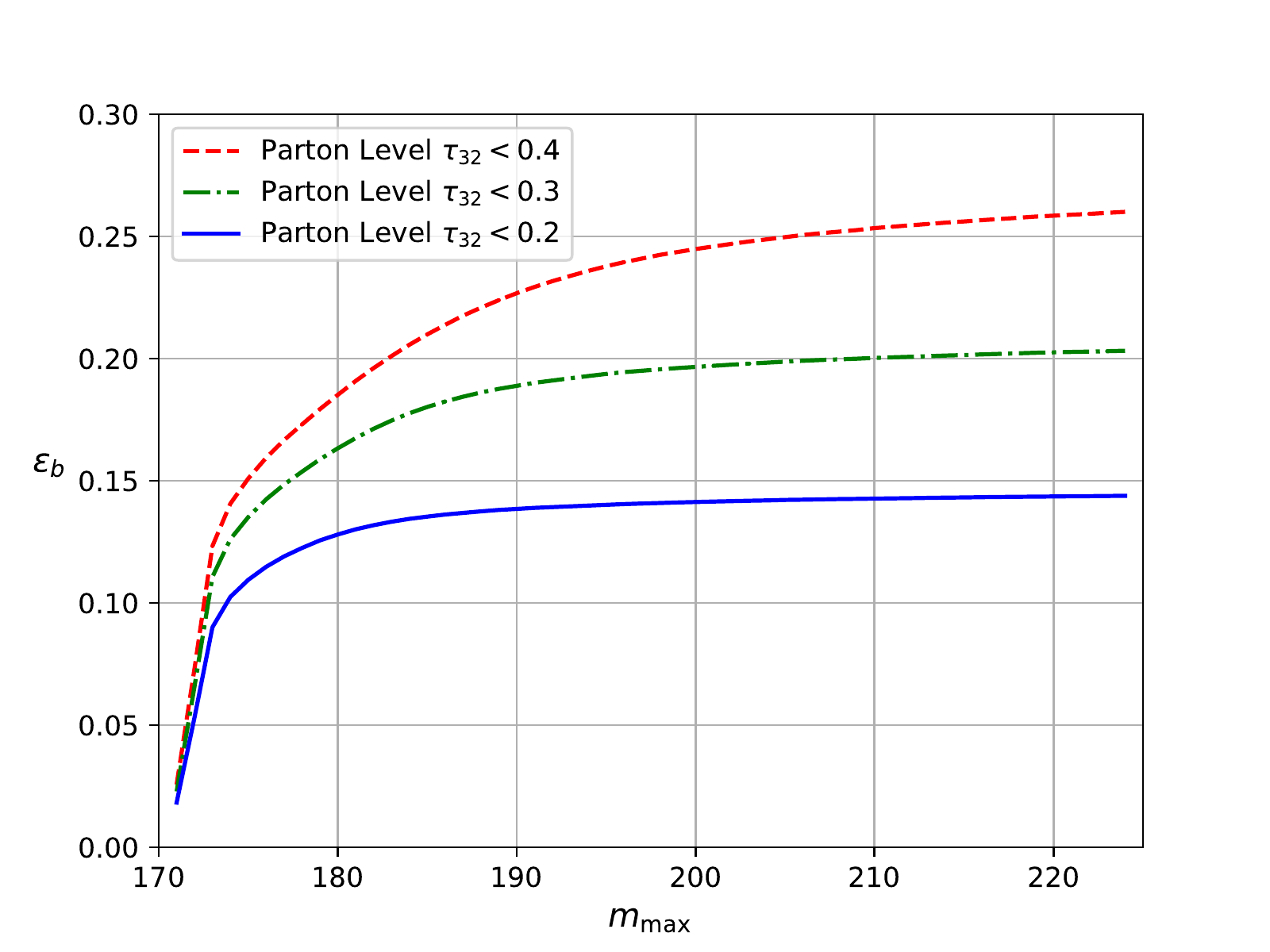}
\caption{Parton level Monte Carlo, un-groomed.}
\end{subfigure}
\begin{subfigure}[t]{0.48\textwidth}
\centering
\includegraphics[width=1\textwidth]{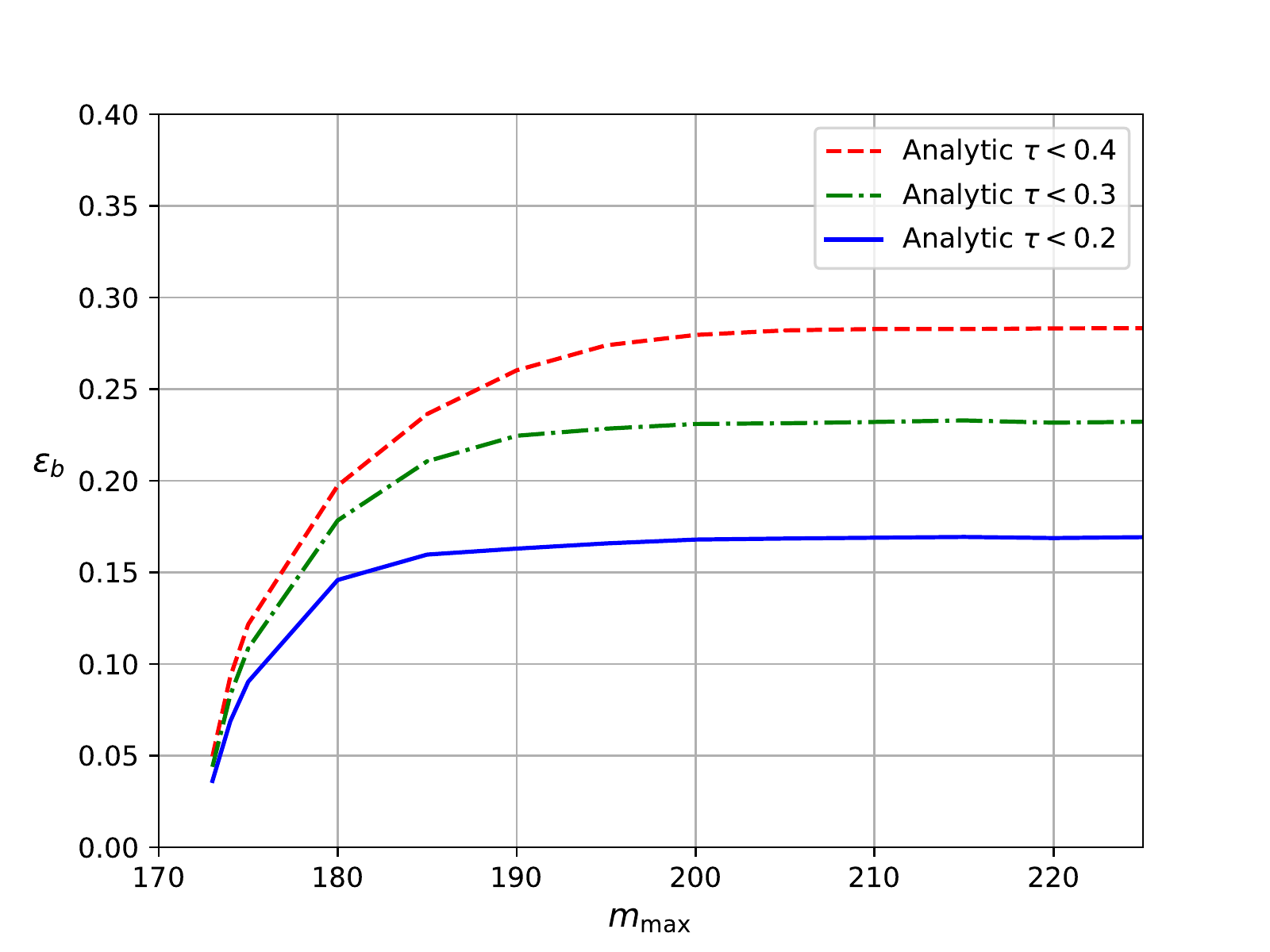}
\caption{Analytic, groomed with Soft Drop $(\beta=2)$.}
\end{subfigure}%
~
\begin{subfigure}[t]{0.48\textwidth}
\centering
\includegraphics[width=1\textwidth]{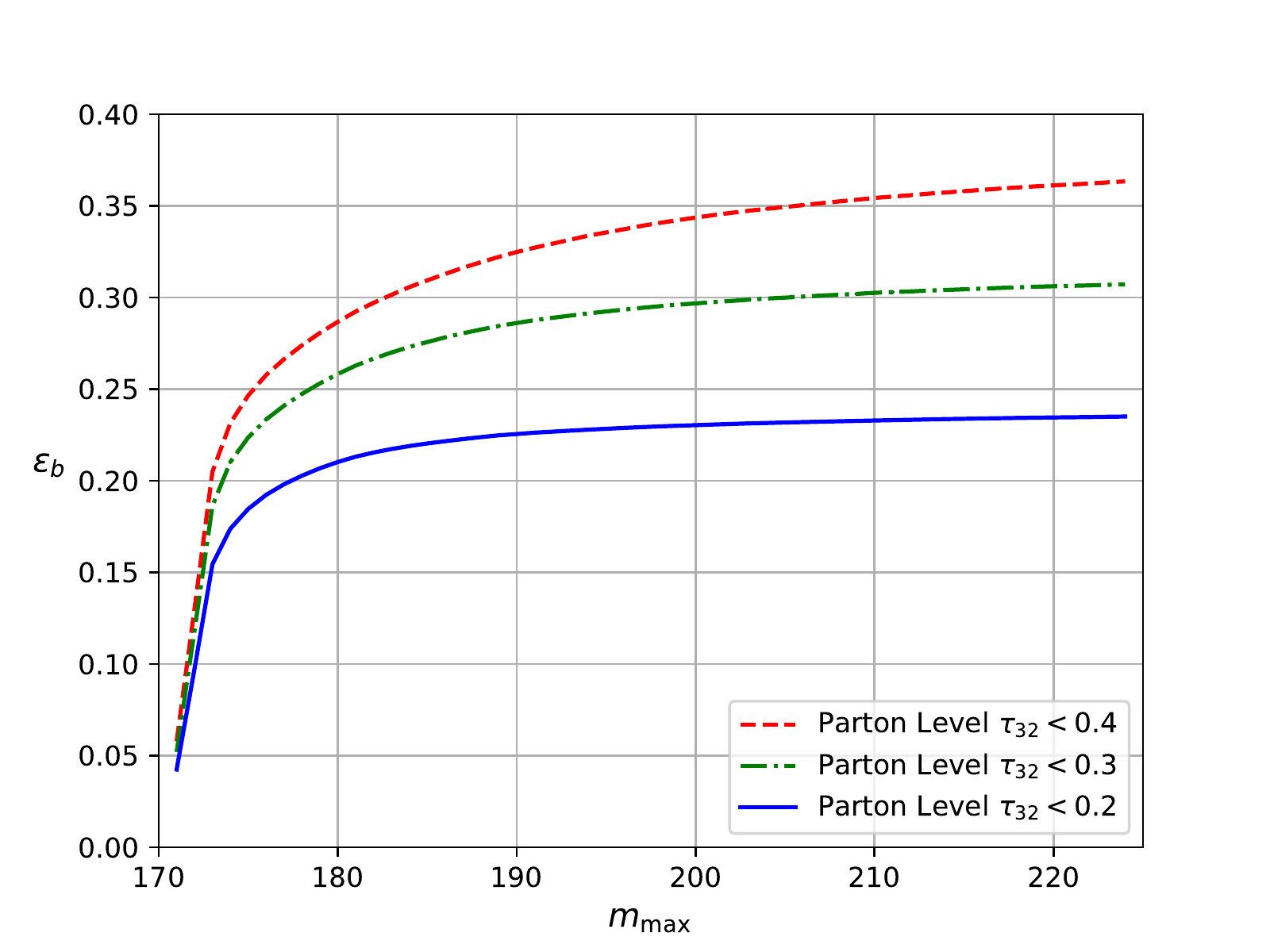}
\caption{Parton level Monte Carlo, groomed with Soft Drop $(\beta=2)$.}
\end{subfigure}
\begin{subfigure}[t]{0.48\textwidth}
\centering
\includegraphics[width=1\textwidth]{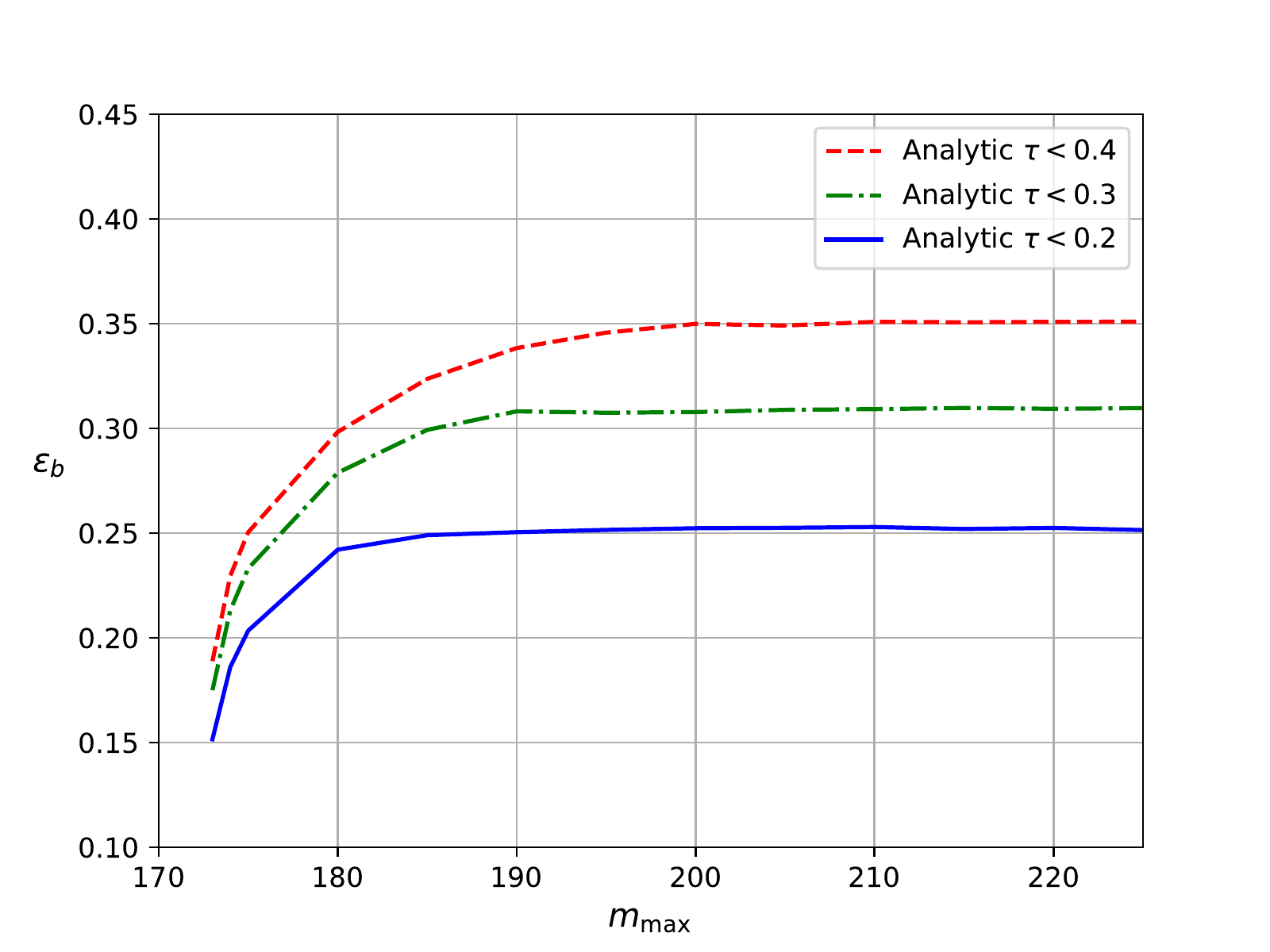}
\caption{Analytic, groomed with mMDT.}
\end{subfigure}%
~
\begin{subfigure}[t]{0.48\textwidth}
\centering
\includegraphics[width=1\textwidth]{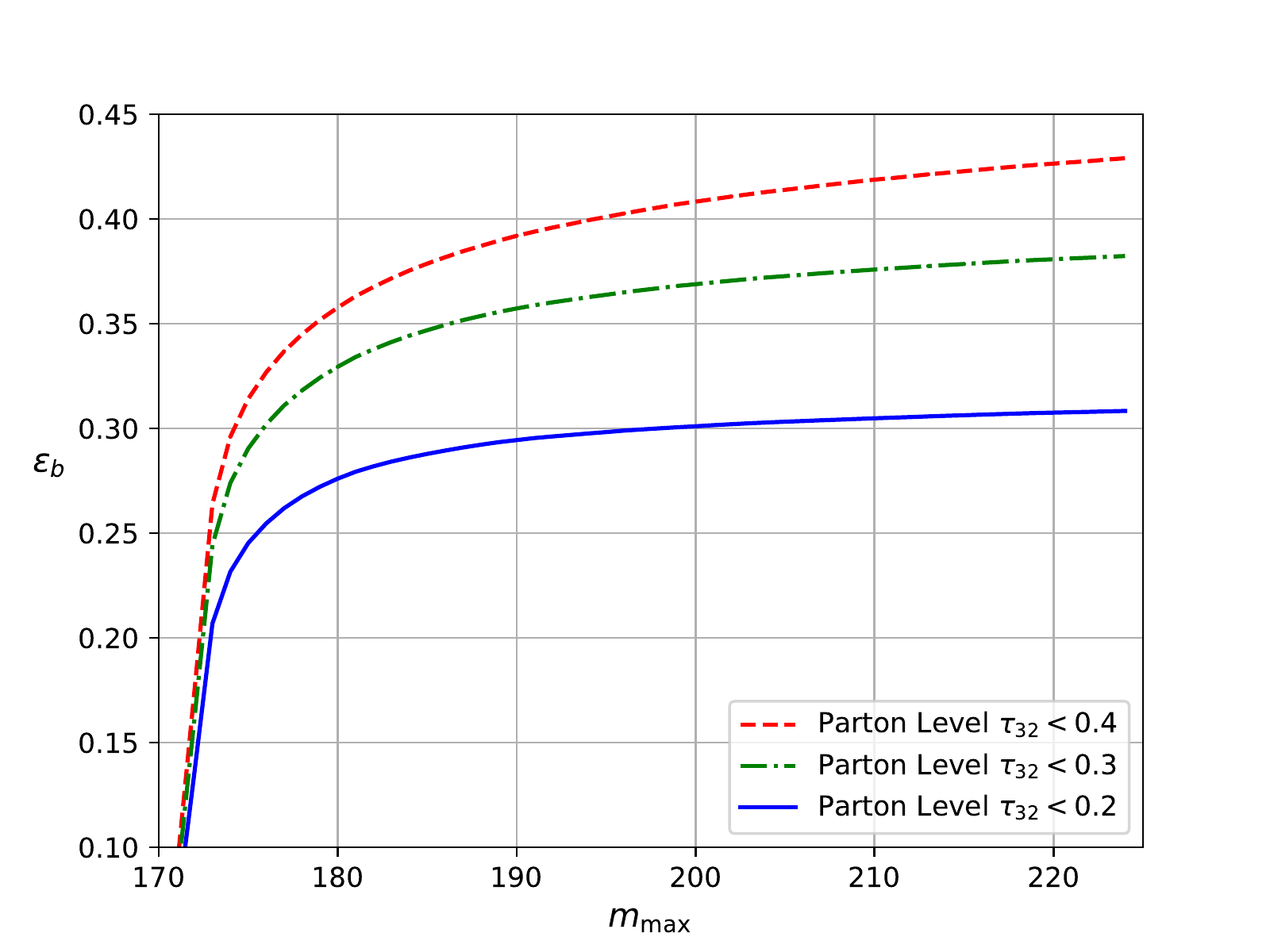}
\caption{Parton level Monte Carlo, groomed with mMDT.}
\end{subfigure}
\caption{Analytic and Monte Carlo parton level curves showing how the signal tagging rate varies with $m_{\text{max}}$.}
\label{fig:AnalyticTopMass}
\end{figure}

In this section we discuss a notable feature of our calculations in terms of the differences between signal and background jets. As suggested by  Eq.~\eqref{eq:TauToMass}, one can reduce the cut on jet mass $m_\text{max}$, without impacting the signal until we reach a critical value depending on $\tau$. Until we reach this point reducing $m_\text{max}$ results in a decrease in the background tagging rate and hence an increase in performance. While our analytic studies are somewhat simplified and in particular neglect subleading terms, it is interesting to study the extent to which our observations may apply to parton shower studies when subleading effects are present. Figure \ref{fig:AnalyticTopMass} shows, using both analytic calculations (left) and parton level MC simulations (right), how the signal tagging rate varies with $m_{\text{max}}$ for several fixed $\tau$ cuts both without grooming and with grooming via Soft Drop and the mMDT.

For the signal distribution the overall shape and dependence on $\tau$ is well described by our calculation, although, as before, there is some difference in the overall normalisation. The difference between our calculation and the distribution derived from MC worsens for smaller values of $m_{\text{max}}$, which should be expected, as non-perturbative effects, which can not be completely removed from parton shower simulations, will start to play more of a role in this region. While the signal tagging rate derived from MC simulations does not flatten off to the same extent as the analytic calculations do as $m_{\text{max}}$ is increased, it is clear that beyond a certain value of $m_{\text{max}}$  the signal efficiency depends only very weakly on  $m_{\text{max}}$ . 

\begin{figure}[ht]
\centering
\begin{subfigure}[t]{0.48\textwidth}
\centering
\includegraphics[width=1\textwidth]{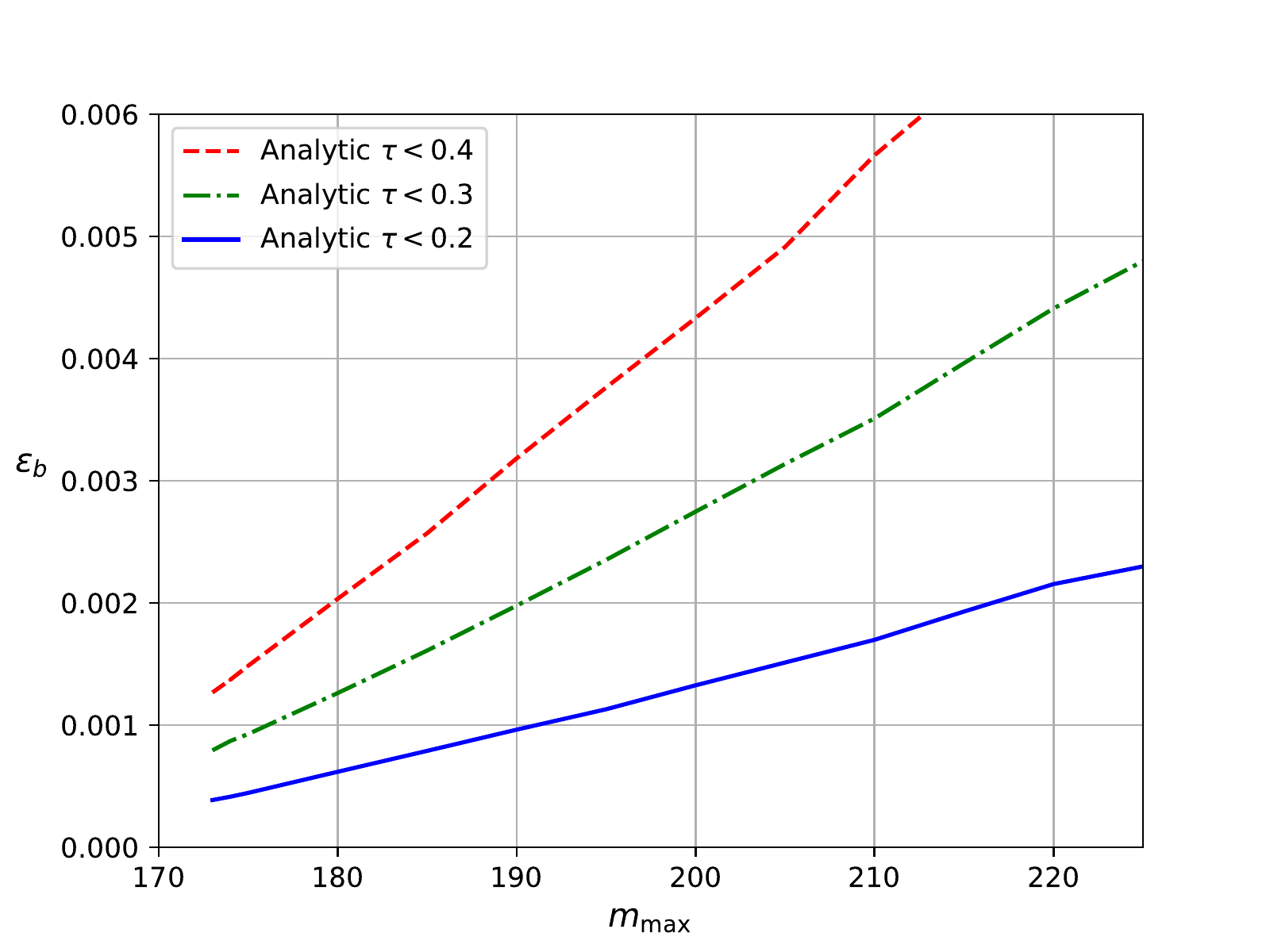}
\caption{Analytic, un-groomed.}
\end{subfigure}%
~
\begin{subfigure}[t]{0.48\textwidth}
\centering
\includegraphics[width=1\textwidth,]{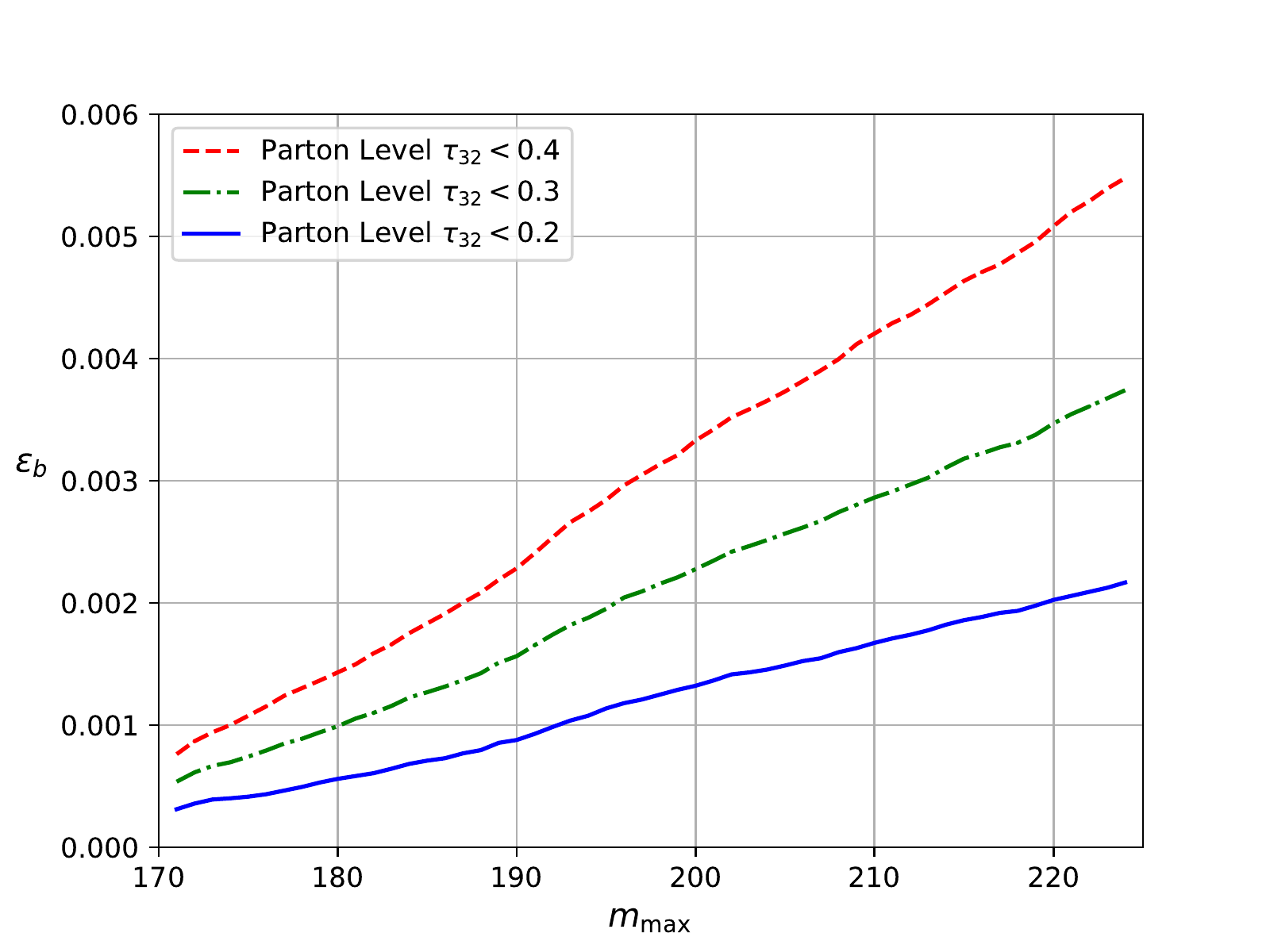}
\caption{Parton level Monte Carlo, un-groomed.}
\end{subfigure}
\begin{subfigure}[t]{0.48\textwidth}
\centering
\includegraphics[width=1\textwidth]{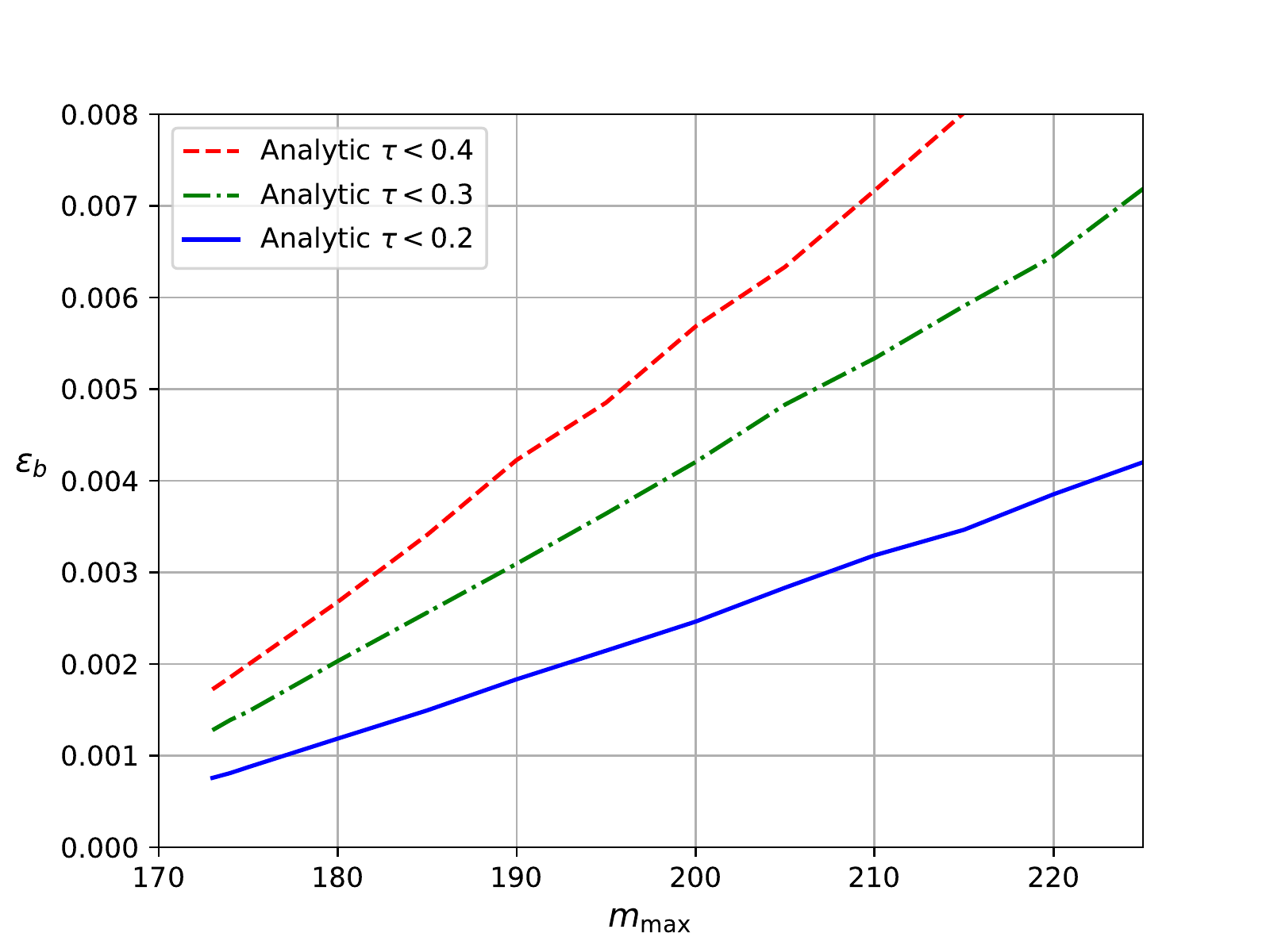}
\caption{Analytic, groomed with Soft Drop $(\beta=2)$.}
\end{subfigure}%
~
\begin{subfigure}[t]{0.48\textwidth}
\centering
\includegraphics[width=1\textwidth]{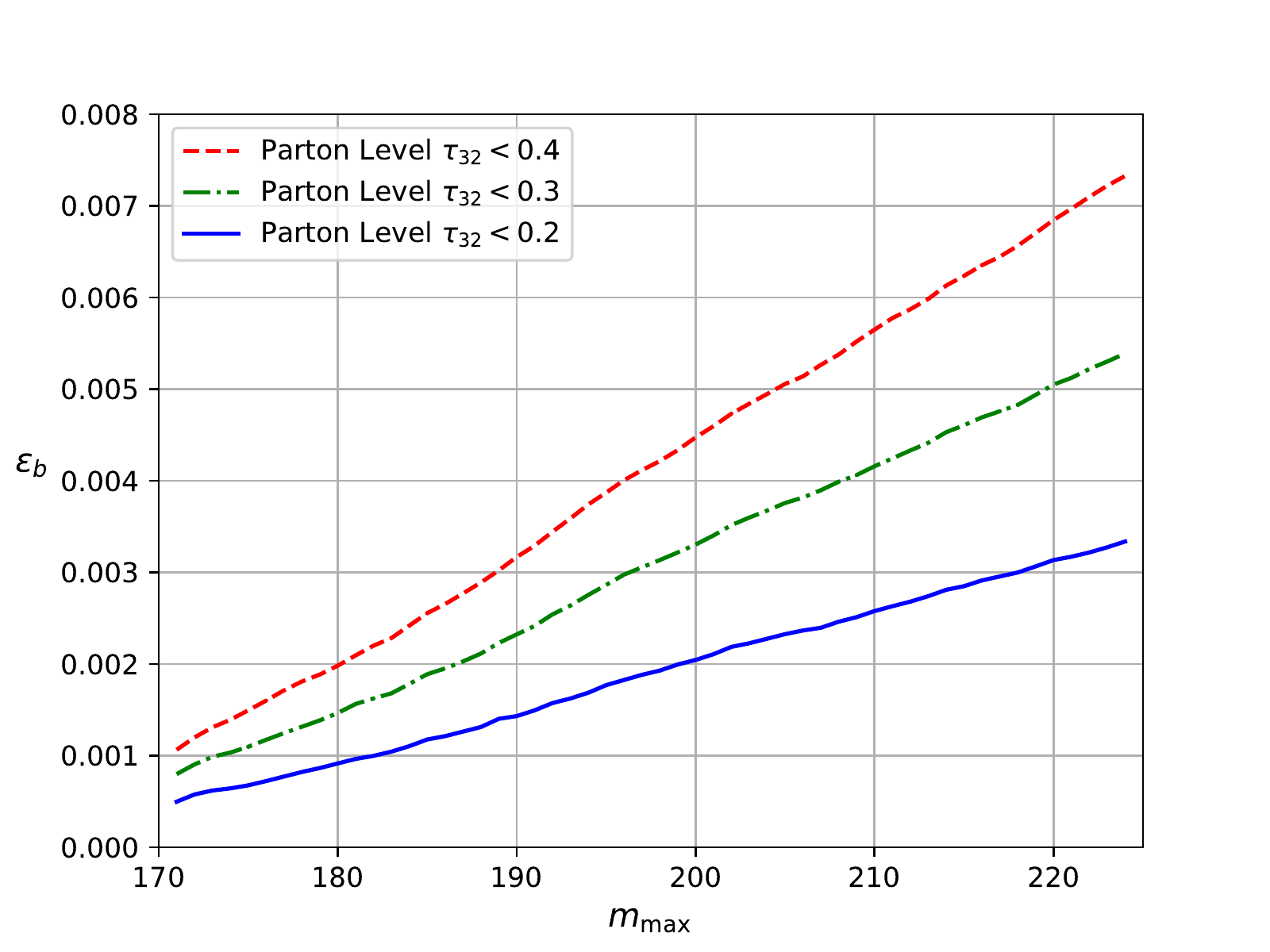}
\caption{Parton level Monte Carlo, groomed with Soft Drop $(\beta=2)$.}
\end{subfigure}
\begin{subfigure}[t]{0.48\textwidth}
\centering
\includegraphics[width=1\textwidth]{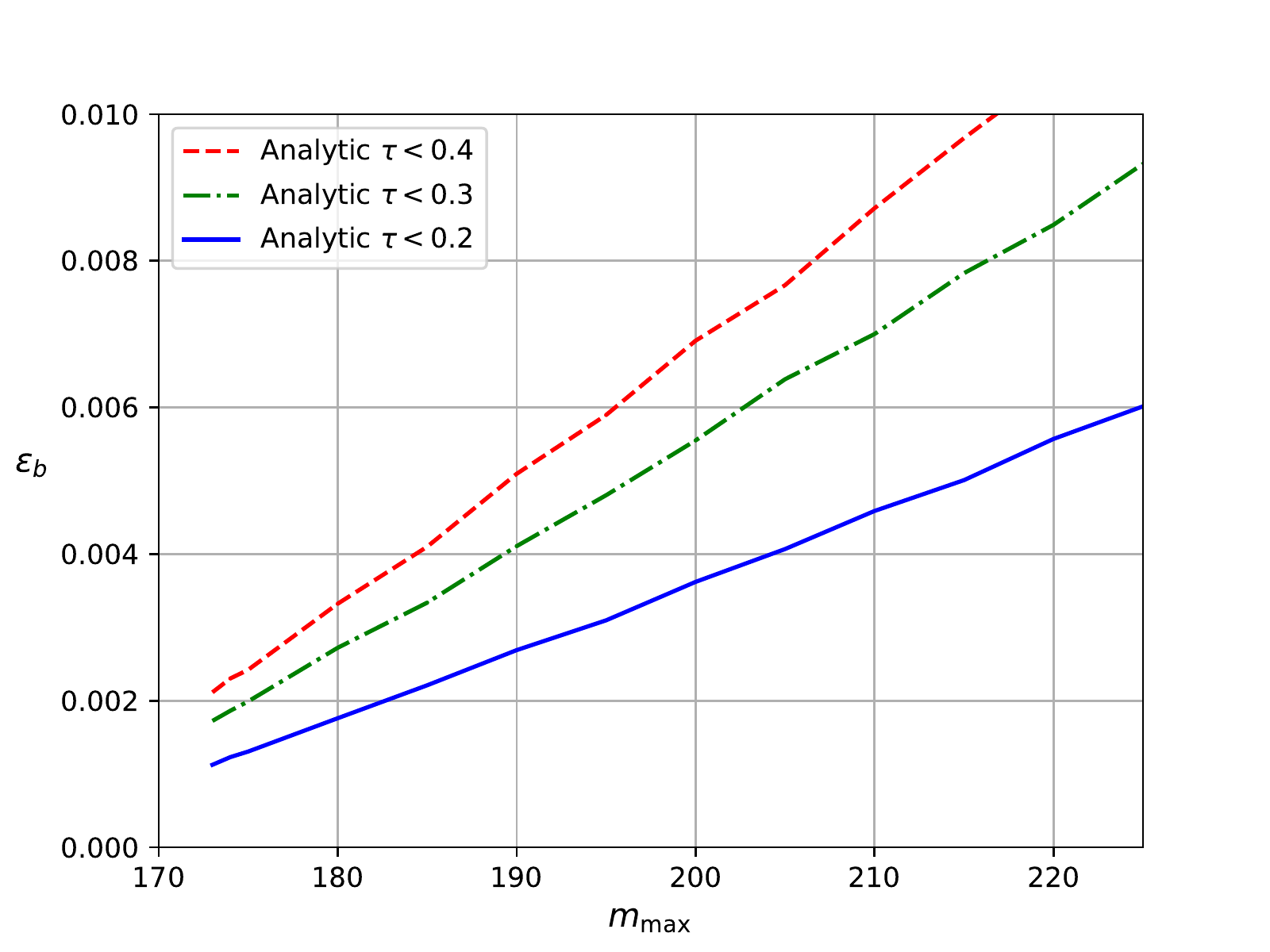}
\caption{Analytic, groomed with mMDT.}
\end{subfigure}%
~
\begin{subfigure}[t]{0.48\textwidth}
\centering
\includegraphics[width=1\textwidth]{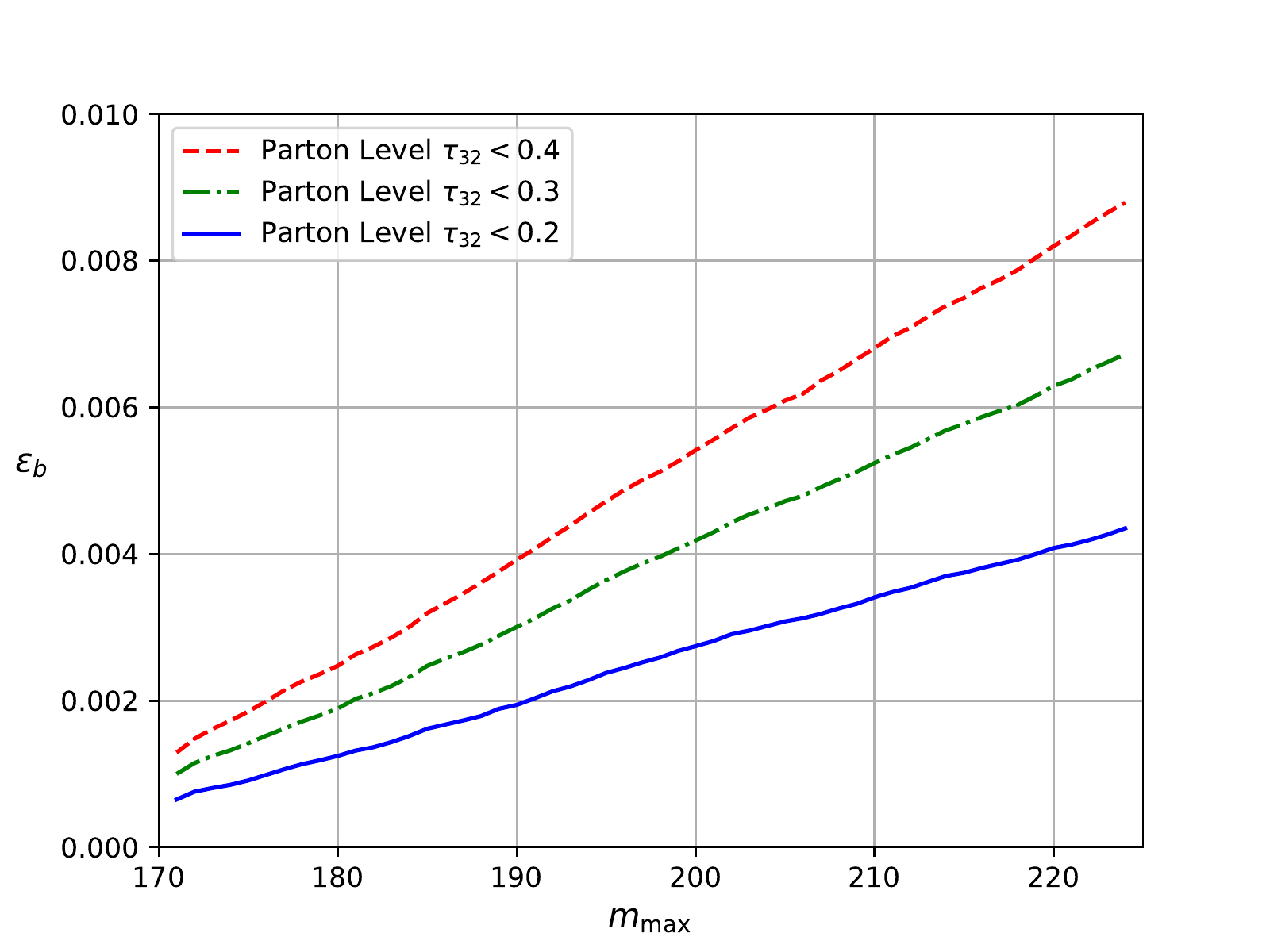}
\caption{Parton level Monte Carlo, groomed with mMDT.}
\end{subfigure}
\caption{Analytic and Monte Carlo (parton level) curves showing how the background tagging rate varies with $m_{\text{max}}$.}
\label{fig:AnalyticQuarkMass}
\end{figure}

Figure \ref{fig:AnalyticQuarkMass} shows similar plots for the case of quark jets. Our analytic predictions are again seen to be in overall good agreement with the Pythia shower capturing the $m_\text{max}$ and $\tau$ dependences. It is notable that the jet mass constraint affects the background tagging rate in the same way for any cut on $\tau_{32}$, as there are not two competing scales in the Sudakov factor. This opens up the possibility to improve the performance of the tagging procedure by reducing $m_{\text{max}}$ so that the signal tag rate remains approximately constant whilst removing a significant portion of the background.

\begin{figure}[h]
\centering
\includegraphics[width=0.7\textwidth]{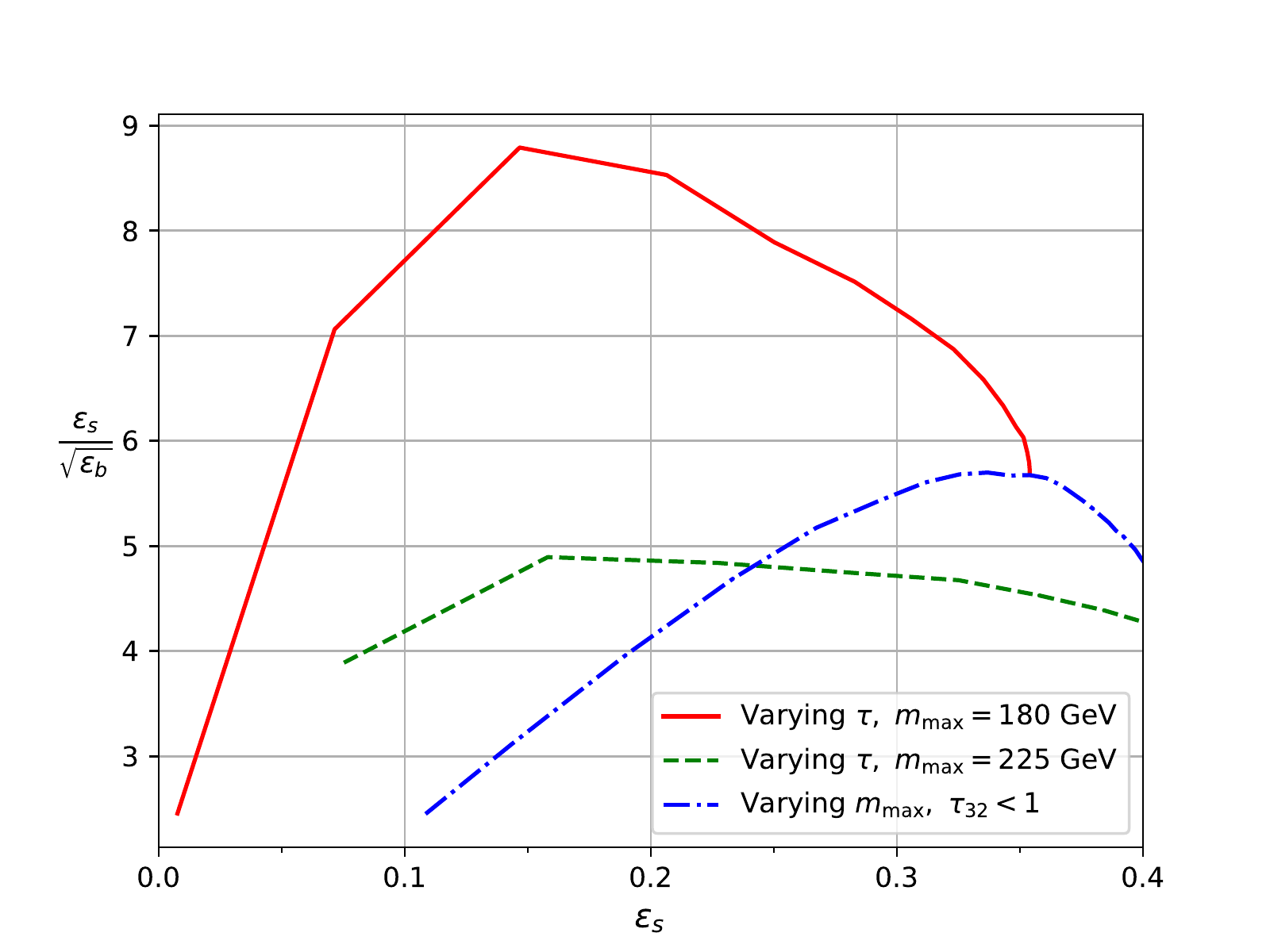}
\caption{Signal significance against efficiency for three variations on the tagging procedure. All jets are groomed with mMDT and tagged with \Ym. Either $\tau$ or $m_{\text{max}}$ is varied with a fixed cut placed on the other. The samples were produced using Pythia with hadronisation and UE activated.}
\label{fig:SignificanceComparison}
\end{figure}

One may wonder, given the effectiveness of a tight cut on the jet mass, what improvement is gained by cutting on $\tau_{32}$ in these circumstance. Figure \ref{fig:SignificanceComparison} also shows a curve generated by varying $m_{\text{max}}$ over the range $173$ GeV to $225$ GeV, but with no cut on $\tau_{32}$. In this case the signal significance is higher than cutting on $\tau_{32}$ with $m_{\text{max}}=225$ GeV, but cutting on $\tau_{32}$ with $m_{\text{max}}=180$ GeV is still the highest performing tagging procedure.

\FloatBarrier

\begin{figure}[h]
\centering
\includegraphics[width=0.7\textwidth]{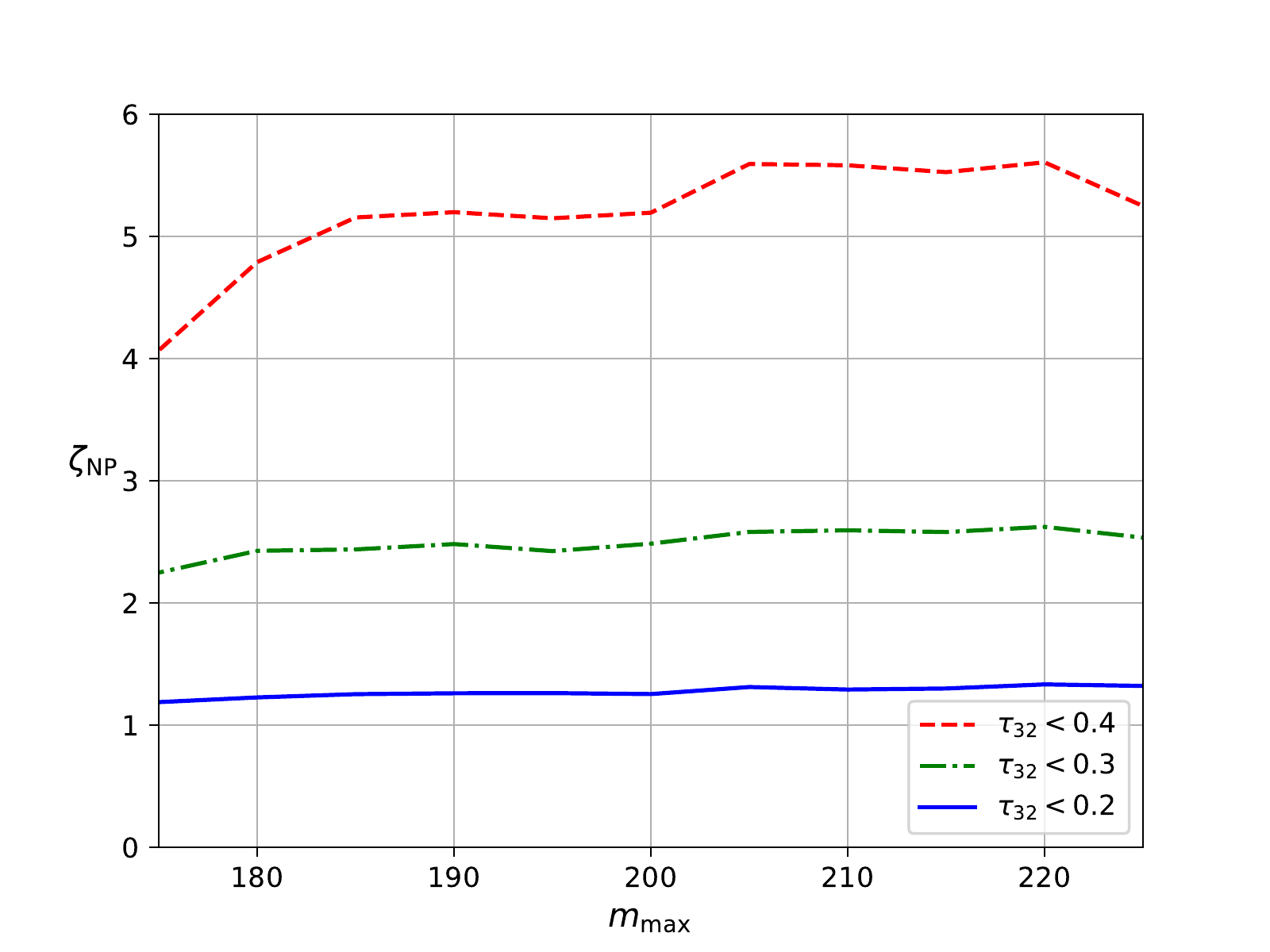}
\caption{A measure of resilience to non perturbative effects as $m_{\text{max}}$ varies for three cuts on $\tau_{32}$. Jets are pre-groomed with mMDT.}
\label{fig:Resilience}
\end{figure}

We now investigate the impact of non-perturbative corrections on this tagging procedure as $m_{\text{max}}$ is varied. Figure \ref{fig:Resilience} shows the resilience \cite{Proceedings:2018jsb} to non-perturbative effects, defined as $\zeta_{\text{NP}}=\left(\frac{\Delta\epsilon_s^2}{\langle \epsilon_s\rangle^2}+\frac{\Delta\epsilon_b^2}{\langle \epsilon_b\rangle^2}\right)^{\sfrac{-1}{2}}$ where $\Delta\epsilon$ is the difference between the parton and hadron level tagging efficiency and $\langle \epsilon\rangle$ is the mean of the two, for jets pre-groomed with mMDT, as $m_{\text{max}}$ is varied, for three different values of $\tau$. To construct the resilience 10 million $q\bar{q}$ events and 1 million $t\bar{t}$ events were generated at both parton and hadron level using Pythia. From figure \ref{fig:Resilience} we see that the resilience to non-perturbative effects does not strongly depend on $m_{\text{max}}$ in the range considered, even with $m_{\text{max}}$ as low as $180$ GeV. By contrast, reducing the cut on $\tau_{32}$ from $0.4$ to $0.2$ results in a marked drop in resilience. It would therefore be beneficial, in terms of reducing the impact of non-perturbative effects, to take $\tau$ not too small, say $\tau=0.4$, while imposing a rather tight cut on the jet mass to provide the discriminating power. These cuts provide a signal significance of around $6$ with a signal efficiency around $0.35$. This is both a higher signal efficiency and significance than was reported in section \ref{sec:pheno} with $m_{\text{max}}=225$ GeV and $\tau=0.2$, the highest significance achieved with the higher value of $m_{\text{max}}$.

\section{Conclusions}\label{sec:conclusions}
In this article we have studied top-tagging from first principles of QCD, as part of a larger program to understand the features of tagging and grooming methods in a model-independent fashion. 
We chose a combination of methods, starting from the application of a prong-finding step aimed at tagging three-pronged decays and rejecting background, followed by a radiation constraining step. We also pre-groom jets with both the mMDT and with Soft Drop with $\beta=2$ to reduce non-perturbative contributions. For prong finding we have used \Ym, an adaptation of Y-splitter introduced in Ref.~\cite{Dasgupta:2016ktv} while as a radiation constraining shape variable we have applied the N-subjettiness ratio $\tau_{32}$ with $\beta=2$. While our specific choices (use of \Ym for prong-finding and $\beta=2$ for $\tau_{32}$) are helpful in somewhat simplifying analytical studies, combinations similar to the ones used here have commonly been employed, including for experimental studies involving top-tagging \cite{Aad:2014xra,ATLAS:2015nkq,CMS:2016tvk}. 

We started by carrying out Monte Carlo studies which provided some of the motivation for what followed in terms of yielding information on performance, resilience to non-perturbative effects, and optimal parameter choices for our combination of methods. Next we turned to studying QCD background jets. Here we have built on previous work \cite{Dasgupta:2018emf} on understanding top-tagging, and in particular \Ym, by including the constraint from $\tau_{32}$. We have derived results for the double differential distribution in jet mass and $\tau_{32}$ as well as for the cumulant where we integrate over $\tau_{32}$ with the condition $\tau_{32} < \tau$. We obtained a result in the limit of small $\tau$ and then included finite $\tau$ corrections along similar lines to the studies in Ref.~\cite{Napoletano:2018ohv}. We also performed studies both with and without pre-grooming with mMDT and Soft Drop ($\beta=2$). We compared our results to those from both Herwig and Pythia showers and in all cases we saw that our analytical calculations are able to capture the essential impact of the tagging, shape-variable and grooming steps.

We  then turned to studying signal jets. We found that in the highly boosted limit a mass window constraint gives rise to a simple Sudakov form factor which is in good agreement with Pythia results. We then added \Ym as in Ref.~\cite{Dasgupta:2018emf}, but improved upon previous calculations by also considering a situation, at order $\alpha_s$ , where a soft gluon can be one of the prongs resolved by \Ym. Including this contribution we found the results to be in significantly  better agreement with Pythia than was the case with previous results where such a correction was not considered \cite{Dasgupta:2018emf}. We then studied the impact of a $\tau$ cut, including finite $\tau$ effects and also considering pre-grooming with mMDT and Soft Drop $\beta=2$. In all cases, in spite of the complexity of the problem, our simplifying approximations were sufficient to capture, the basic behaviour, i.e the $\tau$ dependence over a wide range in $\tau$, seen also with parton showers. Remaining differences with parton showers were at a level that was consistent with our expectations from missing subleading terms.  One immediately exploitable outcome of our analytic results was the suggestion that using a tighter mass cut than our default choice (for a given $\tau$) would reduce the background rather than the signal  while not significantly affecting the resilience to non-perturbative effects. This finding was used to show how a highly performant and resilient method could be developed using our combination of tools.

Finally we would say that although the combination of methods we have considered gives rise to a highly non-trivial observable, we have demonstrated that analytical methods can still give substantial insight into the basic physics mechanisms that control the performance of such tool combinations.  Further systematic improvements on the results we have obtained are possible, with the inclusion of subleading logarithmic terms being one avenue to pursue. Also, while our specific choice of tools is helpful for analytical studies and was taken mainly for convenience, combinations similar to the ones here have been in widespread use, and have not been analytically understood thus far. We believe that our studies should therefore encourage analytical investigations of other similar combinations, including for example variants using $\tau_{32}^{(\beta=1)}$,  and help develop a more complete picture of the distinct role played by the different elements and/or steps that form part of a number of top tagging methods.

\section*{Acknowledgements}
We would like to thank Gregory Soyez and Kiran Ostrolenk for helpful discussions. This work has been funded by the UK Science and Technologies Facilities Council under grant ST/P000800/1. JH thanks the UK Science and Technologies Facilities Council (STFC) for a PhD studentship award.

\section*{Appendix A}
 Here we explain the impact of the different approximations on emissions $\rho_a$ and $\rho_b$, which enter the pre-factor  for the resummed expressions in the small $\tau$ limit and for finite $\tau$.  In order to illustrate this we examine the differential distribution in a fixed-coupling approximation and at order $\bar{\alpha}^3$, which is the first order at which $\tau_{32}$ is non-zero.
 The finite $\tau$ result truncated at order $\bar{\alpha}^3$ is (from Eq.~\eqref{eq:taudist}):
 \begin{equation}
\frac{\rho \tau}{\sigma} \frac{\sd^2\sigma}{\sd \rho\sd\tau}\overset{\tau<1/2}{=} \bar{\alpha}^2 \frac{1}{1-\tau} \int_\zeta^1\frac{\sd z_a}{z_a} \int_\zeta^1\frac{\sd z_b}{z_b}   \int_0^\rho \frac{\sd\rho_a}{\rho_a} \frac{\rho}{\rho-\rho_a} \Theta \left(\rho_a > \frac{1-\tau}{2-\tau}  \rho \right) \Theta_{\rho_\text{min}} R'((\rho-\rho_a)\tau),
\end{equation}
 where in a fixed-coupling approximation, and to leading-logarithmic accuracy, we have,
 \begin{equation}
 R'((\rho-\rho_a)\tau) = \bar{\alpha} \left( \ln \frac{1}{\rho-\rho_a}+\ln \frac{1}{\tau} \right).
 \end{equation}
 We remind the reader that 
 \begin{equation}
 \Theta_{\rho_\text{min}} = \Theta \left(\min  \left((\rho-\rho_a)(1-\tau),z_a z_b\max \left(\frac{\rho_a}{z_a},\frac{(\rho-\rho_a)(1-\tau)}{z_b} \right)\right)>\rho_{\text{min}}\right),
  \end{equation}
 where $\frac{\rho_a}{z_a} = \theta_a^2$  and $ \frac{(\rho-\rho_a)(1-\tau)}{z_b}=\frac{\rho_b}{z_b}=\theta_b^2$.  For our illustrative purposes, let us take a  specific contribution to the pre-factor that arises from the region 
 $\theta_a^2 \gg \theta_b^2$ and $\rho_b =(\rho-\rho_a)(1-\tau) < z_b \rho_a$. We also assume values of the parameters so that $\frac{\rho_\text{min}}{\rho}> \zeta$, again purely as part of our illustrative example \footnote{Strictly speaking the condition is $\frac{\rho_\text{min}}{\rho}> \frac{\zeta}{1+\zeta}$.}. The conclusions we derive will  apply to other  regions of phase-space and parameter ranges too. For the region of phase space considered  we obtain the contribution
 \begin{multline}
\label{eq:taudist2}
\frac{\rho \tau}{\sigma} \frac{\sd^2\sigma}{\sd \rho\sd\tau}\overset{\tau<1/2}{=} \bar{\alpha}^2 \frac{1}{1-\tau} \int_\zeta^1\frac{\sd z_a}{z_a} \int_\zeta^1\frac{\sd z_b}{z_b}   \int_0^\rho \frac{\sd\rho_a}{\rho_a} \frac{\rho}{\rho-\rho_a} \Theta \left(\rho_a > \frac{1-\tau}{2-\tau}  \rho \right) R'((\rho-\rho_a)\tau) \\
\Theta \left( (\rho-\rho_a)(1-\tau) > \rho_{\text{min}} \right) \Theta \left(  \frac{\rho_a}{z_a}>\frac{(\rho-\rho_a)(1-\tau)}{z_b}\right) \Theta \left((\rho-\rho_a)(1-\tau) <z_b  \rho_a   \right).
\end{multline}

The directly computed small $\tau$ limit result, corresponding to the result in section \ref{sec:smalltau}, is 

\begin{equation}
\frac{\rho \tau}{\sigma} \frac{\sd^2\sigma}{\sd \rho\sd\tau}=\bar{\alpha}^2 \int_\zeta^1 \frac{\sd z_a}{z_a} \int_\zeta^1\frac{\sd z_b}{z_b}\int_0^\rho \frac{\sd \rho_b}{\rho_b} \Theta \left (\rho_b > \rho_{\mathrm{min}} \right) \Theta\left(\rho_b < z_b \rho \right) \times \Theta \left(\rho>\frac{z_a}{z_b} \rho_b\right)\bar{\alpha}\left( \ln \frac{1}{\tau}+\ln \frac{1}{\rho_b} \right)
\end{equation}
which gives
\begin{equation}
\label{eq:smalltauacubed}
\frac{\rho \tau}{\sigma} \frac{\sd^2\sigma}{\sd \rho\sd\tau} = \frac{1}{2}\bar{\alpha}^2 \ln \frac{1}{\zeta} \ln^2 \frac{\rho}{\rho_{\text{min}}}  \times \bar{\alpha}\left( \ln  \frac{1}{\rho_{\mathrm{min}}}+ \ln \frac{1}{\tau}+\frac{1}{3} \ln \frac{\rho_{\mathrm{min}}}{\rho} \right).
\end{equation}

Evaluating  the small  $\tau$ limit of the exact result in our chosen configuration i.e. Eq.~\eqref{eq:taudist2}, gives:
\begin{equation}
\frac{\rho \tau}{\sigma} \frac{\sd^2\sigma}{\sd \rho\sd\tau} = \frac{1}{2}\bar{\alpha}^2 \ln \frac{1}{\zeta} \ln^2 \frac{\rho}{\rho_{\text{min}}}  \times \bar{\alpha}\left( \ln  \frac{1}{\rho_{\mathrm{min}}}+ \ln \frac{1}{\tau}+\frac{1}{3} \ln \frac{\rho_{\mathrm{min}}}{\rho} \right)+\frac{3}{4} \bar{\alpha}^3 \zeta(3) \ln \frac{1}{\zeta},
\end{equation}
which differs from Eq.~\eqref{eq:smalltauacubed} by a highly subleading $\bar{\alpha}^3 \zeta(3) \ln \frac{1}{\zeta}$ term.
The full $\tau$ dependent result from  Eq.~\eqref{eq:taudist2} reads 
 \begin{equation}
\frac{\rho \tau}{\sigma} \frac{\sd^2\sigma}{\sd \rho\sd\tau}= \frac{\bar{\alpha}^3}{1-\tau} \left[\frac{1}{2}\ln \frac{1}{\zeta} \ln^2 \frac{\rho}{\rho_{\text{min}}}  \times \left( \ln  \frac{1}{\rho_{\mathrm{min}}} +\ln \frac{1}{\tau}+\frac{1}{3} \ln \frac{\rho_{\mathrm{min}}}{\rho} +\ln \left(1-\tau \right)\right) - \ln \frac{1}{\zeta} \,   \text{Li}_3\left(\frac{1}{\tau-1} \right) \right].
 \end{equation}

We note that retaining the finite $\tau$ effects results in the appearance of three features : Firstly there is the overall $1/(1-\tau)$ multiplicative term, which has a significant impact on the result beyond the small $\tau$ region and  is important to retain. Secondly there is a $\ln(1-\tau)$ term in addition to the large logarithms we resum. Given that we do not resum logarithms of $1-\tau$, and indeed focus on the region $\tau \sim 0.2$, this constitutes a  negligible contribution relative to the logarithms we resum, dominated by the $\ln \frac{1}{\rho_{\mathrm{min}}}$ term. Finally there is a highly subleading $\bar{\alpha}^3 \ln \frac{1}{\zeta}$ term accompanied by a trilogarithm in $1-\tau$ which we can safely neglect. Hence ignoring the $\tau$ dependence in the pre-factor, other than the $1/(1-\tau)$ term, is a valid approximation for our work.

\section*{Appendix B}
Here we provide an alternate derivation of $\rho\frac{\sd \Sigma(\tau)}{\sd\rho}$ that directly derives this distribution as opposed to integrating the double differential. We can start from the standard factorised formula for any number of emissions, similar to Eq. \eqref{eq:finiteTauStart} but instead of fixing $\tau_{32}$ we set an upper bound $\tau_{32}<\tau$:
\begin{multline}
\frac{\rho }{\sigma} \frac{\sd\sigma}{\sd \rho\sd\tau}=\bar{\alpha}^2 \int_\zeta^1\frac{\sd z_a}{z_a}\int_0^1 \frac{\sd\rho_a}{\rho_a}\int_\zeta^1\frac{\sd z_b}{z_b}\int_0^{\rho_a} \frac{\sd\rho_b}{\rho_b}\Theta(\min(\rho_b,z_az_b\max(\frac{\rho_a}{z_a},\frac{\rho_b}{z_b}))>\rho_{min}) \\ \exp\left[-\int_0^1 R(\rho')\frac{\sd \rho'}{\rho'}\right] \sum_{p=1}^{\infty}\frac{1}{p!}\prod_{i=1}^p\int_{0}^{\rho_b}R'(\rho_i)\frac{\sd \rho_i}{\rho_i} \rho\delta(\rho-\rho_a-\rho_b-\sum_{i\neq a,b}\rho_i)\, \Theta((\rho-\rho_a)(1-\tau)<\rho_b).
\end{multline}
The delta function can now be used to do the sum over emissions labelled with $i$, where it is crucial to notice that for any $i$, $\rho_i<\rho-\rho_a-\rho_b$ as implied by the delta function, and that the upper limit on $\rho_i$ of $\rho_b$ is weaker than this for $\tau<\frac{1}{2}$. Using the standard jet mass result for the sum over emissions, as we did before, we arrive at
\begin{multline}
\frac{\rho}{\sigma} \frac{\sd\sigma}{\sd \rho}\overset{\tau<1/2}{=} \bar{\alpha}^2 \int_\zeta^1\frac{\sd z_a}{z_a} \int_\zeta^1\frac{\sd z_b}{z_b}   \int_{\frac{1-\tau}{2-\tau}  \rho}^\rho \frac{\sd\rho_a}{\rho_a} \int_{(1-\tau)(\rho-\rho_a)}^{\min(\rho-\rho_a,\rho_a)} \frac{\sd\rho_b}{\rho_b} \frac{\rho}{\rho-\rho_a-\rho_b} \Theta_{\rho_\text{min}} \left(\rho_a,\rho_b,\tau,  \rho_{\text{min}} ,z_a,z_b\right) \\  R'(\rho-\rho_a-\rho_b)\frac{\exp[-R(\rho-\rho_a-\rho_b)-\gamma_E R'(\rho-\rho_a-\rho_b)]}{\Gamma[1+R'(\rho-\rho_a-\rho_b)]}.
\end{multline}
We now wish to integrate over $\rho_b$ which we note contains two regions, $\rho_a<\rho-\rho_a$ and $\rho_a>\rho-\rho_a$, the former of which vanishes if we neglect the $\tau$ dependence of the $\rho_a$ integral limits as in section \ref{sec:tau_Cumulative} \footnote{The neglected term is proportional to $\alpha^3L^2L_{\rho}$ where $L_{\rho}$ is a log of $\rho$ or $\rho_{\text{min}}$ while $L$ is a log of the ratio or $\zeta$. This is clearly beyond our accuracy.}. Enforcing the condition $\rho_b>\rho_{\text{min}}$, which is embodied in $\Theta_{\rho_\text{min}}$ gives an upper limit on $\rho_a$ of $\rho-\frac{\rho_{\text{min}}}{1-\tau}$, which, within our accuracy we can approximate as $\rho$. To carry out this integral within single logarithmic accuracy, we can expand the radiator about some fixed $\rho_b$ which we take as $(\rho-\rho_a)(1-\tau_0)$so that:
\begin{equation}
R(\rho-\rho_a-\rho_b)\simeq R((\rho-\rho_a)\tau_0)-R'((\rho-\rho_a)\tau_0)\ln\left(\frac{\rho-\rho_a-\rho_b}{(\rho-\rho_a)\tau_0}\right) + \mathcal{O}(R''),
\end{equation} 
where $\tau_0$ should be chosen close to $\tau$ as values of $\rho_b$ close to $(\rho-\rho_a)(1-\tau)$ are expected to dominate the integral. The integral can then be carried out to give 
\begin{multline}\label{eq:alternateDerivation}
\frac{\rho}{\sigma} \frac{\sd\sigma}{\sd \rho}\overset{\tau<1/2}{=} \bar{\alpha}^2 \int_\zeta^1\frac{\sd z_a}{z_a} \int_\zeta^1\frac{\sd z_b}{z_b}   \int_{\frac{1}{2} \rho}^{\rho} \frac{\sd\rho_a}{\rho_a}\frac{\rho}{\rho-\rho_a}  \Theta_{\rho_\text{min}} \left(\rho_a,\rho_{\text{min}} ,z_a,z_b\right) \left(\frac{\tau}{\tau_0} \right)^{R'((\rho-\rho_a)\tau_0)}\\  {}_2F_1(1,R'(\rho-\rho_a)\tau_0,1+R'(\rho-\rho_a)\tau_0,\tau) \frac{\ \exp[-R((\rho-\rho_a)\tau_0)-\gamma_E R'((\rho-\rho_a)\tau_0)]}{\Gamma[1+R'((\rho-\rho_a)\tau_0)]},
\end{multline}
in perfect agreement with Eq. \eqref{eq:tauint}. 

Although less convenient for making contact with the result reported in Eq. \eqref{eq:tauint}, we could equally well have integrated over $\rho_a$, leaving the $\rho_b$ integral to be done numerically, as we could have done in section \ref{sec:tau_differential}. To do this one would expand $R(\rho-\rho_a-\rho_b)$ around $\rho_a=\rho-\frac{\rho_b}{1-\tau}$ which would lead to:
\begin{multline}
\frac{\rho}{\sigma} \frac{\sd\sigma}{\sd \rho}\overset{\tau<1/2}{=} \bar{\alpha}^2 \int_\zeta^1\frac{\sd z_a}{z_a} \int_\zeta^1\frac{\sd z_b}{z_b}   \int_{\rho_{min}}^{\frac{\rho}{2}} \frac{\sd\rho_b}{\rho_b}\frac{\rho}{\rho-\rho_b} \left( \frac{\tau(1-\tau_0)}{(1-\tau)\tau_0}\right)^{R'(\rho_b\frac{\tau_0}{1-\tau_0})} \Theta_{\rho_\text{min}} \left(\rho_b,\rho_{\text{min}} ,z_a,z_b\right) \\  {}_2F_1(1,R'(\rho_b\frac{\tau_0}{1-\tau_0}),1+R'(\rho_b\frac{\tau_0}{1-\tau_0}),\frac{\rho_b\tau_0}{(1-\tau_0)(\rho-\rho_b)}) \frac{\ \exp[-R(\rho_b\frac{\tau_0}{1-\tau_0})-\gamma_E R'(\rho_b\frac{\tau_0}{1-\tau_0})]}{\Gamma[1+R'(\rho_b\frac{\tau_0}{1-\tau_0})]},
\end{multline}
where again, $\tau_0$ should be taken close to $\tau$, and any $\tau$ dependence in the leading order pre-factor has been neglected.

\bibliographystyle{JHEP}
\bibliography{top-paper-2021.bib}

\end{document}